\documentclass[12pt]{emulateapj}
\usepackage{natbib}
\usepackage{color}
\usepackage{hyperref} 
\hypersetup{colorlinks=true,citecolor=blue}

\def \bea {\begin{eqnarray}}
\def \ena {\end{eqnarray}}               
\def \bee {\begin{equation}}
\def \ene {\end{equation}}
\def    \simlt  {\lower.5ex\hbox{$\; \buildrel < \over \sim \;$}}
\def    \simgt  {\lower.5ex\hbox{$\; \buildrel > \over \sim \;$}}
\def	\gtsim	{\simgt}

\newcommand     \mum    {\,\mu{\rm m}}  

\def	\cm		{\,{\rm {cm}}}

\def	\km		{\,{\rm {km}}}
\def	\B		{{\rm B}}

\def	\erg		{\,{\rm {erg}}}
\def	\eV		{\,{\rm {eV}}\,}
\def    \exp 		{\,{\rm {exp}}}
\def	\g		{\,{\rm g}}

\def	\Hz		{\,{\rm {Hz}}}

\def	\K		{\,{\rm K}}
\def    \kB    		{k_{\rm B}}
\def	\AU		{\,{\rm {AU}}}
\def	\pc		{\,{\rm {pc}}}

\def	\s		{\,{\rm s}}
\def	\sr		{\,{\rm {sr}}}
\def    \ln  		{\,{\rm {ln}}}

\def	\IP		{\rm {IP}}

\def	\H		{\rm H}

\def    \Bv     	{\bf  B}

\def 	\bE		{{\bf E}}

\def	\gas		{\rm {gas}}

\def \bol {\rm {bol}}

\def	\gr		{\rm gr}
\def	\d		{\rm d}
\def	\rad		{\rm {rad}}

\def    \abs     	{\rm {abs}}
\def    \sca     	{\rm {sca}}

\def    \ext    	{\rm {ext}}

\def	\dep		{\rm {dep}}

\def	\tot		{\rm {tot}}

\def    \coll        	{\rm {coll}}

\def	\IGM		{\rm {IGM}}

\def	\sp		{\rm {sp}}
\def	\ion		{\rm {ion}}
\def	\coll		{\rm {coll}}

\begin{document}
\title{On origin and destruction of relativistic dust and its implication for ultrahigh energy cosmic rays}
\author{Thiem Hoang\altaffilmark{1,}\altaffilmark{2,}\altaffilmark{3}, A Lazarian\altaffilmark{4}, and R Schlickeiser\altaffilmark{1}}

\altaffiltext{1}{Institut f$\ddot{\rm u}$r Theoretische Physik, Lehrstuhl IV: Weltraum- und 
Astrophysik, Ruhr-Universit$\ddot{\rm a}$t Bochum, D-44780 Bochum, Germany}
\altaffiltext{2}{Canadian Institute for Theoretical Astrophysics, University of Toronto, 60 St. George Street, Toronto, ON M5S 3H8, Canada;\href{mailto:hoang@cita.utoronto.ca}{hoang@cita.utoronto.ca}}
\altaffiltext{3}{Current address: Institute of Theoretical Physics, Goethe  Universit$\ddot{\rm a}$t Frankfurt, D-60438 Frankfurt am Main, Germany}
\altaffiltext{4}{Department of Astronomy, University of Wisconsin-Madison, Madison, WI 53705, USA}

\begin{abstract}
Dust grains may be accelerated to relativistic speeds by radiation pressure, diffusive shocks, and other acceleration mechanisms. Such relativistic grains have been suggested as primary particles of ultrahigh energy cosmic rays (UHECRs). In this paper, we first revisit the problem of acceleration by radiation pressure and calculate maximum grain velocities achieved. We find that grains can be accelerated to relativistic speeds with Lorentz factor $\gamma < 2$ by powerful radiation sources, which is lower than earlier estimates showing that $\gamma$ could reach $\sim 10$. We then investigate different destruction mechanisms for relativistic grains traversing a variety of environments. In solar radiation, we find that the destruction by thermal sublimation and Coulomb explosions is important. We also quantify grain destruction due to electronic sputtering by ions and grain-grain collisions. Electronic sputtering is found to be rather inefficient, whereas the evaporation following grain-grain collisions is shown to be an important mechanism for which the $a \sim 0.01- 1\mum$ grains would be destroyed after sweeping a gas column $N_{\rm Coul}\sim 5\times 10^{19}-5\times 10^{20}\cm^{-2}$. Relativistic dust in the interstellar medium and intergalactic medium (IGM) would be disrupted by Coulomb explosions due to collisional charging after traversing a gas column $N_{\rm Coul} \sim 10^{17}\cm^{-2}$ unless grain material is very strong. We show that photoelectric emission by optical and ultraviolet background radiation is also significant for the destruction of relativistic dust in the IGM. The obtained results indicate that relativistic dust from galaxies would be destroyed before reaching the Earth's atmosphere and unlikely to account for UHECRs. 

\end{abstract}

\section{Introduction}\label{sec:intro}
During the last several years, we have witnessed significant progress in the understanding of ultrahigh energy ($E>10^{18} \eV$) cosmic rays (UHECRs) thanks to large collaborative research, including Pierre Auger Observatory (\citealt{2007Sci...318..938P}), HiRes \citep{2008APh....30..175A}, and Telescope Array \citep{2013EPJWC..5206002S}. However, the origin and nature of primary particles of UHECRs remains unclear (see \citealt{2006PhRvD..74d3005B} for a review). The idea of primary protons for UHECRs is challenged by the fact that highest energy protons, presumably coming from extragalactic sources, are likely decelerated via pion photoproduction due to their interactions with cosmic microwave background (CMB) photons, such that the cutoff of cosmic ray (CR) spectrum, namely GZK cutoff (\citealt{Greisen:1966ec}; \citealt{1966JETPL...4...78Z}), occurs at $\sim 6\times 10^{19}$ eV. This idea also contradicts with available measurements of CR compositions, which show a significant fraction of heavy nuclei.

Observational results from {\it Auger} array reveal that active galactic nuclei (AGNs), which are believed to be able to accelerate protons and nuclei to $E\sim 10^{21} \eV$, would be a potential source of UHECRs. Indeed, the predicted spectra based on AGNs are also in a good agreement with observational data (see \citealt{2006PhRvD..74d3005B}). Early Auger data \citep{2007Sci...318..938P} show a strong correlation of the direction of highest CR events with the position of AGNs, which implies some degree of anisotropy in the direction of UHECRs. However, \cite{2014arXiv1411.6111P} claimed that none of their tests shows a statistically significant evidence of the anisotropy previously reported. Thus, it is timely to revisit the original hypothesis of relativistic dust (i.e., dust moving with velocity $v> 0.1c$ where $c$ is the speed of light) as an origin of primary UHECRs.

\cite{Spitzer:1949bv} first pointed out that dust grains could be accelerated to speeds close to that of light by radiation pressure from supernovae. These relativistic grains are thought to gradually disintegrate into heavy nuclei (e.g., Mg, O, Si, Fe, C) due to collisions with gaseous atoms in the interstellar medium (ISM; see also \citealt{1954Tell....6..232A}). Later on, motivated by the detection of an extensive air shower (EAS) that requires primary particles of energy $E>10^{20}\eV$ \citep{1971PhRvL..27.1604S}, \cite{1972Ap&SS..16..238H} suggested that relativistic dust grains might be a potential candidate for UHECRs above $\sim 10^{20} \eV$. Relativistic dust was once proposed to explain cosmic gamma-ray bursts \citep{1974ApJ...187L..93G}.

Hayakawa's idea seems straightforward. A relativistic, spherical grain of size $a$ moving with Lorentz factor $\gamma=(1-\beta^{2})^{-1/2}$ with $\beta=v/c$ has kinetic energy $E_{\rm gr}(a,\gamma)=(\gamma-1) m_{\rm gr}c^{2} \approx 7.05\times 10^{19}(\gamma-1)/10\hat{\rho}a_{-5}^{3} \eV$, where $a_{-5}=a/10^{-5}\cm$, $\hat{\rho}=\rho/3\g\cm^{-3}$ with $\rho$ mass density within the grain, and $m_{\gr}=(4\pi/3) \rho a^{3}\approx 1.26\times 10^{-14}\hat{\rho}a_{-5}^{3} \g$ is the grain mass. Therefore, a relativistic grain with $a=0.1\mum$ and $\gamma \sim 10$ carries energy $E\sim 6.34\times 10^{19} \eV$, which is sufficient to account for superGZK particles.

Yet the hypothesis of relativistic dust as primary particles of UHECRs has not been well received, possibly, due to the following problems. First, relativistic grains might be destroyed in the solar radiation field before reaching the Earth's atmosphere by a process so-called Coulomb explosion (\citealt{1956Tell....8..268H}). This phenomenon occurs when the tensile strength caused by grain positive charge exceeds the maximum limit that the grain material can support. \cite{1973Ap&SS..21..475B} found that relativistic dust with $\gamma>30$ would be disrupted by Coulomb explosions in the solar radiation field due to photoelectric emission. {Collisional charging by the interstellar gas was also suggested by the authors to be important for Coulomb explosions for subrelativistic grains moving in the ISM.} \cite{1977Ap.....13..432E} proposed a survival mechanism for relativistic dust based on ion field emission from the grain surface. They estimated that iron grains between $a\sim 0.03-0.06\mum$ and $\gamma<360$ would not be disrupted by Coulomb explosions and could reach the Earth. {However, both \cite{1973Ap&SS..21..475B} and \cite{1977Ap.....13..432E} considered Coulomb explosions assuming a constant, low photoelectric yield ($Y_{\rm pe}= 0.1$), which did not take into account the emission of Auger and secondary electrons, and the dependence of yield on grain charge. This paper {considers} these effects.

Second, relativistic grains may be destroyed via radiative heating by strong solar radiation. \cite{1993Ap&SS.205..355M} showed that iron grains with $E\sim 10^{19}$ eV and $\gamma <360$ would melt but carbonaceous grains likely survive this radiation field. They also suggested that the iron grains with $E< 10^{19} \eV$ if coming from the direction opposite to the Sun would survive solar radiation. Their study adopted a constant, high emission efficiency ($\sim 0.5$) for all grain sizes, which is much higher than the exact value given by the Planck-averaged emission efficiency. As a result, the grain equilibrium temperature is largely underestimated. The present paper accounts for this effect. In addition, because it is unclear whether melting has any significant direct effect on the destruction of refractory dust, we will focus on a more general process so-called thermal sublimation.

\cite{1999APh....12...35B} revisited the problem of grain charging by photoelectric emission using inverse Compton scattering theory and derived a critical value $\gamma_{\rm cr}\sim 10^{4}$ below which relativistic grains would not be disrupted by Coulomb explosions. This critical value appears to be much higher than earlier estimates by \cite{1977ICRC....2..358B} and \cite{1977Ap.....13..432E}. It is noted that the Compton treatment in \cite{1999APh....12...35B}, which treats all electrons of atoms in the dust grain as free electrons, is only valid for $\gamma h\nu$ much larger than the bonding energy of atomic electrons. For K-shell electrons of iron grains with $E\sim 7000$ eV, this requires $h\nu > 10^{4} \eV$, or $\gamma \ge 10^{4}$ with solar photons. {\it Thus, an improved treatment of grain charging for $\gamma< 10^{4}$ is needed before any reliable conclusion on the survival of these grains can be made.} 

While {\it radiative destruction} by solar photons has been studied by a number of aforementioned authors, {\it collisional destruction} by collisions of ions (neutrals) from ambient plasma with relativistic grains is not yet studied in detail. In the frame of reference fixed to the relativistic grain, ions can have kinetic energy $E>1$ GeV. Such energetic ions can swiftly pass through an interstellar grain (size likely $a\le 1 \mum$) while transferring part of their energy to the grain's atoms. Most of the energy loss of ions in the grain is expended to produce electronic excitations and ionizations (\citealt{1963ARNPS..13....1F}; \citealt{1971RvMP...43..297I}). Energetic heavy ions can produce a large number of ionizations along the ion path, for which the Coulomb repulsive force between nearby ionized atoms of the grain can produce a damage track in the grain (\citealt{1965JAP....36.3645F}; \citealt{2009JPCM...21U4205I}). Heat {deposition by} secondary electrons can impulsively produce a hot cylinder track, and some atoms in the hot track near the grain surface can be ejected (sputtering) (see e.g., \citealt{2004ApJ...603..159B}), which may be important for grain destruction.

The destruction of relativistic grains by sputtering in the Galaxy was briefly discussed in \cite{1979Ap&SS..63..517D}. The study extrapolated the sputtering yield obtained from the {\it knock-on} sputtering regime (the low-energy regime $E<1$ MeV) for the $E>1$ GeV range. They obtained a rather low sputtering yield ($Y_{\sp}\sim 10^{-6}$) and concluded that sputtering is inefficient for the destruction of relativistic grains. However, modern understanding on sputtering indicates that, in the high-energy regime ($E>1$MeV), the {\it electronic} sputtering induced by electronic excitations is more important (see \citealt{Sigmund:2005tw} for a review). Thus, it is worth to quantify the efficiency of electronic sputtering and its destructive consequence for relativistic dust.

When a relativistic grain comes into the Earth's atmosphere, it will produce a huge EAS corresponding to the superposition of $N_{n}\sim {m_{\gr}}/{m_{\H}}\approx 6.0\times 10^{9}\hat{\rho}a_{-5}^{3}$ shower events initiated by nucleons of mass $m_{\H}$. This is because the mutual interaction between atoms within the grain is negligible compared to their kinetic energy, such that the grain's atoms can interact independently with atomic nuclei in the atmosphere. Observational measurements for the depth of maximum shower $X_{\max}$ (measured by the air mass traversed by the primary cosmic ray in units of $\g \cm^{-2}$ from above the atmosphere to the location of maximum number of secondary particles produced) suggest that relativistic grains could not reproduce air showers as observed \citep{1980ApJ...235L.167L}. {\cite{2000PhRvD..61h7302A} used Monte Carlo simulations to study the development of dust grain initiated air showers (DGAS), aiming to make comparison with the event recorded by Yakutsk array on May 7, 1989 that is considered the best candidate of DGAS. The study found that the air shower at Yakutsk might be explained by relativistic dust with energy $E=36-38$ EeV and $\log(\gamma)=4-3.8$. Appealing to the uncertainty in the survival of relativistic dust, the study suggested that relativistic dust not to be completely ruled out. Therefore, it is important to revisit this problem.}

Our major improvements for the relativistic dust idea in this paper include the followings. First, we calculate maximum grain velocities ($\gamma$) accelerated by radiation pressure, accounting for the redshift of radiation spectrum in the frame of relativistic grains. Second, we study in detail the interactions of photons and ions with relativistic grains moving with arbitrary $\gamma$, taking into account effects of incident photons, secondary and Auger electrons. For {\it radiative destruction}, we compute equilibrium temperatures for grains heated by the full spectrum of the radiation fields, taking into account {\it radiative, evaporative cooling}, and collisional heating by energetic ions passing through the grain. For Coulomb explosions, we calculate grain charges using improved photoelectric emission models and collisional ionization by ambient gas. Third, we investigate the new {\it collisional destruction} mechanisms due to electronic sputtering and grain-grain collisions, which are disregarded in earlier works. Last, we consider the destruction processes of relativistic grains moving in different astrophysical environments, including the solar system, ISM, and intergalactic medium (IGM).

The structure of the paper is as follows. In Section \ref{sec:source} we summarize the Lorentz transformations of radiation energy and photon density from a stationary to a comoving frame fixed to the dust grain and revisit the problem of radiation pressure acceleration that was suggested as a mechanism for producing relativistic dust. In Section \ref{sec:trans} we briefly discuss the interaction between photons and relativistic grains. Various interaction processes between relativistic grains and ions, electrons, and neutrals in plasma are discussed in Section \ref{sec:coll}. Grain heating by solar radiation and grain destruction by sublimation are investigated in Section \ref{sec:heat}. Collisional charging, photoelectric emission, and grain destruction by Coulomb explosions are studied in Section \ref{sec:charging}. The destruction of relativistic grains by electronic sputtering and grain-grain collisions is quantified in Section \ref{sec:sputt}. An extended discussion and summary are presented in Section \ref{sec:discussion} and \ref{sec:sum}.

\section{Origin of relativistic dust: Radiation Pressure Acceleration}\label{sec:source}

\subsection{Lorentz Transformations}\label{sec:Lorentz}

Let us begin with a short summary of the Lorentz transformations from a stationary frame (hereafter SF) to a comoving frame attached to the grain (hereafter GF) for some radiative quantities (see Appendix \ref{apdx:trans} for more details), which are needed for studying relativistic effects of dust both during the acceleration stage and their subsequent interactions with photons and gas atoms.

Assuming that a dust grain is moving with Lorentz factor $\gamma $ in a radiation field. Let $\mu=\cos\theta$ be the cosine angle between the direction of grain motion and that of photon propagation. Let $\nu, u({\nu})$ and $n({\nu})$ be the photon frequency, spectral energy density of radiation, and number density of photons in the SF. In the GF, these radiative quantities are denoted by $\nu', u(\nu')$, and $n(\nu')$, respectively. Throughout this paper, the prime denotes physical quantities in the comoving GF as usual. 

The photon frequency in the GF is shifted as follows:
\bea
\nu'=\gamma\nu(1-\beta \mu).\label{eq:nuGF}
\ena

For an {\it isotropic radiation field}, the spectral and total energy density in the GF are respectively given by
\bea
u'(\nu') d\nu' &=& u(\nu)d\nu \gamma^{2}\left(1+\frac{\beta^{2}}{3}\right),\label{eq:unu_iso}\\
u'_ {\rad}&=& u_{\rad} \gamma^{2} \left(1+\frac{\beta^{2}}{3}\right),\label{eq:urad_iso}
\ena
where $u_{\rad}=\int d\nu u(\nu)$.

Similarly, for a {\it point source of radiation}, we have the followings
\bea
u'(\nu')d\nu' &=& u(\nu)d\nu \gamma^{2}(1-\beta\mu_{\rm gr})^{2},\label{eq:unu_uni}\\
u'_ {\rad}&=& u_{\rad} \gamma^{2} (1-\beta\mu_{\rm gr})^{2}\label{eq:urad_uni},
\ena
where $\mu_{\rm gr}=1$ and $-1$  for the cases grains leaving and approaching the radiation source in the radial direction, respectively.

\subsection{Dust survival near luminous radiation sources}

Assuming that a central radiation source of bolometric luminosity $L_{\bol}$ suddenly ignites, dust grains in the surrounding medium are rapidly heated to temperature $T_{d}$ and may be destroyed by thermal sublimation.

{\cite{1989ApJ...345..230G} investigated the sublimation of dust grains using detailed balance and derived the sublimation rate for a grain of radius $a$:}
\bea
\frac{da}{dt'}=-n_{d}^{-1/3}\nu_{0}\exp\left(\frac{-B}{\kB T_d}\right),\label{eq:dasdt}
\ena
where $n_{d}$ is the atomic number density of dust, $B$ is the sublimation energy per atom, $\nu_0 = 2\times 10^{15} \s^{-1}$ and $B/\kB=68100 -20000N^{-1/3}\K $ for silicate grains,  $\nu_0 = 2\times 10^{14} \s^{-1}$ and $B/\kB=81200-20000N^{-1/3} \K $ for carbonaceous grains with $N$ being the total number of atoms of the grain (\citealt{1989ApJ...345..230G}; \citealt{2000ApJ...537..796W}).

Since the radiation flux decreases with increasing distance $r$ from the central source as $L_{\bol}/r^{2}$, grains close to the source can be heated above the sublimation temperature and then rapidly destroyed. The closest distance that dust can survive the sublimation is equal to
\bea
r_{\rm sub}=\left(\frac{L_{\rm UV}}{5\times 10^{12}L_{\odot}}\right)^{1/2}\left(\frac{T_{\rm sub}}{180
0\K}\right)^{-5.6/2} \pc,\label{eq:rsub}
\ena
where  $L_{\rm UV}$ is the luminosity in the optical and UV, which is roughly one half of the bolometric luminosity, $L_{\odot}$ is the solar luminosity, $T_{\rm sub}$ is the dust sublimation temperature (see \citealt{1995ApJ...451..510S}; also Appendix \ref{apdx:sub}). Above, the attenuation of radiation by dust extinction, which is small for dust near the sublimation zone, has been disregarded. In the following, we adopt $T_{\rm sub}=1500\K$ and $1800$K for silicate and graphite grains (see \citealt{1989ApJ...345..230G}). {With this choice of $T_{\rm sub}$, the $0.1\mum$ silicate and graphite grains located at $r_{\rm sub}$ can survive for $\sim 0.4$ yr and $\sim 3000$ yr, respectively (see Equations \ref{eq:tausub_sil} and \ref{eq:tausub_gra}).}

Luminous radiation sources (e.g., quasars and Seyfert galaxies) are frequently strong emitters of X-rays, which are expected to be important for grain destruction via Coulomb explosions. Nevertheless, it was estimated in \cite{1995ApJ...451..510S} that only the small $<0.01\mum$ grains are destroyed while larger grains would survive at $r\sim 1$pc.

\subsection{Maximum Grain Velocities}
Below we discuss grain acceleration by radiation pressure for two classes of most powerful radiation sources, including supernovae and active galactic nuclei (AGNs).
\subsubsection{Supernovae}

Supernovae (type Ia and type II) have typical luminosity $L_{\bol}\sim 10^{8}L_{\odot}$, for which grains are unlikely accelerated to $v\sim c$. In this {\it non-relativistic} case, the radiation pressure force acting on a stationary grain at distance $r$ from the central source is given by the usual formula:
\bea
F_{\rm rad}=\int d\nu c\frac{u({\nu})}{h\nu} \frac{h\nu}{c} Q_{\rm pr,\nu}\pi a^{2} =
u_{\rad}\langle Q_{\rm pr}\rangle \pi a^{2},\label{eq:Frad}
\ena
where $u({\nu})=L_{\nu}/(4\pi r^{2}c)$, $u_{\rad}=L_{\bol}/(4\pi r^{2}c)$, $Q_{\rm pr,\nu}$ is the radiation pressure efficiency by photon of frequency $\nu$, and 
\bea
\langle Q_{\rm pr}\rangle = \frac{\int Q_{\rm pr, \nu} u(\nu)d\nu}{\int u(\nu)d\nu},\label{eq:Qpr_avg}
\ena
is the radiation pressure efficiency averaged over the radiation spectrum seen by the grain (Appendix \ref{sec:Qabs_Qpr}).

Disregarding the minor effect of gravitational force on the grain, the increase of grain radial velocity in time interval $dt$ for the non-relativistic case is governed by:
\bea
\frac{m_{\rm gr}dv}{dt}=F_{\rm rad}=\frac{L_{\bol}}{4\pi r^{2}c}\langle Q_{\rm pr}\rangle \pi a^{2},
\label{eq:dvdt_rad}
\ena
where $F_{\rad}$ from Equation (\ref{eq:Frad}) has been used.

Using $dt=dr/v$ and integrating the above equation from the initial distance $r_{i}$ to $r$, we obtain
\bea
v^{2}=\frac{L_{\bol}}{2\pi c m_{\rm gr}}\langle Q_{\rm pr}\rangle \pi a^{2}\left(\frac{1}{r_{i}}-\frac{1}{r}\right),\label{eq:vRmax}
\ena
where the grain initially is assumed to be at rest. Equation (\ref{eq:vRmax}) shows that the grain velocity at $r\gg r_{i}$ is mainly determined by its initial distance $r_{i}$ and $L_{\bol}$.

Dust grains with $T_{\rm sub}=1800\K$ can survive beyond $r_{\rm sub}\sim 10^{16}\cm$ for $L_{\bol}\sim 10^{8}L_{\odot}$ (see Eq. \ref{eq:rsub}). Plugging $r_{i}=r_{\rm sub}$ and $L_{\bol}$ into Equation (\ref{eq:vRmax}), we obtain: 
\bea
v\simeq 0.1c\left(\frac{L_{\bol}}{10^{8}L_{\odot}}\right)^{1/2}\left(\frac{\langle Q_{\rm pr}\rangle}{1.5}\right)^{1/2} \left(\frac{r_{i}}{10^{16}\cm}\right)^{-1/2}a_{-5}^{-1/2}.~~~\label{eq:vRSN}
\ena

\subsubsection{Active galactic nuclei: Quasars or Seyfert Galaxies}
Quasars or Seyfert galaxies have typical luminosity $L_{\bol} \sim 10^{13} L_{\odot}$, which yield $r_{\rm sub}\sim 1$pc. Equation (\ref{eq:vRSN}) reveals that dust grains can achieve speeds $v=1.1c$! Such unphysical result directly arises from using Equation (\ref{eq:vRSN}) derived for the {\it non-relativistic} case.

In the {\it relativistic case}, the radiation pressure force acting on a grain is equal to 
\bea
F_{\rad} = \int d\nu' c\frac{u({\nu'})}{h\nu'}\frac{h\nu'}{c} Q_{\rm pr,\nu'}\pi a^{2} =u'_{\rad}\langle Q_{\rm pr}\rangle_{\gamma}\pi a^{2},~~~\label{eq:radforce}
\ena
where $\nu'$ is the frequency of photon and $u'_{\rad}=L'_{\bol}/(4\pi r^{2}c)$ is the energy density in the GF. Here $\langle Q_{\rm pr}\rangle_{\gamma}$ is given by Equation (\ref{eq:Qpr_avg}) with $\nu$ replaced by $\nu'$, which results in its dependence on $\gamma$.

The equation of motion for the radial velocity component of {\it relativistic particles} derived by \cite{1937MNRAS..97..423R} takes the following form:
\bea
\frac{m_{\rm gr}cd\tilde{u}}{dt'}&= &F_{\rad} w\left(1-w\tilde{u}\right),\label{eq:duds}
\ena
where $dt'$ is the proper time measured in the GF, $\tilde{u}=\gamma \beta$, and $w=\gamma-\tilde{u}=\gamma(1-\beta)$. Above, the isotropic reemission of dust in the GF has been assumed.

With $dr= c\beta dt = c\tilde{u} dt'$, Equation (\ref{eq:duds}) can be rewritten as  (cf. \citealt{1971Ap&SS..13...70N})
\bea
m_{\rm gr}cd\tilde{u}=\frac{L'_{\bol}}{4\pi r^{2}c}\langle Q_{\rm pr}\rangle_{\gamma}\pi a^{2}(\gamma-\tilde{u})\left(1-\gamma \tilde{u} +\tilde{u}^{2}\right)\frac{dr}{c\tilde{u}},\nonumber
\ena
or
\bea
\frac{\tilde{u}d\tilde{u}}{\left(\gamma-2 \tilde{u} -2\tilde{u}^{3} + 2\gamma \tilde{u}^{2}\right)}=\frac{L'_{\bol}}{4\pi m_{\rm gr}c^{3}}\langle Q_{\rm pr}\rangle_{\gamma}\pi a^{2}\frac{dr}{r^{2}}.\label{eq:dudr}
\ena


\begin{figure*}
\centering
\includegraphics[width=0.33\textwidth]{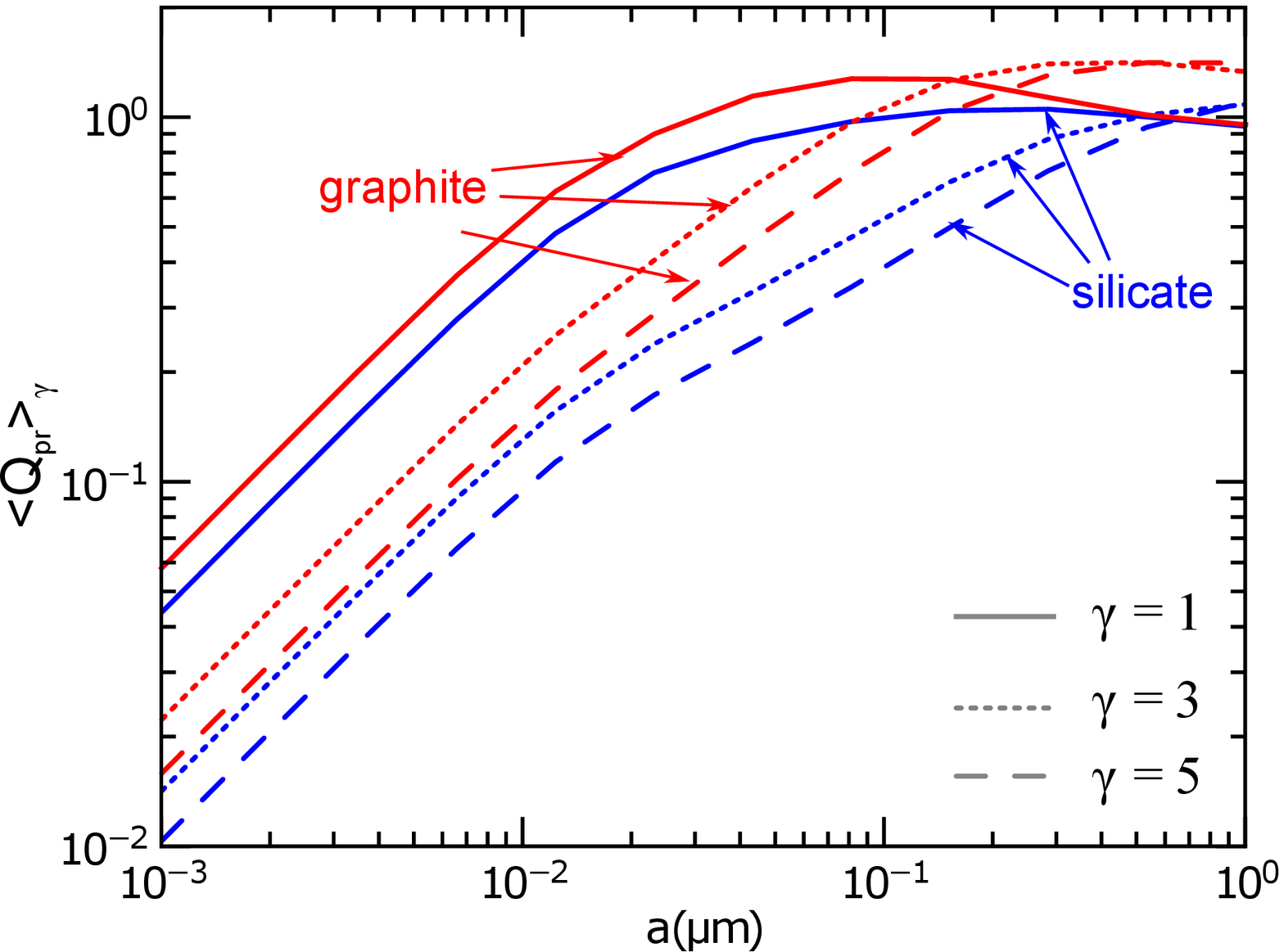}
\includegraphics[width=0.33\textwidth]{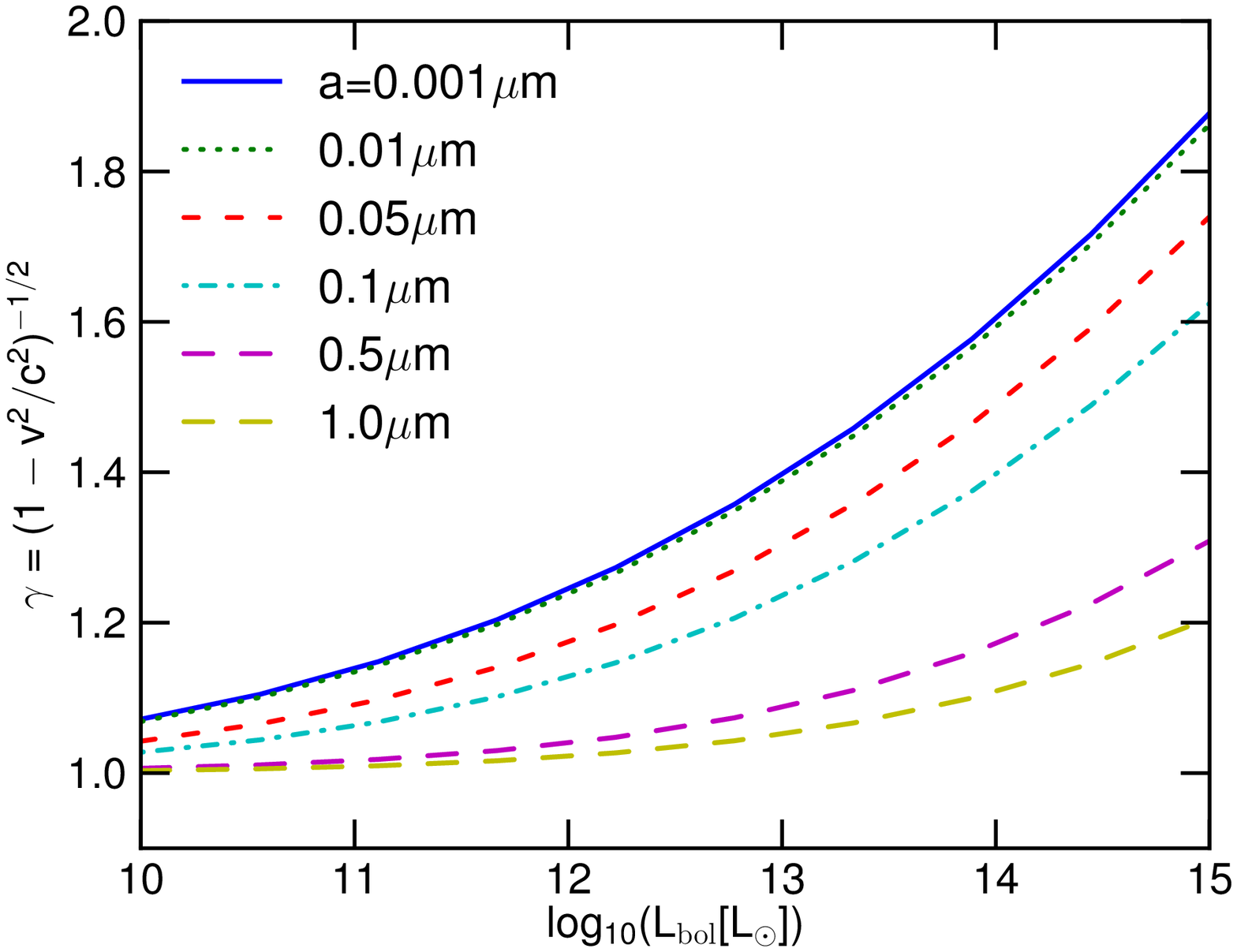}
\includegraphics[width=0.33\textwidth]{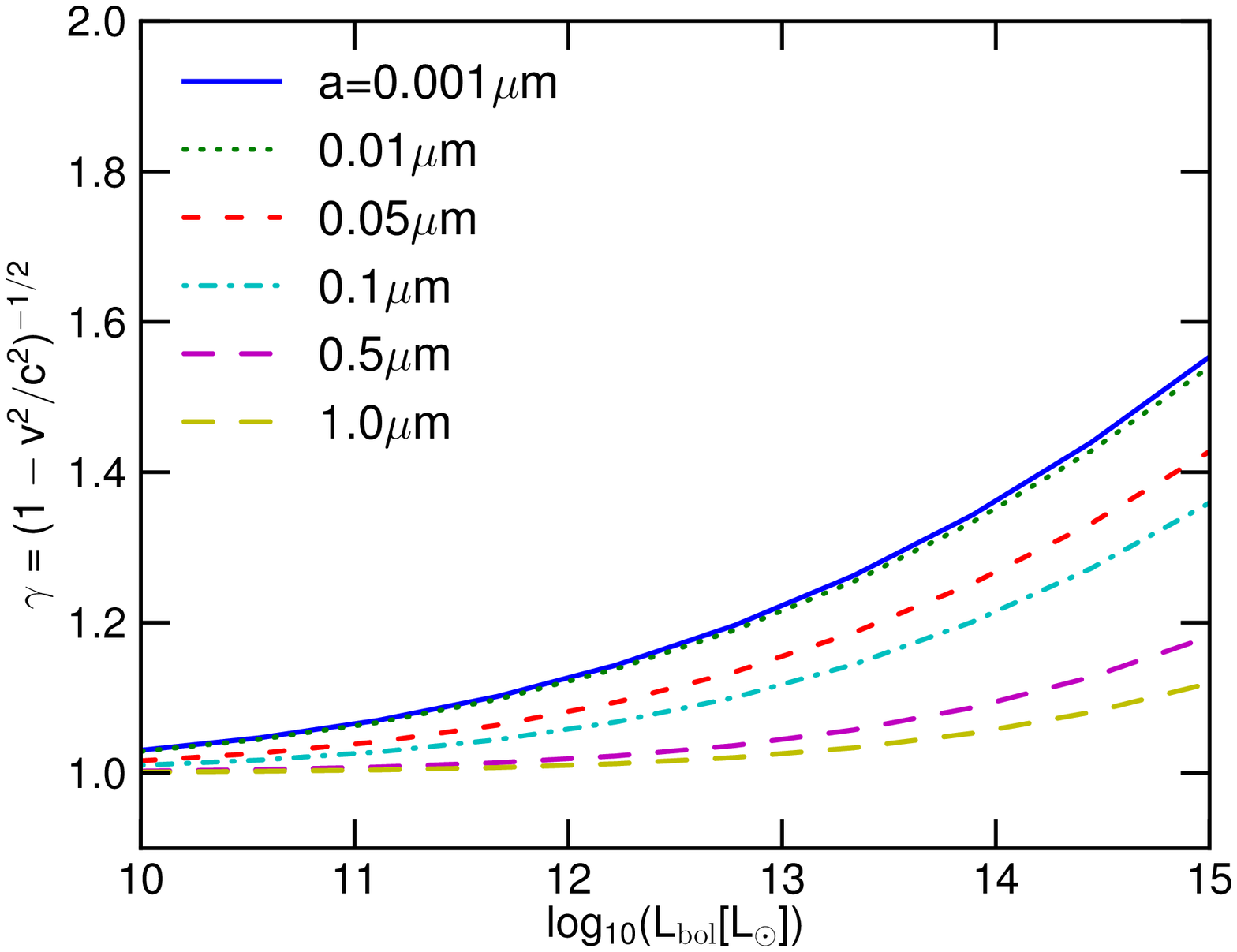}
\caption{{Left panel:} Averaged radiation pressure efficiency $\langle Q_{\rm pr}\rangle_{\gamma}$ versus grain size for different Lorentz factor $\gamma =\left(1-v^{2}/c^{2}\right)^{-1/2}$. {\it Middle and Right panels}: $\gamma$ of grains accelerated by radiation pressure as a function of the luminosity of the radiation source for different grain sizes for graphite (middle) and silicate (right).}
\label{fig:gamrad}
\end{figure*}

To find grain terminal velocities using Equation (\ref{eq:dudr}), first we substitute $L'_{\bol}=L_{\bol}\gamma^{2}(1-\beta)^{2}$ (see Equation \ref{eq:urad_uni}). We also compute $\langle Q_{\rm pr}\rangle_{\gamma}$ for different $\gamma$ using a typical radiation spectrum of unobscured AGNs (see \citealt{2006ApJ...645.1188W}, hereafter WDB06). Then, we solve Equation (\ref{eq:dudr}) numerically for $\tilde{u}$ as a function of $r$ with initial radius $r_{i}=r_{\rm sub}$ and final radius $r_{f}=20r_{\rm sub}$. We found that $\tilde{u}$ rapidly increases with $r$ and becomes saturated after several $r_{\rm sub}$. The terminal velocities represented through $\gamma$ for different grain sizes are shown in the middle and right panels of Figure \ref{fig:gamrad}. It can be seen that, for bright quasars or Seyfert galaxies of $L_{\bol}\sim 10^{14}-10^{15}L_{\odot}$, radiation pressure can accelerate dust grains to relativistic speeds with $\gamma <  2$. This terminal $\gamma$ is much lower than $\gamma \sim 10$ obtained by simple estimates in previous papers (e.g., \citealt{1972Ap&SS..16..238H}).

\subsection{Effects of Gas Drag During the Acceleration Stage}
Maximum grain velocities presented in the preceding subsection were obtained by disregarding the drag force arising from the ambient gas. To see its importance on grain deceleration, let us estimate the acceleration time by radiation pressure and compare it with the drag time.

The acceleration time by radiation pressure is given by
\bea
\tau_{\rm acc}&=&\frac{v}{|dv/dt|} =\frac{m_{\rm gr}v}{F_{\rm rad}}=\frac{m_{\rm gr}v 4\pi r^{2}c}{\langle Q_{\rm pr}\rangle \pi a^{2}L_{\bol}},\nonumber\\
&\simeq &1.1\times 10^{6}\hat{\rho}a_{-5}r_{\pc}^{2}\frac{1.0}{\langle Q_{\rm pr}\rangle}\left(\frac{v}{0.01c}\right)\left(\frac{L_{\bol}}{10^{13}L_{\odot}}\right)^{-1}\s,~~~~~\label{eq:tacc}
\ena
$r_{\pc}$ is the distance of the grain in units of $\pc$. Thus, grains can rapidly be accelerated to $v\sim 0.01c$ in $\sim 10^{6}\s$ for $L_{\bol}\sim 10^{13}L_{\odot}$.

For $v$ much larger than the thermal velocity of gaseous atoms under our interest, gas drag due to Coulomb collisions by ions is subdominant, and the drag force can be given by (see \citealt{2011ApJ...732..100D}) 
\bea
F_{\rm drag} =n_{\gas}\mu_{\gas} m_{\H}v^{2}\pi a^{2},\label{eq:dragcoll}
\ena
where $n_{\gas}$ and $\mu_{\gas}$ is the number density and mean molecular weight of the gas, respectively. This drag force is obtained by assuming that the collision between ions/atoms with the grain is completely inelastic such that incident particles transfer their entire momenta to the grain.

The drag time is equal to:
\bea
\tau_{\rm drag}&=&\frac{m_{\gr}v}{F_{\rm drag}}=\frac{m_{\gr}v}{n_{\gas}\mu_{\gas} m_{\H}v^{2}\pi a^{2}},\nonumber\\
&\simeq&7.9\times 10^{6}\hat{\rho}a_{-5}\mu_{\gas}^{-1}\left(\frac{n_{\gas}}{10^{4}\cm^{-3}}\right)^{-1}\left(\frac{v}{0.01c}\right)^{-1}\s~~~~~,\label{eq:tdrag}
\ena
which is simply equal to the time needed for the grain to collide with an amount of gas equal to the grain's mass. 

Equations (\ref{eq:tacc}) and (\ref{eq:tdrag}) yield $\tau_{\rm drag}/\tau_{\rm acc} \simeq 7.1(n_{\gas}/10^{4}\cm^{-3})^{-1}(v/0.01c)^{-2}(L_{\bol}/10^{13}L_{\odot})$. For the gas with $n_{\gas}<10^{4}\cm^{-3}$, $\tau_{\rm drag}/\tau_{\rm acc}\gg 1$, thus gas drag can be negligible during the acceleration stage up to $v\sim 0.01$c. Above $v\sim 0.01$c (energy $\sim 50$ keV),\footnote{Using the range of H ion $R_{\H}\simeq 0.01\hat{\rho}^{-1}(E/1{\rm keV})\mum$ (\citealt{1979ApJ...231...77D}) one can estimate the energy required for ion passing through the grain of size $a$ to be $E>10a_{-5}\hat{\rho}$ keV.} gas atoms simply pass through the grain (see Section \ref{sec:coll}), which results in drag force rapidly falling with increased velocity. In this case, grains can be accelerated to $v\sim c$ by radiation pressure as shown in Figure \ref{fig:gamrad}.  For $n_{\gas} > 10^{4}\cm^{-3}$, the gas drag force is dominant over the radiation force at $v\ll 0.01c$, and grains are obviously not accelerated to the relativistic speeds.\footnote{For realistic conditions of AGNs narrow line region (NLR) has typical density $n_{\gas} \sim 10^{2}-10^{5}\cm^{-3}$, and broad line region has higher density $n_{\gas}\sim 10^{6}-10^{10}\cm^{-3}$. Therefore, grains initially located in low density regions of NLR have chance to be accelerated to $v\sim c$ by radiation pressure.} Lastly, it is shown in Appendix (\ref{apdx:sputt}) that the sputtering by ions is negligible during the acceleration stage.

\section{Interactions of relativistic dust with radiation}\label{sec:trans}

\subsection{General Consideration}
\subsubsection{Low-energy photons}
An incident, low-energy photon ($h\nu< 20 \eV$) will be absorbed by the dust grain. When the photon energy is larger than the work function of the solid $W$ (see Appendix \ref{apdx:pe}), an electron from the band structure can be excited and potentially becomes a photoelectron. The kinetic energy of the excited electron is equal to $E_{e}=h\nu-W$. To escape from the grain surface, the excited electron first must travel from the photoabsorption site to the surface during which it transfers a significant fraction of its energy to the grain. If the excited electron has sufficient energy to overcome the surface Coulomb barrier (in case of positive grain charge), then it will escape from the grain surface to become a photoelectron (see \citealt{2001ApJS..134..263W} for more details). For $h\nu < W$, the photon energy is solely transferred to the dust through electronic excitations without producing a photoelectron. 

\subsubsection{High-energy photons}
For high-energy photons ($h\nu>20 \eV$), their interaction with the dust grain is considered as the interaction of photons with individual atoms within the grain. After the absorption of an energetic photon, electrons from {\it inner shells} can be excited and become photoelectrons.

When a photoelectron is liberated from the inner shell of low-energy level $i$, it also leaves a hole. This hole will be filled by an electronic transition from a higher energy level $j$. Such an electronic transition can be done through radiativeless (Auger) process or radiative (fluorescent) process. The former process is accompanied by the ejection of a second electron (namely Auger electron) from some allowed energy level $k$. The Auger electron due to the $(i,j,k)$ transition deposits some energy to the grain when moving from the emitting place to the grain surface and becomes a photoelectron if it has sufficient energy to overcome the surface Coulomb barrier. After an Auger transition, the atom is doubly ionized, with one hole in $j$ shell and another in $k$ shell. These $j$ and $k$ holes now can be filled by radiativeless transitions from $j'$ and $k'$ shells, producing secondary Auger electrons. This process continues until all Auger transitions have been used, and the atom is multiply ionized. Theory indicates that, for X-ray photon energy, the Auger transition process is dominant over the radiative process (see \citealt{1996ApJ...459..686D}, hereafter DS96). 


\subsubsection{Very high-energy photons}
For very high-energy photons ($h\nu> 10^{4} \eV$), the dust grain appears transparent to the radiation field, and the interaction of photons with the dust grain can be treated as the scattering of photons by free electrons. During each scattering, the photon transfers part of its energy to the electron via Compton effect, resulting in the recoil of the electron. The scattered electron will become a free photoelectron if it has sufficient energy to reach the grain surface and overcome the potential barrier (see, e.g., \citealt{1999APh....12...35B}). 

For photons with energy above $1$ MeV ($\ge 2m_{e}c^{2}$), the pair production starts to become important. Each nucleus $X$ in the dust grain interacts with an energetic photon $\gamma_{\rm ph}$ to produce an {electron-positron pair}, which is given by $X+\gamma_{\rm ph}\rightarrow X + e^{+} + e^{-}$. This additional pair production process might help in explaining the observed high positron fraction in primary cosmic rays by the AMS collaboration (\citealt{2014PhRvL.113l1101A}; \citealt{2014PhRvL.113l1102A}).

The inverse process to pair production, namely annihilation, can happen when an electron $e^{-}$ of the dust grain collides with a positron from the ambient plasma. The process is followed by the emission of a photon via $ e^{-}+ e^{+} \rightarrow 2\gamma_{\rm ph}$.

\section{Interactions of relativistic dust with ambient gas}\label{sec:coll}
\subsection{General Consideration}

Interaction processes between an incident charge particle and atoms of the dust grain (hereafter target atoms) depend on its initial velocity (energy). When the particle velocity is much smaller than the Bohr velocity $v_{0}=e^{2}/\hbar$ (i.e., energy $E\ll 25~ {\rm keV}/{\rm amu}$ with amu being atomic mass unit), {\it elastic} collisions between the particle and target atomic nuclei (hereafter low-energy regime) are crucial. When the particle velocity increases, {\it electronic} interactions between the particle and atomic electrons (hereafter high-energy regime) become increasingly important and reach their maximum at $v\sim v_{0}$. 



Below we summarize well-known formulas from stopping theory of swift ions and electrons passing through matter. Classical reviews can be found in \cite{1963ARNPS..13....1F} and \cite{1999JAP....85.1249Z}. 

\subsection{Incident Ions}
Consider the interaction of an incident ion (projectile) of charge $Z_{P}e$ and kinetic energy $E$ (velocity $\beta c$) with an electron of the target atom in the grain. Let $Z_{T,i}$ be the atomic number of target element $i$. 

The total energy loss of the incident ion per pathlength, which is usually referred to as {\it stopping power} of the material, is given by
\bea
\frac{dE_{\ion}}{dx}=\sum_{i} n_{i}S_{i},
\ena 
where $n_{i}$ is the atomic number density of element $i$, $S_{i}$ is the electronic stopping cross-section of element $i$ in units of $\eV \cm^{2}$, and the sum is taken over all elements $i$ present in the grain. The total atomic number density in the grain is thus $n_{d}=\sum_{i}n_{i}$.

The Bethe-Bloch theory yields the following (see \citealt{1963ARNPS..13....1F}):
\bea
S_{i} &=&\frac{4\pi(Z_{P}e^{2})^{2}}{m_{e}c^{2}\beta^{2}}Z_{T,i}\nonumber\\
&&\times \left[\frac{1}{2}\ln \frac{2\gamma^{2}m_{e}c^{2}\beta^{2}T_{\max}}{I_{i}^{2}}-\beta^{2}-\frac{C}{Z_{T,i}}-\frac{\delta}{2}\right],\label{eq:dE-dx}~~~
\ena
where $T_{\max}$ is the maximum energy transferred from the ion to the atomic electron, and $I_{i}$ is the mean excitation energy of element $i$. Here $C/Z_{T,i}$ is the shell correction term, and $\delta/2$ is the density correction term.

In binary collisions, $T_{\max}$ is given by
\bea
T_{\max}=\frac{2\gamma^{2}M_{P}^{2}m_{e}c^{2}\beta^{2}}{m_{e}^{2}+M_{P}^{2}+2\gamma m_{e}M_{P}},\label{eq:Tmax}
\ena
where $M_{P}$ is the atomic mass of the projectile.

In the case $2\gamma m_{e}/M_{P} \ll 1$ (heavy ion or not very high $\gamma$) we have $T_{\max}=2\gamma^{2}\beta^{2}m_{e}c^{2}$. Thus Equation (\ref{eq:dE-dx}) can be rewritten as
\bea
S_{i}=\frac{4\pi(Z_{P}e^{2})^{2}}{m_{e}c^{2}\beta^{2}}Z_{T,i}\left[\ln \frac{2m_{e}c^{2}\beta^{2}}{I_{i}}+\mathcal{L}(\beta)\right],\label{eq:dEidx}
\ena
where
\bea
\mathcal{L}(\beta)=\ln \frac{1}{1-\beta^{2}}-\beta^{2}-\frac{C}{Z_{T,i}}-\frac{\delta}{2}.\label{eq:Lbeta}
\ena

The Bethe-Bloch formula (Eq. \ref{eq:dE-dx}) with its correction terms is of high accuracy for energetic ions with $E>1$ MeV considered in this paper. The shell correction term $C/Z_{T,i}$ is maximum at $\sim 0.3$ and becomes negligible for $E >10$ MeV/amu. The density correction term $\delta/2$ is important and increases to above $1$ for $E> 5$ GeV/amu  (see \citealt{1999JAP....85.1249Z} for more details). Here we calculate $\delta/2$ using an analytical fit from \cite{1984ADNDT..30..261S}.
 
\subsection{Incident Electrons}
\subsubsection{Electronic Excitation}

The energy loss of an electron with kinetic energy $E$ per pathlength in the material due to electronic excitations is equal to (see e.g., \citealt{2011hea..book.....L})
\bea
\frac{dE_{{\rm el}}}{dx}=\sum_{i} n_{i}S_{{\rm el},i}\
\ena
where $S_{{\rm el},i}$ is the energy loss by atom $i$ having $Z_{T,i}$ electrons, which reads
\bea
S_{{\rm el},i}=\frac{4\pi e^{4}}{m_{e}c^{2}\beta^{2}}Z_{T,i}\left[\frac{1}{2}\ln \frac{2\gamma^{2}m_{e}c^{2}\beta^{2}T_{\max}}{I_{i}^{2}}+\frac{\mathcal{F}(\gamma)}{2}\right].~~~\label{eq:dEeldx}
\ena

Here $T_{\max}$ is the maximum transfer energy to the target electron as usual, {the first term in the square bracket is analogous to that in Equation (\ref{eq:dE-dx}) for incident ions}, and
\bea
\mathcal{F}(\gamma)=-\left(\frac{2}{\gamma}-\frac{1}{\gamma^{2}}\right)\ln 2 +\frac{1}{\gamma^{2}}+\frac{1}{8}\left(1-\frac{1}{\gamma}\right)^{2}.\label{eq:Fgamma}
\ena


For electron-electron collisions, we have $T_{\max}=E/2= (\gamma-1)m_{e}c^{2}/2$ (see \citealt{1954PhRv...93...38R}), and Equation (\ref{eq:dEeldx}) becomes
\bea
S_{{\rm el},i}=\frac{4\pi e^{4}}{m_{e}c^{2}\beta^{2}}Z_{T,i}\left[\frac{1}{2}\ln \left(\frac{E^{2}}{I_{i}^{2}}\frac{\gamma+1}{2}\right)+\frac{\mathcal{F}(\gamma)}{2}\right].\label{eq:dEedx}
\ena

\subsubsection{Bremsstrahlung radiation}
In addition to electronic excitations, while passing the grain, energetic electrons are accelerated in the Coulomb force field of atomic nuclei and emit continuous radiation (namely free-free emission) due to the Bremsstrahlung effect. \cite{Bethe:1934cv} derived the total energy loss per pathlength by radiation for an electron:
\bea
\frac{dE_{\rm Brems}}{dx} =\sum_{i}n_{i}\frac{Z_{T,i}^{2}r_{e}^{2}}{137}E\left[4\ln\left(\frac{2E}{m_{e}c^{2}}\right)-\frac{4}{3}\right]
\ena
for	$m_{e}c^{2}<E <137m_{e}c^{2}Z_{T,i}^{-1/3}$, and
\bea
\frac{dE_{\rm Brems}}{dx} =\sum_{i}n_{i}\frac{Z_{T,i}^{2}r_{e}^{2}}{137}E\left[4\ln\left(183Z_{T,i}^{-1/3}\right)
+\frac{2}{9}\right]
\ena
for $E >137m_{e}c^{2}Z_{T,i}^{-1/3}$. {Here $r_{e}=e^{2}/m_{e}c^{2}$ is the classical electron radius}.

The spectrum of Bremsstrahlung emission is broad, independent of frequency, ranging from zero frequency to $\nu_{\max}=E/h\approx 511{~\rm keV}(\gamma-1)/h$. {Because most of the Bremsstrahlung radiation energy is concentrated at highest frequency photons that easily escape from the grain, we can assume that the radiative energy loss of electrons does not contribute to grain heating.}

\subsection{Incident Neutrals}
When an atom is moving at relativistic speeds, its kinetic energy is much larger than the orbital energy of atomic electrons. The incident atom can be considered to be a pair of the ion (nucleus) and electrons that independently interact with target atoms. The energy loss of the atom is then the sum of the energy loss from its nucleus and electrons:
\bea
\frac{dE}{dx}=\left( \frac{dE_{\ion}}{dx}\right)+\left(\frac{dE_{\rm el}}{dx}\right)+\left(\frac{dE_{\rm Brems}}{dx}\right).\label{eq:dEdx}
\ena

\subsection{Total energy loss, range and mean ionization length}\label{sec:lamion}

We calculate the energy loss of incident ions/electrons colliding with relativistic grains of different Lorentz factor $\gamma$. To facilitate latter comparison with available experimental data of sputtering yield, we consider the quartz (SiO$_{2}$) for silicate with mass density $\rho=3.3 \g\cm^{-3}$, which yields $n_{\rm Si}=n_{\rm O}/2=3.3\times 10^{22}\cm^{-3}$ and $n_{d}\sim 10^{23}\cm^{-3}$.

Figure \ref{fig:dEdx} (left panel) shows the energy loss of a H atom (i.e., a proton and electron) by different processes and its total energy loss. The energy loss of the electron via electronic excitation is dominant until the radiative loss via Bremsstrahlung effect becomes important for $\gamma > 300$. The right panel shows the total energy loss of several atoms (H, He, C, O, and Fe) in the silicate material. As expected, heavier atoms have much higher energy loss, which can transfer higher amount of energy to the target atoms. The energy loss decreases rapidly with increasing $\gamma$ up to a minimum at $\gamma \sim 2$ due to the decrease in the Coulomb ionization cross-section. As $\gamma$ continues to increase, $dE/dx$ rises slowly due to the relativistic correction for the transversal electric field. For $\gamma> 300$, $dE/dx$ rises rapidly due to the radiative loss of electrons. 

The range that a projectile of initial energy $E_{0}$ penetrates in solid before completely stopped is defined as
\bea
R(E_{0})=\int_{E_{0}}^{0}\left(\frac{dE}{dx}\right)^{-1}dE.\label{eq:RE}
\ena

It is also convenient to define the mean length between two successive ionizations as $\lambda_{\ion}=\frac{W}{dE/dx}$
where $W$ is the average energy required to create an ionization in the grain.

\begin{figure*}
\centering
\includegraphics[width=0.4\textwidth]{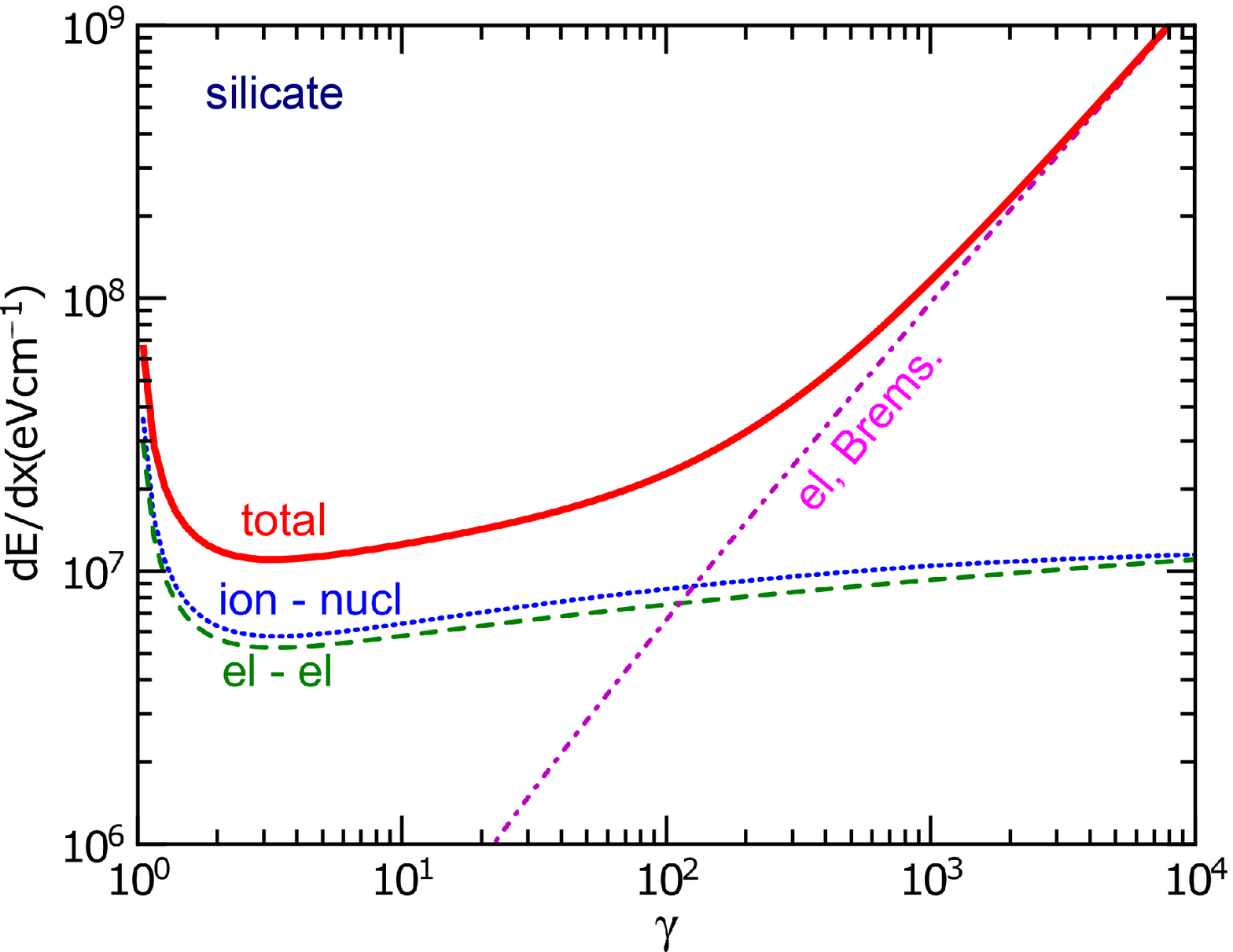}
\includegraphics[width=0.4\textwidth]{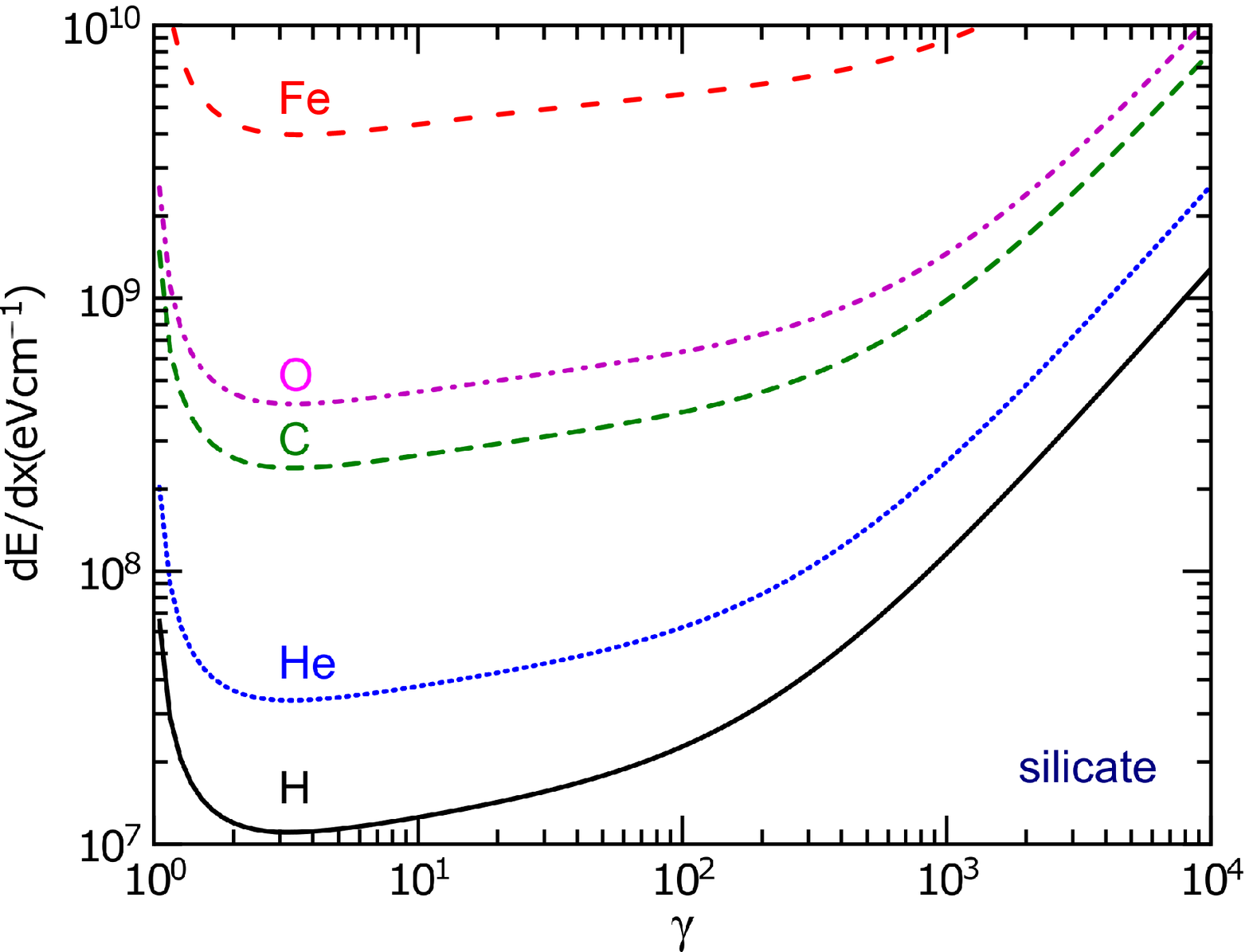}
\caption{Left panel: energy loss per pathlength ($\eV\cm^{-1}$) of a H atom in the silicate material as a function of the Lorentz factor $\gamma$. Individual interaction processes between the ion/electron with the target nucleus/electron (ion-nucleus, ion-electron, electron-electron, and Bremsstrahlung radiation loss of electron) are indicated. Right panel: total energy loss per pathlength for different atoms.}
\label{fig:dEdx}
\end{figure*}
 
\begin{figure*}
\centering
\includegraphics[width=0.4\textwidth]{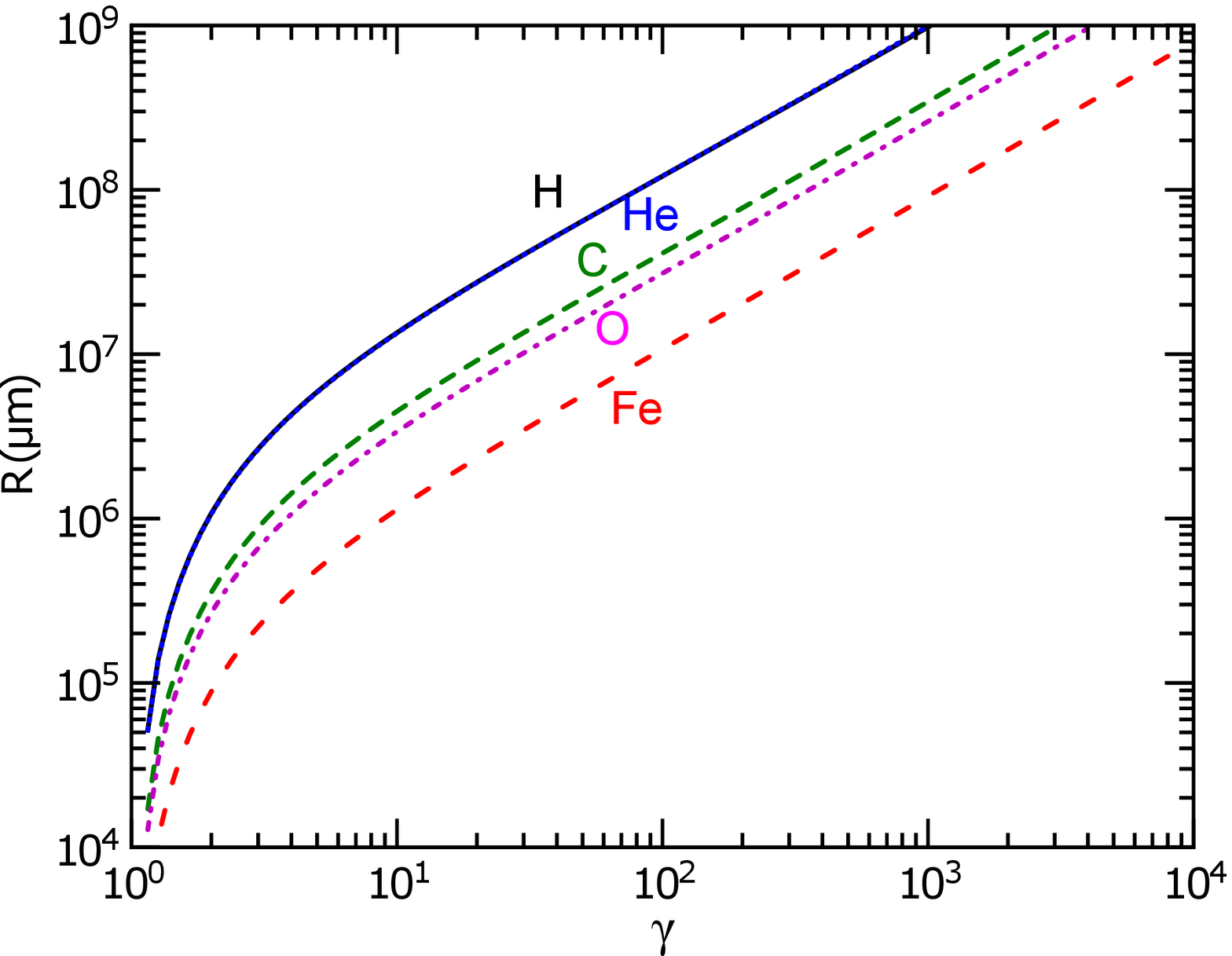}
\includegraphics[width=0.4\textwidth]{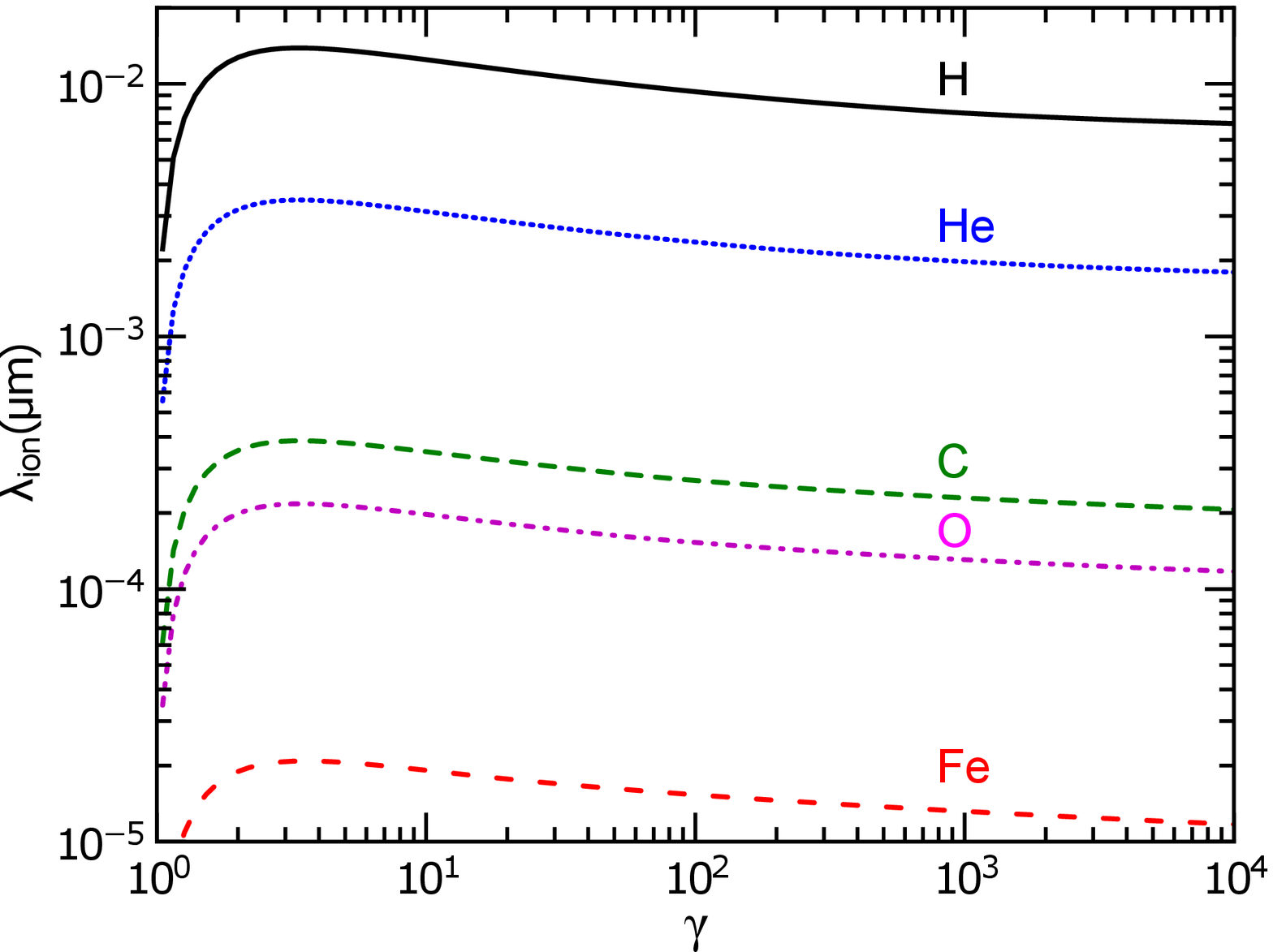}
\caption{Range $R$ (left) and mean ionization length $\lambda_{\ion}$ (right) for different ions in the silicate material with $W=8 \eV$.}
\label{fig:ionmfp}
\end{figure*}

Figure \ref{fig:ionmfp} shows $R$ and $\lambda_{\ion}$ as functions of $\gamma$ for the different ions bombarding the silicate material. It can be seen that relativistic ions have $R\gg 1\mum$, i.e., they easily pass through interstellar grains. In addition, light ions (i.e., H, He) can create $2a/\lambda_{\ion}\sim 10a_{-5}-100a_{-5}$ ionizations along their entire path through the grain. Moreover, a heavy Fe ion can create $\sim 20$ ionizations per atomic layer, which forms a track of dense ionizations in the grain referred to as {\it ionization track}.

\section{Destruction of relativistic dust by thermal sublimation}\label{sec:heat}
\subsection{Grain Heating}
\subsubsection{Collisional Heating}
To see the effect of incident particles on grain heating, it is useful to estimate the average energy transferred to an atom {along the track}:
\bea
\Delta E_{1}= l \frac{dE}{dx}\simeq n_{\d}^{-1/3}\frac{dE}{dx}
\simeq 21.54 n_{23}^{-1/3}\left(\frac{dE/dx}{10^{9}\eV\cm^{-1}}\right) \eV,~~~~~\label{eq:DE1}
\ena
where $l$ is the mean interatomic distance and $n_{23}=n_{\d}/10^{23}\cm^{-3}$. 

Using $dE/dx$ from Figure \ref{fig:dEdx}, we find that a H atom with $dE/dx\sim 10^{7}\eV\cm^{-1}$ can transfer $\Delta E_{1}\sim 0.21n_{23}$ eV to a target atom. Thus, the energy loss of the H atom is mostly spent to grain heating. An Fe atom with $dE/dx \sim 4\times 10^{9}\eV\cm^{-1}$ can transfer $\Delta E_{1}\sim 86n_{23}\eV$ to a target atom. This energy is sufficient to ionize atoms from the band structure and outer shells, providing secondary electrons with kinetic energy $E_{e}< 86\eV$. The range of the secondary electrons is $R_{e}\simeq 118 \hat{\rho}^{-0.85}(E_{e}/1{~\rm keV})^{1.5}$\AA~(see \citealt{1979ApJ...231...77D}), which is less than $4$\AA~for $E_{e}< 86\eV$. Therefore, the secondary electrons produced by impinging Fe atoms quickly loose their energy to the lattice and heat the grain up. Similarly, secondary electrons produced by other incident ions (e.g., He, C and O) transfer most of their energy to the grain heating.


Let $f_{\coll}$ be the fraction of the energy loss $dE/dx$ that is expended to grain heating. Then, the rate of collisional heating in the GF can be written as
\bea
\frac{dE_{\rm h, \coll}}{dt'} = \left(n'_{\gas}\beta c\right)\left(f_{\coll}\frac{dE}{dx}\right)\left(\frac{4\pi a^{3}}{3}\right),
\ena
where $n'_{\gas}=\gamma n_{\gas}$ is the gas density in the GF.

Taking the typical numeric values for $dE_{\coll}/dt$, we can obtain
\bea
\frac{dE_{\rm h, \coll}}{dt'} &\simeq& 2\times 10^{-7} \left(\frac{\gamma n_{\gas}}{1\cm^{-3}}\right)\beta a_{-5}^{3}\left(\frac{f_{\coll}}{1.0}\right)\nonumber\\
&&\times\left(\frac{dE/dx}{10^{9}\eV\cm^{-1}}\right)\erg \s^{-1}.\label{eq:collheat}~~~~
\ena

It is adequate to adopt $f_{\coll}\approx 1$ for all incident ions, whereas for electrons we only take the energy loss due to electronic excitations (i.e., not including the radiative loss). 

\subsubsection{Radiative Heating}

The heating rate for a grain of size $a$ by photons coming from direction $\mu'$ is given by
\bea
\frac{dE_{\rm h, rad}}{dt'}=\int d\nu' \int d\mu' \pi a^{2}Q_{\abs,\nu'} f_{\dep}(\nu')c u'(\nu',\mu'),~~~
\label{eq:dEhdt}
\ena
where $u(\nu',\mu')$ is the specific spectral energy density, $Q_{\abs,\nu'}$ is the absorption efficiency (see Appendix \ref{sec:Qabs_Qpr}), and $f_{\dep}(\nu')$ is the fraction of photon energy deposited to the grain taken from Table 5.1 and 5.2 of DS96.

For the point source of the solar radiation field and assuming that the grain is arriving in the radial direction (i.e., $\mu_{\gr}=-1$), one has $\nu'=\nu\gamma(1+\beta)$. Through the Lorentz transformations (see Appendix \ref{apdx:trans}), the integral over $\mu$ can be analytically carried out, and one obtains:
\bea
\frac{dE_{\rm h, rad}}{dt'}=\int d\nu\pi a^{2}Q_{\abs,\nu'} f_{\dep}(\nu')c u(\nu)\gamma^{2}\left(1+\beta\right)^{2}.~~~~~
\label{eq:dEhdt_uni}
\ena

For the isotropic radiation (e.g., ISRF, CMB), $\nu'$ is a function of both $\nu$ and $\mu$. Equation (\ref{eq:dEhdt}) is rewritten as
\bea
\frac{dE_{\rm h, rad}}{dt'}=\int d\nu\int_{-1}^{1}d\mu\pi a^{2}Q_{\abs,\nu'} f_{\dep}(\nu')c \frac{u(\nu)}{2}\gamma^{2}\left(1-\beta\mu\right)^{2},~~~~~
\label{eq:dEhdt_iso}
\ena
where the term $\left(1-\beta\mu\right)^{2}$ describes the focusing effect of radiation due to the relativistic motion of the grain. The integral over $\mu$ is numerically computed using a grid of $32$ directions for $\mu$ {from $-1$ to $1$}.

\subsection{Cooling and Equilibrium Temperature}\label{sec:cooling}

The grain emits thermal radiation isotropically in the GF, which results in radiative cooling at rate:
\bea
\frac{dE_{\rm c, rad}}{dt'}=\int d\nu' 4\pi a^{2}Q_{\abs, \nu'}B_{\nu'}(T_{\d})
=4\pi a^{2}\langle Q_{\abs}\rangle_{T_{\d}} \sigma T_{\d}^{4},~~~~
\label{eq:dEcdt}
\ena
where
\bea
\langle Q_{\abs}\rangle_{T_{d}}=\frac{\int d\nu' Q_{\abs,\nu'}B_{\nu'}(T_{\d})}{\int d\nu' B_{\nu'}(T_{\d})} \label{eq:Qabsavg}
\ena
 is the Planck-averaged emission efficiency.

The sublimation/evaporation of atoms from the grain surface also results in grain cooling, so-called evaporative cooling, at rate
\bea
\frac{dE_{\rm c, evap}}{dt'}= 4\pi a^{2}n_{d} \frac{da}{dt'} U_{0},
\ena
where $da/dt'$ is given by Equation (\ref{eq:dasdt}), and $U_{0}$ is the energy carried away by each atom, which is equal to the binding energy of the dust atoms (see \citealt{2000ApJ...537..796W}. The evaporative cooling is expected to be dominant for $T_{d}\ge T_{\rm sub}$.
 
The grain equilibrium temperature is obtained by setting the heating rate equal to the cooling rate:
\bea
\frac{dE_{\rm h, rad}}{dt'}+\frac{dE_{\rm h, \coll}}{dt'}=\frac{dE_{\rm c, rad}}{dt'} + \frac{dE_{\rm c, evap}}{dt'}.\label{eq:Teq}
\ena

\subsection{Thermal Sublimation in the Solar Radiation Field}

To study the thermal destruction of relativistic dust in the solar radiation field, we calculate the grain equilibrium temperature by solving Equation (\ref{eq:Teq}) numerically for the full spectrum of solar radiation described by the Planck function $B_{\nu}(T_{\odot})$. The photon energy $h\nu'$ and energy density $u'(\nu')$ in the GF are obtained through the Lorentz transformations in Appendix \ref{sec:trans}. The grain temperature $T_{d}(r)$ as a function of solar distance $r$ is calculated for $r=1\AU$ to $r=10\AU$. The latter distance is chosen such that the grain temperature is sufficiently low and the sublimation is negligible.

\begin{figure*}
\centering
\includegraphics[width=0.4\textwidth]{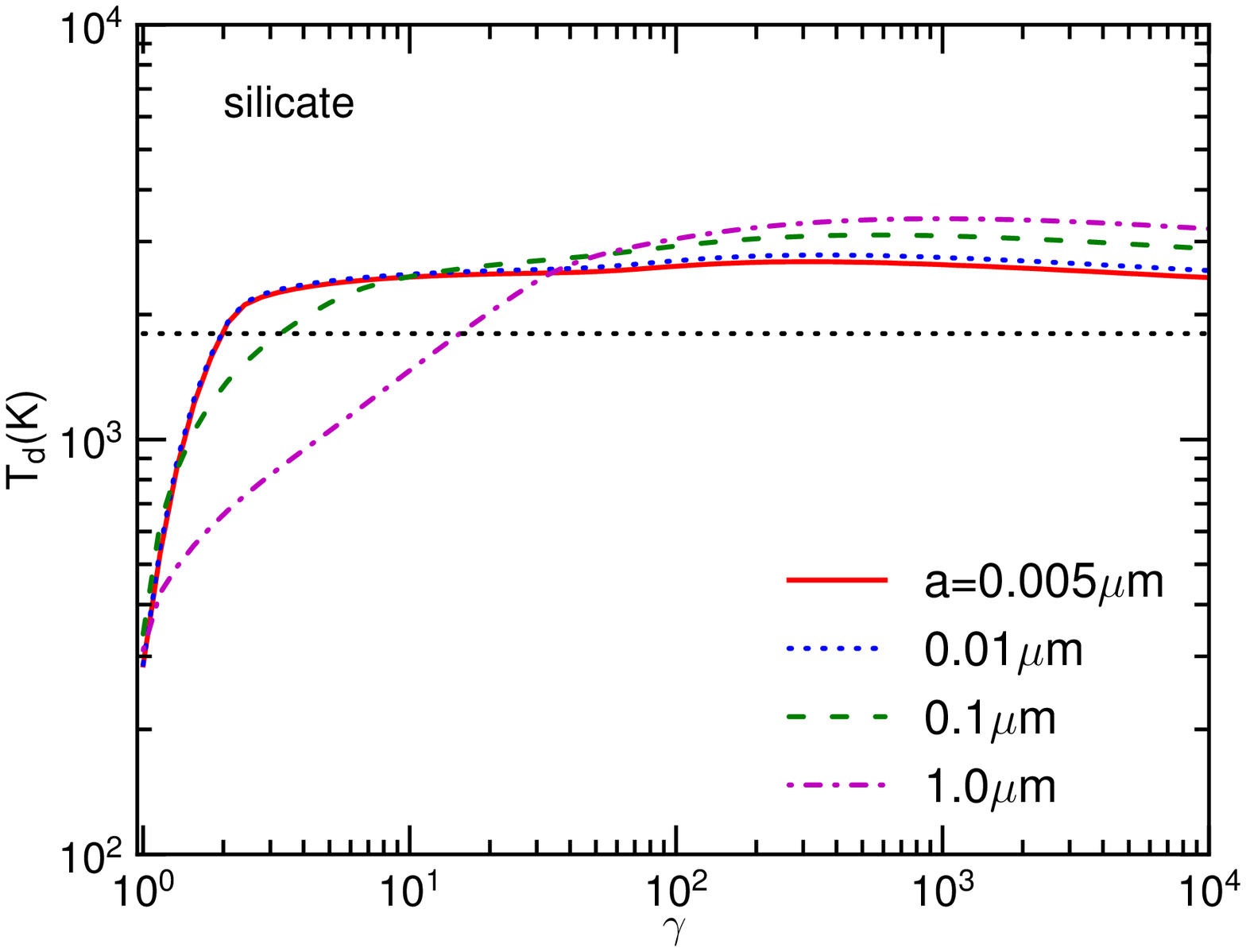}
\includegraphics[width=0.4\textwidth]{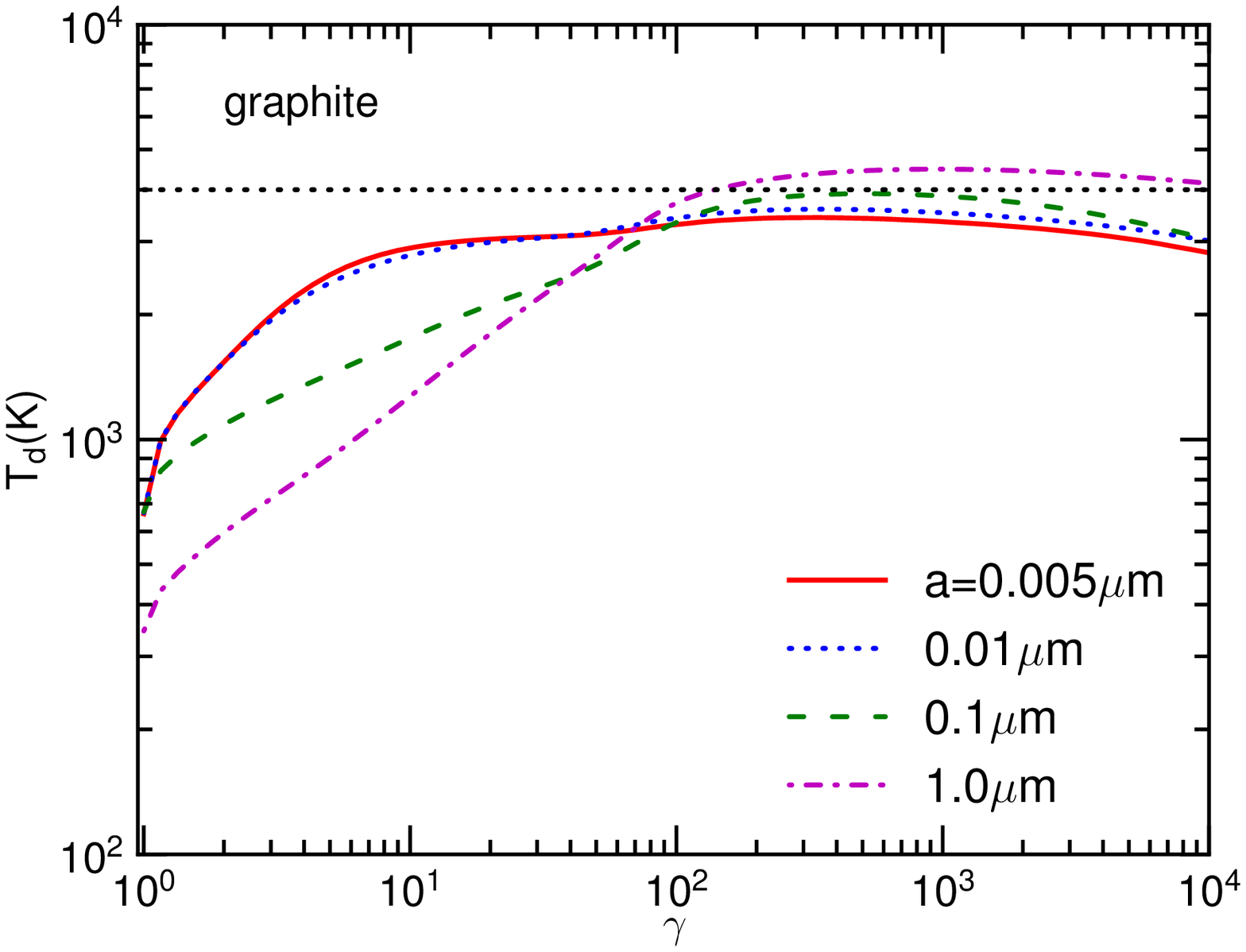}
\caption{Grain equilibrium temperature as a function of the Lorentz factor $\gamma$ for silicate (left) and graphite (right) grains located at solar distance of 1 AU. Horizontal dotted lines show the melting temperature, $T_{\rm m}=1800\K$ for silicate grains and $T_{\rm m}=4000\K$ for graphite grains.}
\label{fig:Tdfull}
\end{figure*}

Figure \ref{fig:Tdfull} shows $T_{d}(r=1\AU)$ for silicate and graphite grains of different sizes. The silicate material with $\gamma>10$ will undergo {\it melting} (phase transition) when $T_{d}=T_{\rm m}=1800\K$ while approaching the Sun, but the graphite with $T_{\rm m}=4000\K$ will hardly melt.

Because it is unclear whether melting has any direct significant effect on the destruction of relativistic dust,\footnote{In terms of grain kinetic energy, a solid grain is identical to the drop of the same volume.} here we will focus on a general effect of thermal sublimation that is expected to efficiently occur before the melting point. In this scenario, whether grains can survive solar heating depends on the sublimation time and arrival time into the Earth's atmosphere. 

The sublimation time of a dust grain with size $a$ in the SF is defined as
\bea
\tau_{\rm sub}(T_d)=-\gamma \frac{a}{da/dt'}
= a\gamma n_{d}^{1/3}\nu_{0}^{-1}\exp\left(\frac{B}{\kB T_d}\right),\label{eq:tausub}
\ena
where $da/dt'$ from Equation (\ref{eq:dasdt}) has been used.

Plugging the numerical parameters into the above equation, we obtain
\bea
\tau_{\rm sub}(T_d)=6.36\times 10^{3}\gamma a_{-5}\exp\left[68100\K\left(\frac{1}{T_d}-\frac{1}{1800\K}  \right)\right] \s\label{eq:tausub_sil}~~~~~
\ena
for silicate grains, and
\bea
\tau_{\rm sub}(T_d)=1.36\gamma a_{-5}\exp\left[81200\K\left(\frac{1}{T_d}-\frac{1}{3000\K}  \right)\right]\s\label{eq:tausub_gra}~~~~~
\ena
for graphite grains.\footnote{The sublimation rates above are for uncharged grains, and the presence of positive charge is expected to increase the sublimation (see \citealt{1989ApJ...345..230G}).}

{It can be seen that, at just above $\sim 1$ AU, the sublimation time of silicate grains with $T_{d}\sim 1800\K $ is sufficiently long compared to the time needed to reach the Earth $\tau_{\rm arr}=1\AU/(\beta c)=500\s$ for $\beta\sim 1$. As a result, the relativistic grains that survive until $R \sim 1-2\AU$ would enter the Earth's atmosphere as primary UHECRs. 

{To see the survival of a relativistic grain along its journey to the Earth, we integrate Equation (\ref{eq:dasdt}) to obtain the final grain size at $r_{f}=1\AU$:
\bea
a_{f} - a_{i} = \int_{r_{i}}^{1\AU} n_{d}^{-1/3}\nu_{0}\exp\left(-\frac{B}{\kB T_{d}(r)}\right) \frac{dr}{\beta c},~~~
\ena
where the increase of $T_{d}$ with decreasing $r$ is included, and we disregard the weak dependence of $T_{d}$ on the instantaneous grain size during its arrival. 

Let $\delta a_{\rm sub}/a=(a_{i}-a_{f})/{a_{i}}$. Then, grains are considered to be completely destroyed by sublimation when $\delta a_{\rm sub}/a\approx 1$, and they may survive for $\delta a_{\rm sub}/a< 0.1$.

The obtained results ${\delta a_{\rm sub}}/{a}$ are shown in Figure \ref{fig:asub} for silicate (left) and graphite (right) grains. There exists a survival window for relativistic grains with parameters $(a,\gamma)$ satisfying the condition $\delta a_{\rm sub}/a< 0.1$ (blue areas). For graphite grains, there exists two regions of survival at $\gamma<10^{2}$ and $\gamma>10^{3}$. The latter arises from the usage of $f_{\rm dep}$ computed for neutral grains in DS96, which have $f_{\rm dep} < 1$ for highest photon energy. However, as shown in our next section, grains of large $\gamma$ have highly positive charge for which entire photon energy can transfer to the grain (i.e., $f_{\rm dep}\sim 1$). As a result, the survival chance at highest $\gamma$ seems extremely rare.} 

\begin{figure*}
\centering
\includegraphics[width=0.4\textwidth]{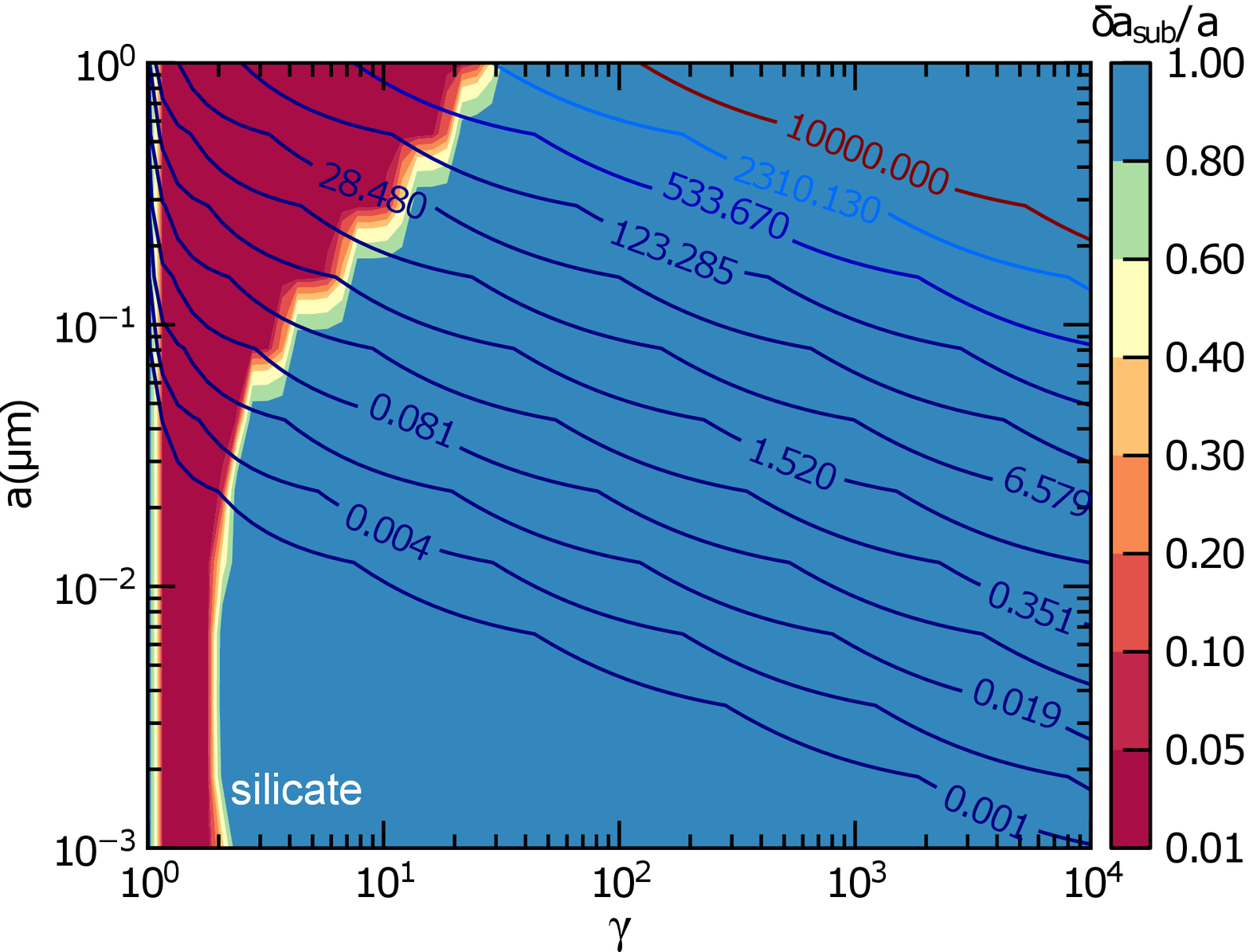}
\includegraphics[width=0.4\textwidth]{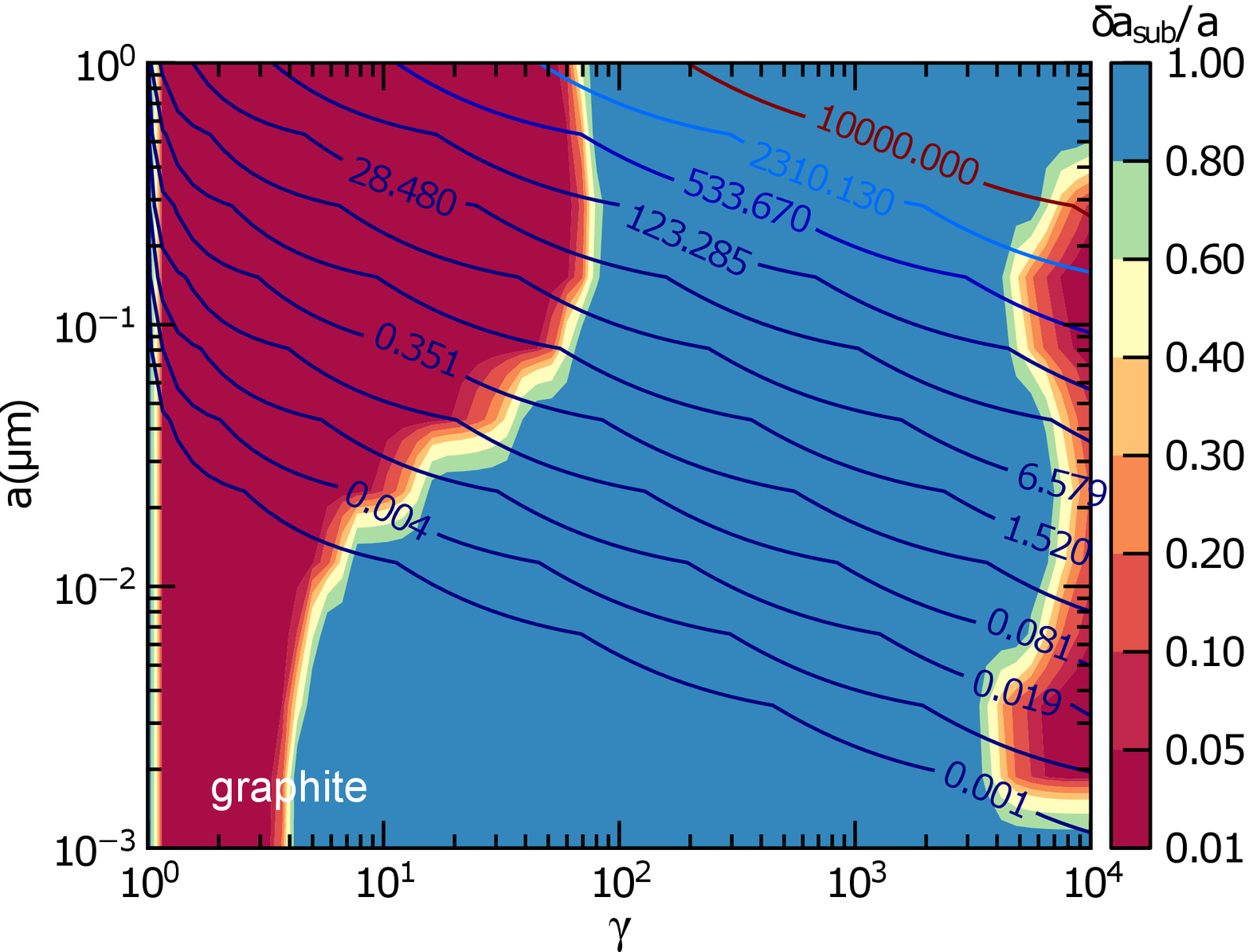}
\caption{Map of $\delta a_{\rm sub}/a$ at 1 AU in the plane $(\gamma, a)$ for silicate (left panel) and graphite (right panel). Regions with $\delta a_{\rm sub}/a\ge 0.1$ indicate grain sublimation is significant, and the regions with $\delta a_{\rm sub}/a= 1$ correspond to the complete destruction of grains. Contours show the levels of grain kinetic energy $E_{\rm gr}$ in units of $10^{20}\eV$.}
\label{fig:asub}
\end{figure*}

\section{Destruction of relativistic dust by Coulomb explosions}\label{sec:charging}

\subsection{Collisional charging}

In the high-energy regime, incident electrons and ions will not be stopped within the grain or recombine to affect the grain charge as in the low-energy regime. {These energetic particles will instead produce secondary electrons, and some of them can have sufficient kinetic energy to escape from the grain surface, increasing the grain charge.} The sputtering may also eject some ions from the grain, but we will see in Section \ref{sec:sputt} that this process is very inefficient in the high-energy regime. 

The total charging rate for a dust grain having $4\pi a^{3}n_{d}/3$ atoms by a flow of charge particles ($P$) with density $n_{P}$ arriving at velocity $\beta c$ is given by
\bea
J_{\rm coll}(a, Z) = \sum_{P=e, {\rm H, He}}\gamma n_{P}\beta c\sigma_{I}(Z_{P}, a, Z)\frac{Z_{d}n_{d}4\pi a^{3}}{3},~~~\label{eq:dJi_dt}
\ena
where $\sigma_{I}$ is the ionization cross-section, and $Z_{d}$ is the mean atomic number of dust, which is equal to $6$ for graphite and $10$ for quartz  (see Appendix \ref{apdx:Jcoll}). 

\subsection{Photoelectric Emission}

Following the absorption of an energetic photon, the relativistic grain becomes positively charged as a result of three basic processes: emission of primary photoelectrons, emission of Auger electrons, and emission of secondary electrons excited by high-energy primary photoelectrons and Auger electrons (see WDB06 for more details; also Appendix \ref{apdx:pe}). 

Let $Y_{\rm p},Y_{\rm A}$ and $Y_{\rm sec}$ be the photoelectric yields for primary, Auger, and secondary electrons, which are functions of photon energy $h\nu'$, grain size $a$ and charge $Z$. 

The charging rate by primary photoelectric emission is given by
\bea
J_{\rm pe}(a,Z)=\int _{\nu_{\rm pet,b}}^{\nu'_{\max}}d\nu' \int_{-1}^{1} d\mu'\pi a^{2}Q_{\abs,\nu'}\left(Y_{{\rm p};b}+\sum_{i,s}Y_{{\rm p};i,s}\right)c n' ,~~~~~\label{eq:Jpe}
\ena
where $\nu'$ is the frequency and $n'=n'(\nu',\mu')$ is the density of photons coming from direction $\mu'$ in the GF, $\nu_{\rm pet,b}$ denotes the frequency threshold of photoelectric emission from the band structure (see Appendix \ref{apdx:pe}), and $\nu'_{\max}$ is the maximum frequency of the radiation spectrum in the GF. Here, $Y_{{\rm p};b}$ is the yield of primary photoelectrons from the band structure, $Y_{{\rm p};i,s}$ is the partial yield from electronic shell $s$ of element $i$, and the sum is taken over all inner shells $s$ and elements $i$. In addition, the dependence of $Q_{\abs}$ and $Y$ on $a$ and $\nu'$ has been omitted for simplicity.

The charging rate due to the emission of Auger electrons is
\bea
J_{\rm A}(a,Z)=\int _{I_{is,\min/h}}^{\nu'_{\max}}d\nu' \int_{-1}^{1} d\mu' \pi a^{2}Q_{\abs,\nu'}\sum_{i,s,j}Y_{{\rm A};i,s,j}c n',~~~~~\label{eq:JA}
\ena
where $I_{is,{\min}}$ is the ionization potential of the shell $is$, and the sum is over all electronic shells $s$, elements $i$, and Auger transitions $j$.

Similarly, the charging rate for secondary effects arising from the excitation of primary photoelectrons and Auger electrons is
\bea
J_{\rm sec}(a,Z)=\int _{\nu_{\rm pet,b}}^{\nu'_{\max}}d\nu' \int_{-1}^{1} d\mu' \pi a^{2}Q_{\abs,\nu'}\sum_{k}Y_{{\rm sec};k}c n',~~~~~\label{eq:Jsec}
\ena
where the sum runs over all primary photoelectrons and Auger electrons. 

The charging rate due to photoelectric emission can be simplified as
\bea
J_{\rm chrg}(a,Z)=\int d\nu \pi a^{2}Q_{\abs,\nu'}Y_{\rm chrg} c \frac{u(\nu)}{h\nu}\gamma\left(1+\beta\right),~~~ \label{eq:dZdt_uni}
\ena
for the point radiation source, and
\bea
J_{\rm chrg}(a,Z)=\int d\nu\int_{-1}^{1} d\mu \pi a^{2}Q_{\abs,\nu'}Y_{\rm chrg} c \frac{u(\nu)}{2h\nu}\gamma\left(1-\beta\mu\right),~~~~~ \label{eq:dZdt_iso}
\ena
for the isotropic radiation field where ${\rm chrg}$ denotes the {\rm pe, A}, and {\rm sec} processes.

The total charging rate in the GF due to photoelectric emission is then given by
\bea
 J_{\rm phe}(a,Z)=\sum_{i={\rm pe, A, sec}} J_{\rm i}(a,Z).\label{eq:dZdt}
\ena

\subsection{Maximum grain charge}
\subsubsection{Coulomb explosions}
Efficient grain charging by photoelectric emission and collisional ionization can rapidly increase grain positive charge, which results in an increased electric surface potential $\phi=Ze/a$ and tensile strength $\mathcal{S}=(\phi/a)^2/4\pi$. When the tensile strength exceeds the maximum limit that the material can support $\mathcal{S}_{\max}$, the grain will be disrupted by Coulomb explosions.

Setting $\mathcal{S}=\mathcal{S}_{\max}$, we can derive the maximum surface potential and charge that the grain still survives:
\bea
\phi_{\max}\simeq 1.06\times 10^{3}\left(\frac{\mathcal{S}_{\max}}{10^{10} {\rm dyn} \cm^{-2}}\right)^{1/2}a_{-5} {\rm V},\label{eq:phimax}\\
Z_{\max}\simeq 7.4\times 10^4 \left(\frac{\mathcal{S}_{\max}}{10^{10}{\rm dyn} \cm^{-2}}\right)^{1/2}a_{-5}^{2}.\label{eq:Zmax}
\ena

The value $\mathcal{S}_{\max}$ is uncertain due to the uncertainty in the grain material. Experimental measurements for ideal material provide $\mathcal{S}_{\max}\sim 10^{11} {\rm dyn} \cm^{-2}$. Throughout this paper we take a typical material $\mathcal{S}_{\max}\sim 10^{10} {\rm dyn} \cm^{-2}$ for our numerical considerations unless stated  otherwise. 

\subsubsection{Ion field emission}
When a grain is positively charged to a sufficiently strong electric field, the emission of individual ions (ion field emission) from the grain surface can take place. Experiments show that with an electric field $\phi/a \sim 3\times 10^{8}V\cm^{-1}$, ion field emission already occurs for some metals (see Table 1 in \citealt{1970PSSAR...1..513T}). Thus, grains may gradually be destroyed by ion field emission without Coulomb explosions in the case of ideal material with $\mathcal{S}_{\max}\sim 10^{11} {\rm dyn}\cm^{-2}$ (i.e., $\phi_{\max}/a \sim 3\times 10^{8}{\rm V}\cm^{-1}$).

\subsection{Photoelectric Yield}
\begin{figure*}
\centering
\includegraphics[width=0.4\textwidth]{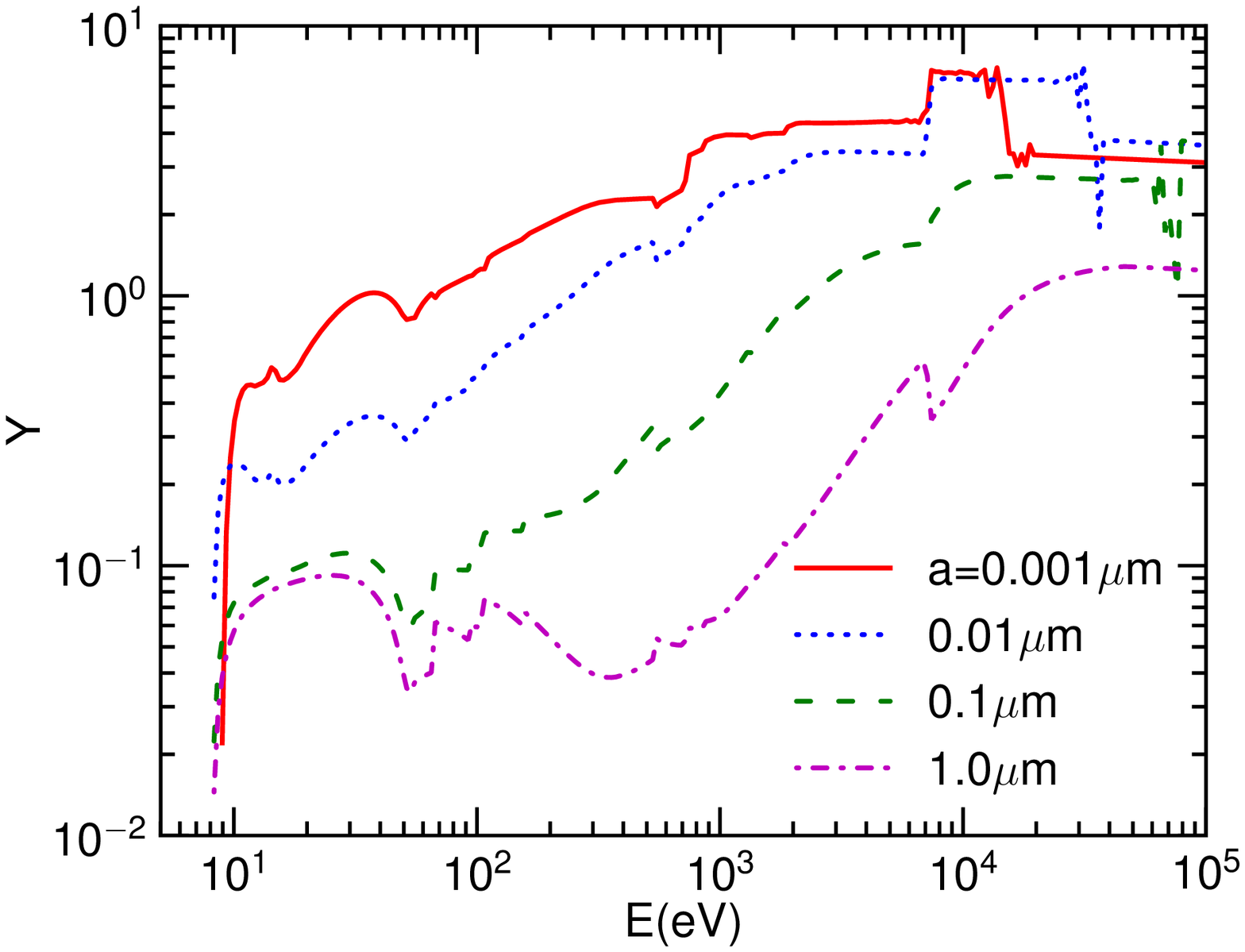}
\includegraphics[width=0.4\textwidth]{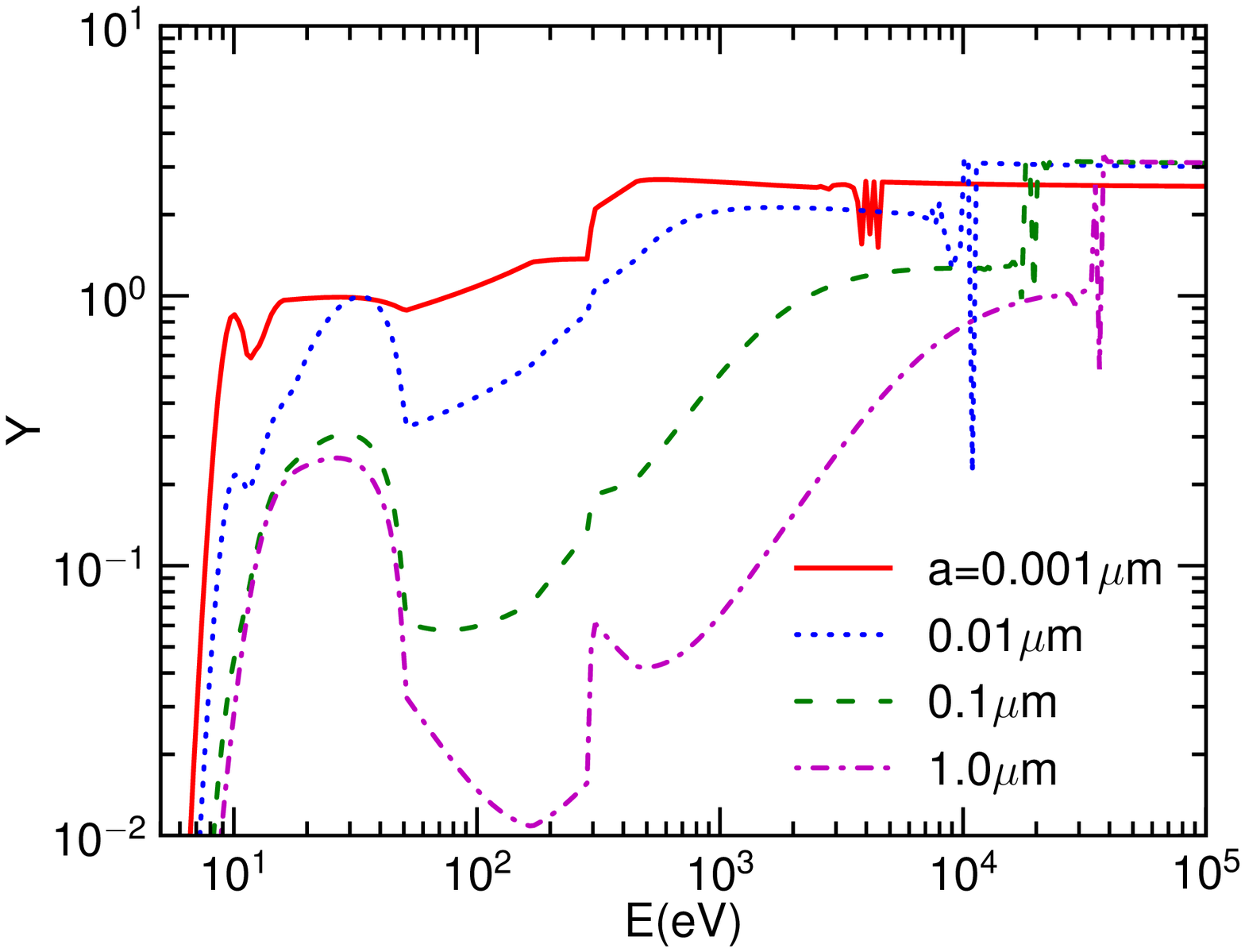}
\caption{Total photoelectric yield due to the emission of primary, Auger and secondary electrons as a function of photon energy for silicate (left) and graphite (right). Neutral grains ($Z=0$) and different grain sizes are considered.}
\label{fig:Y}
\end{figure*}

\begin{figure*}
\centering
\includegraphics[width=0.4\textwidth]{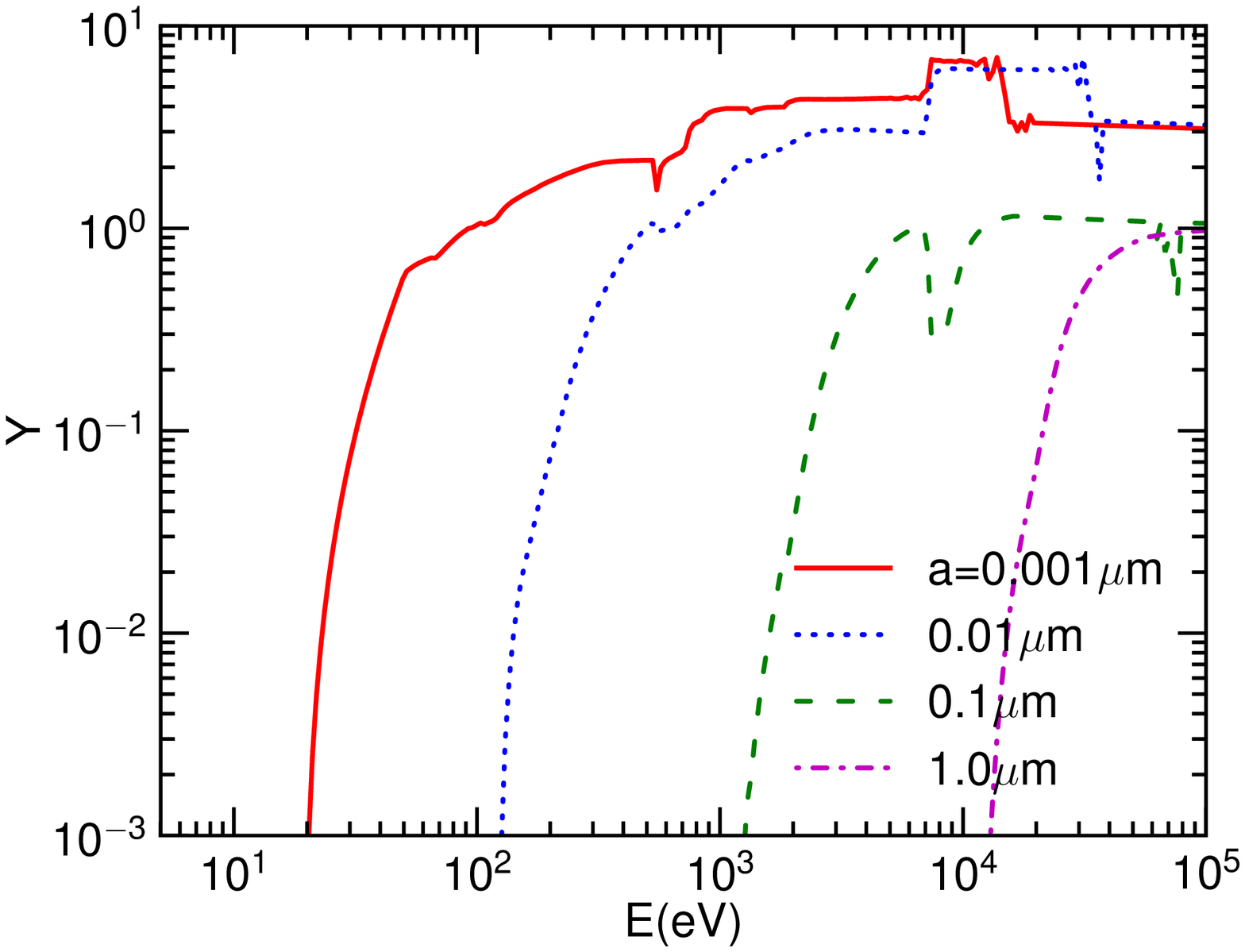}
\includegraphics[width=0.4\textwidth]{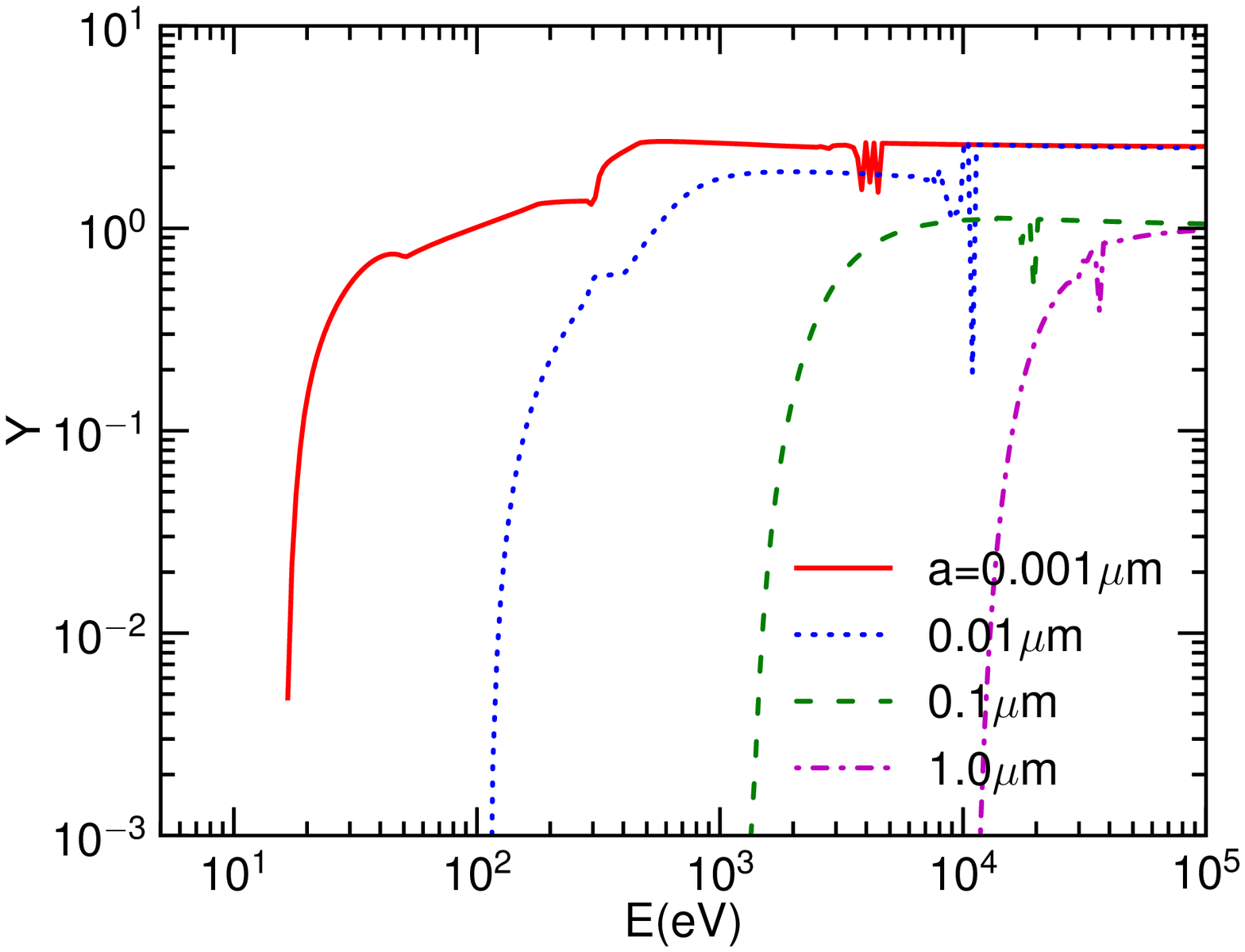}
\caption{Same as Figure \ref{fig:Y} but for grains with $Z=Z_{\max}$. The yield for the low-energy part is significantly suppressed due to the increased potential barrier determined by $Z_{\max}$.}
\label{fig:YZmax}
\end{figure*}

To calculate photoelectric yields, we use the methods from WDB06 for high-energy photons ($E>50 \eV$) and \cite{2001ApJS..134..263W} for low-energy photon ($E<20 \eV$). For $20 \eV < E<50 \eV$, interpolation between two energy ranges is performed. We assume that silicate material is made of MgFeSiO$_{4}$ while graphite material purely consists of carbons. The mass density is $\rho=3.5\g\cm^{-3}$ and $2.2\g\cm^{-3}$ for silicate and graphite, respectively. The ionization potentials $I_{i,s}$ for different shells and elements are similar to those in Table 1 of WDB06. The energy and averaged number of Auger electrons following a primary transition are taken from Table 4.1 and 4.2 in DS96.

Figure \ref{fig:Y} shows the total photoelectric yield (sum over the yields of primary photoelectrons, Auger and secondary electrons) for a neutral silicate (left panel) and graphite (right panel) grain of different sizes. The yield is higher for smaller grains and can be more than unity (i.e., one energetic photon can produce more than one free electron).

Figure \ref{fig:YZmax} shows the photoelectric yield computed for a grain of charge $Z=Z_{\max}$ given by Equation (\ref{eq:Zmax}). The yield $Y$ tends to zero when the photon energy becomes lower than the ionization potential $\IP(Z=Z_{\max})$. Moreover, for photon energy higher than $\IP(Z_{\max})$, $Y$ is also decreased compared to that for the neutral grain due to the higher potential ionization.

\subsection{Coulomb explosions in the solar radiation field}
{Let us assume for now that relativistic dust can enter the solar system and study grain destruction by Coulomb explosions in the solar radiation field with spectrum $B_{\nu}(T_{\odot})$. Discussion on whether relativistic dust can survive traversing the ISM to reach the solar system is presented in Section \ref{sec:ISM}.}

Since the solar energy density varies with solar distance as $u_{\rad}\propto 1/r^{2}$, the charging rate (Eq. \ref{eq:dZdt}) can be rewritten as:
\bea
\frac{dZ}{dt'}=J_{\rm phe}(a,Z)_{\AU}\left(\frac{\AU}{r}\right)^{2},
\ena
where $J_{\rm phe}(a,Z)_{\AU}$ is the charging rate calculated at distance $r=1\AU$. Hence,
\bea
\frac{dZ}{J_{\rm phe}(a,Z)_{\AU}}=\AU^{2}\frac{dt'}{r^{2}}=\frac{\AU^{2}}{c\beta\gamma}\frac{dr}{r^{2}},\label{eq:dZdt_sun}
\ena
where $dr=c\beta dt=c\beta \gamma dt'$ has been used.

To determine the solar distance at which $Z=Z_{\max}$, we integrate Equation (\ref{eq:dZdt_sun}) from initial charge $Z_{i}$ to $Z_{\max}$, which correspond to $r=R_{i}$ to $r=R_{\max}$. It yields
\bea
\left(\frac{1}{R_{\max}}-\frac{1}{R_{i}}\right) = \frac{c\beta\gamma}{\AU^{2}}\times \int_{Z_{i}}^{Z_{\max}}
\frac{dZ}{J_{\rm phe}(a,Z)_{\AU}}.\label{eq:Rinv}
\ena

Assuming that relativistic grains enter the solar system from the ISM ($R_{i}\gg 1$ AU) with $Z_{i}$ equal to the mean grain charge in the ISM, then it seems adequate to take $Z_{i}=0$ because $Z_{\max}$ is essentially much larger than the mean charge of interstellar grains (see \citealt{2001ApJS..134..263W}). 

To numerically compute the integral (\ref{eq:Rinv}), we first calculate the photoelectric yield $Y(a,Z)$ and $J_{\tot}(a,Z)_{\AU}$ from $Z=0$ to $Z=Z_{\max}$.

\begin{figure*}
\centering
\includegraphics[width=0.4\textwidth]{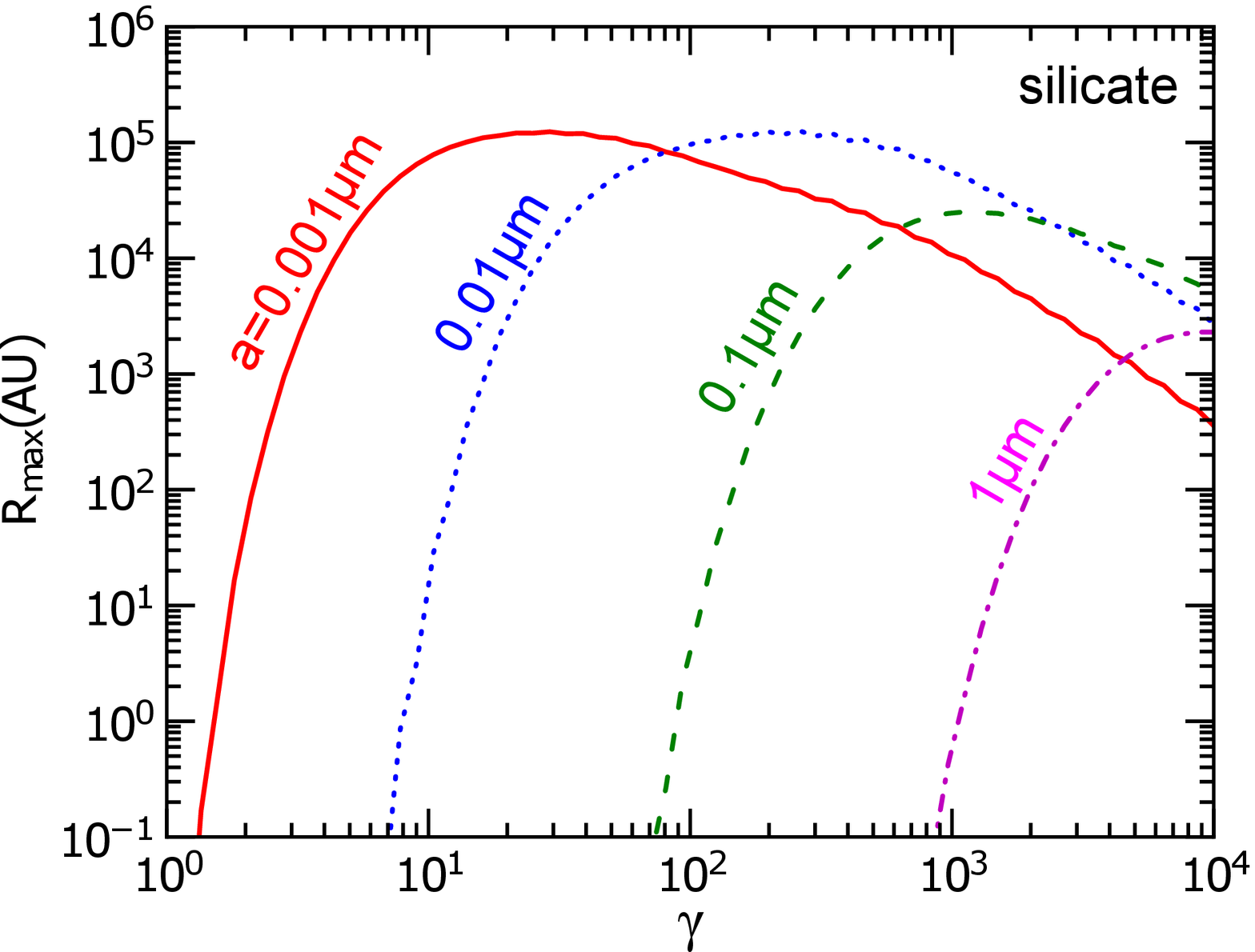}
\includegraphics[width=0.4\textwidth]{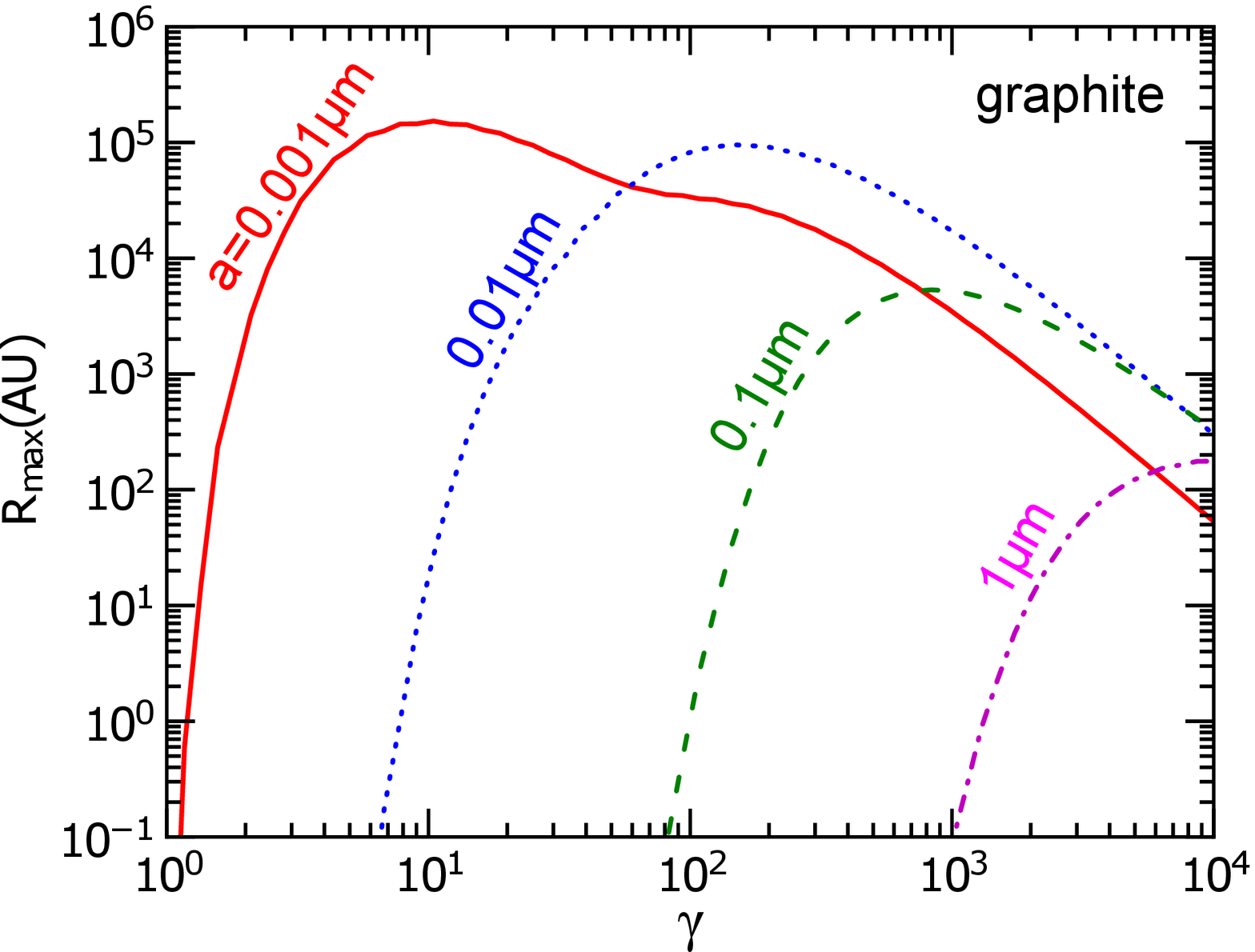}
\caption{Solar distance $R_{\rm max}$ that a relativistic grain coming from the ISM is electrically charged to $Z_{\max}$ for silicate grains (left) and graphite grains (right panel). Typical maximum tensile strength $\mathcal{S}_{\max}=10^{10} {\rm dyn} \cm^{-2}$ is adopted. All grains considered with $\gamma>2\times 10^{3}$ explode in solar radiation at large distance from the Sun, while some grains with lower $\gamma$ can reach the Earth's atmosphere.}
\label{fig:Rmax}
\end{figure*}

Figure \ref{fig:Rmax} shows the obtained results for $R_{\rm max}$ as a function of $\gamma$ for different grain sizes. As shown, some grains have $R_{\rm max}<1\AU$, i.e., they can survive Coulomb disruptions and reach the Earth's atmosphere. For instance, the $a=0.1\mum$ grains can reach the Earth for $\gamma<10^{2}$ and are destroyed above this value. For all grain sizes considered, $R_{\max}$ tends to fall after reaching its peak, which directly arises from the decrease of $Q_{\abs}$ at highest photon energy (see Figure \ref{fig:Qabs}) that decreases the charging rate. 

For $\gamma>10^{4}$, the charging rate was derived using the Compton scattering theory by \cite{1999APh....12...35B}. Using their result, we can estimate grain charge at distance $r= 1$ AU as the following:
\bea
Z_{\rm C}(r=1\AU)&\simeq &8.8\times 10^{10}\left(\frac{\AU}{c\beta}\right)a_{-5}^{3} \exp\left(-\frac{0.4a_{-5}^{1/4}\hat{\rho}^{1/4} }{\gamma_{5}}\right),\nonumber\\
&\simeq&4.4\times 10^{13} a_{-5}^{3}\beta^{-1} \exp\left(-\frac{0.4a_{-5}^{1/4}\hat{\rho}^{1/4} }{\gamma_{5}}\right).~~~
\ena

Comparing $Z_{\rm C}(r=1\AU)$ and $Z_{\max}$ it can be seen that relativistic grains with $\gamma > 10^{4}$ are completely destroyed by Coulomb explosions in solar radiation.

\section{Destruction of relativistic dust by Sputtering and Grain-Grain Collisions}\label{sec:sputt}

\subsection{Introduction to Electronic Sputtering}
The ejection of atoms from dust grains (i.e., sputtering) in hot gas and interstellar shocks for the low-energy regime has been well studied (\citealt{1979ApJ...231..438D}; \citealt{1994ApJ...431..321T}). This {\it knock-on} sputtering arises from elastic collisions in which the kinetic energy of incident ions is directly converted to the motion of target atoms (see \citealt{1981spb1.book....9S}). When the velocity of incident ions is sufficiently large (a fraction of the Bohr velocity $v_{0}$), electronic excitations can result in the ejection of atoms. This {\it electronic sputtering} process has been studied extensively both by numerical simulations and experiments. 

Two leading models have been proposed to explain electronic sputtering, including {\it thermal spike} and {\it Coulomb explosion} (see \citealt{Sigmund:2005tw} and \citealt{2003ssac.proc..357B} for reviews). 

In the {\it thermal spike} model, electronic sputtering is believed to occur as follows. After the passage of a swift heavy ion, on a timescale $\sim 10^{-17}-10^{-13}\s$, electronic interactions can produce a large number of secondary electrons in a narrow cylinder along the ion's path through the grain. In $10^{-13}-10^{-11}\s$, secondary electrons rapidly transfer their energy to surrounding atoms to create a hot ionization track from which some highly excited atoms can gain sufficient kinetic energy to break atomic bonds and escape from the grain surface. 

In the {\it Coulomb explosion} model, electrostatic repulsion between transiently ionized atoms in the ionization track (\citealt{1965JAP....36.3645F}) directly converts electrostatic energy to atomic motion, which can result in the ejection of atoms near the grain surface \citep{1982NIMPR.198..103J}. In both models, the formation and characteristics of the ionization track are a key factor for computing the yield of electronic sputtering.

\subsection{Track Radius and Sputtering Yield}\label{subsec:sputt}
\subsubsection{Track Radius}
The radius of the ionization track created by an energetic heavy ion, $r_{\rm cyl}$, can be calculated using the bond-breaking model \citep{1994NIMPB..94..424T}:
\bea
r_{\rm cyl}=\left(\frac{\eta}{2\pi e_{c}}\frac{dE}{dx}\right)^{1/(2-\eta)}R_{\perp}^{-\eta/(2-\eta)},\label{eq:rcyl}
\ena
where $e_{c}$ is the critical energy density for bond breaking, $0<\eta<1$ is a constant obtained from fitting to experimental data, and $R_{\perp}=(840\epsilon/\rho)$ \AA~ is the maximum radial range of secondary electrons in the material with $\epsilon$ being the ion kinetic energy per mass units in MeV/amu. 

The track radius obtained from Equation (\ref{eq:rcyl}) was found in good agreement with the experimental data (see e.g., \citealt{2007ApJ...662..372B}). For our calculations for silicate grains, we use the good fit parameters from \cite{1994NIMPB..94..424T} $e_{c}=0.024 \eV$\AA$^{-3}$ and $\eta=0.27$ for Fe ion bombarding SiO$_{2}$. For carbonaceous grains, we adopt $e_{c}=0.014\eV$\AA$^{-3}$ for polysterine (C$_{8}$H$_{8}$).

For heavy energetic ions, all atoms in the hot ionization track of height $l$ can be excited, and it is useful to define an average excitation energy per atom by
\bea
E_{\rm exc} = \frac{ldE/dx}{n_{d}l\pi r_{\rm cyl}^{2}}=\frac{dE/dx}{n_{d}\pi r_{\rm cyl}^{2}},\label{eq:Eexc}
\ena
where $n_{d}$ is the atomic number density in the grain. 

\subsubsection{Sputtering Regimes}

Let $U_{0}$ be the binding energy of the grain material and $Y_{{\sp},i}$ be the sputtering yield induced by incident ion/atom $i$. 

For $E_{\rm exc}> U_{0}$, namely {\it high-excitation energy} regime, the sputtering yield can be described by:
\bea
Y_{{\sp},i}=C_{\rm high}\left(\frac{r_{\rm cyl}dE/dx}{U_{0}}\right),\label{eq:Ylinear}
\ena
where $C_{\rm high}$ is a coefficient that describes the fraction of $dE/dx$ converted to atomic motion and the effects of incident angle of ions for the high-excitation regime (see \citealt{2000SurSc.451..108B}). 

For $E_{\rm exc}<U_{0}$, so-called {\it low-excitation energy} regime, the sputtering yield depends nonlinearly on $dE/dx$: 
\bea
Y_{{\sp},i}=C_{\rm low} \left(\frac{ f ldE/dx}{U_{0}}\right)^{q},\label{eq:Ysplow}
\ena
where $f$ is the fraction of $dE/dx$ converted to atomic motion, $C_{\rm low}$ is a coefficient for this low-excitation regime, and $q\ge 3$ (see \citealt{2007NIMPB.256..333M} for experiments and \citealt{1994PhRvB..49..786U} and \citealt{1999NIMPB.152..267B} for simulations).

\subsubsection{Sputtering Yield}
{Figure \ref{fig:Ysp_SiO2} shows the sputtering yield as a function of $dE/dx$ for SiO$_{2}$ from experiments \citep{2002NIMPB.193..830M}, and the yield calculated for Fe ion using Equation (\ref{eq:Ylinear}) with $C_{\rm high}=0.02$ (solid line) and Equation (\ref{eq:Ysplow}) with $C_{\rm low}=0.02, f=0.05$ and $q=3.3$ (dashed line). We can see that the yield obtained in the former case fits well to the experimental data for $dE/dx>8\times 10^{10}\eV\cm^{-1}$. For lower energy loss $dE/dx$, the yield calculated in the latter case better fits to the experimental data.} 

The figure also shows the experimental results from \cite{Khan:2013jg} for graphite material. For the same $dE/dx$, $Y_{\rm sp}$ is significantly lower than that for SiO$_{2}$ and in good agreement with $Y_{\rm sp}$ calculated by the threshold regime with $C_{\rm low}=0.005, f=0.02$ and $q=3.3$ (dotted line). The lower $f$ value required to fit the data for graphite perhaps is due to its high conductivity for which photoelectron energy is rapidly transferred to the entire grain rather than being localized within the ionization track.
 
\begin{figure}
\centering
\includegraphics[width=0.4\textwidth]{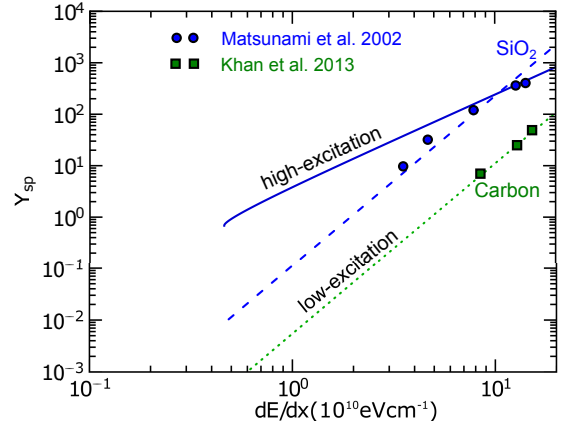}
\caption{Electronic sputtering yield $Y_{\sp}$ vs. stopping power $dE/dx$ from experiments for SiO$_{2}$ (filled circles) and graphite (carbon solid; square symbols). Yields for SiO$_{2}$ calculated in the high- and low-excitation energy regimes are shown in solid and dashed lines. Dotted line shows calculations for graphite in the low-excitation energy regime.}
\label{fig:Ysp_SiO2}
\end{figure}

Results in Section \ref{sec:coll} (e.g., Figure \ref{fig:dEdx}, right panel) indicate that relativistic atoms have energy loss $dE/dx < 2\times 10^{10} \eV\cm^{-1}$, for which the sputtering can be described by the low-excitation energy regime (see Figure \ref{fig:Ysp_SiO2}). Therefore, we use Equation (\ref{eq:Ysplow}) with the obtained parameters $C_{\rm low}, f, q$ to compute the sputtering yield for different atoms. Since the electronic sputtering is determined by ionizations arising from both electron-electron and ion-electron interactions, the energy loss $dE/dx$ adopted for calculations of sputtering yield is then equal to the total energy loss of the atom (nuclei and electrons) minus the radiative energy loss of electrons.

The sputtering yields by different atoms are shown in Figure \ref{fig:Ysp} for silicate (left panel) and graphite (right panel) grains. We adopt $U_{0}=6.4$ eV for silicate and $U_{0}=4$ eV for graphite grains, respectively (see \citealt{1994ApJ...431..321T}). 
We can see that the electronic sputtering is very sensitive to ion atomic mass, which is determined by their energy loss in the grain. The sputtering from heavy Fe is the most important, which can reach $Y_{\rm sp}\sim 10^{-2}$ atoms/ion for silicate material. The sputtering by abundant, light elements (H, He) is negligible. The sputtering yield for graphite material is two order of magnitudes smaller.

\subsection{Destruction by Electronic Sputtering}

Let $X_{i}$ be the abundance of element $i$ relative to hydrogen. The net sputtering yield is obtained by summing $Y_{{\sp},i}$ over all atomic elements in the ISM with their respective abundances:
\bea
Y_{\sp}=\sum_{i} X_{i}Y_{{\sp},i},
\ena
where $X_{i}$ are taken to be solar abundances given in \cite{2011piim.book.....D}, although the abundance of elements in the ISM varies with host galaxies.

The sputtering of atoms from the grain surface results in the decrease in the grain mass at rate:
\bea
\frac{dM_{\rm sp}}{dt}=-\frac{mdN_{\sp}}{dt}=n_{\gas}\beta c\pi a^{2}mY_{\sp},
\ena
where $m$ is the mean atomic mass of dust.

The sputtering rate is given by
\bea
\frac{da}{dt} &=& \frac{1}{4\pi a^{2}\rho}\frac{dM_{\rm sp}}{dt}=- \left(\frac{n_{\gas}\beta c}{4\rho}\right) m Y_{\sp},\nonumber\\
&\simeq &-0.01 \hat{n}_{\gas}\frac{m}{m_{\rm C}}\left(\frac{\beta}{0.1}\right)\left(\frac{Y_{\sp}}{10^{-6}}\right)\mum {\rm~ Myr}^{-1},~~~\label{eq:dadt}
\ena
where $\hat{n}_{\gas}=n_{\gas}/10\cm^{-3}$ and $m_{\rm C}$ carbon atomic mass.

Assuming that the slowing down of relativistic dust is negligible, we can estimate the total column gas density that the grain has collided with until its complete destruction by sputtering as follows:
\bea
N_{\coll}&=&\int n_{\gas} dr  =\frac{4\rho a}{mY_{\sp}},\nonumber\\
&\simeq& 6\times 10^{24}\hat{\rho}a_{-5}\frac{m_{\rm C}}{m}\left(\frac{10^{-6}}{Y_{\sp}}\right)\cm^{-2},\label{eq:Ncoll}~~~~~~
\ena
where $dr = \beta c dt=4\rho da/(n_{\gas}mY_{\sp})$ and Equation (\ref{eq:dadt}) have been used.

{Figure \ref{fig:dadtsp} shows sputtering rate $|da/dt|$ induced by the different atoms in the diffuse medium with $n_{\gas}=10\cm^{-3}$.} As shown $|da/dt|$ rapidly declines with $\gamma$ to its minimum at $\gamma\sim 2$ and starts to rise with increasing $\gamma$. It follows that the sputtering time is $\tau_{\sp}= a/|da/dt| > 2 a_{-5}{\rm Myr}$ for silicate grains and $\tau_{\sp}> 200 a_{-5}{\rm Myr}$ for graphite grains with $\gamma \gg 1$.

Relativistic silicate grains having $Y_{\sp}\sim 10^{-6}$ would be destroyed after sweeping up a gas column density $N_{\coll}\sim 10^{25}a_{-5}\cm^{-2}$, while graphite grains would be destroyed after $N_{\coll}\sim 10^{27}a_{-5}\cm^{-2}$ with $Y_{\sp}\sim 10^{-8}$.

\begin{figure*}
\centering
\includegraphics[width=0.4\textwidth]{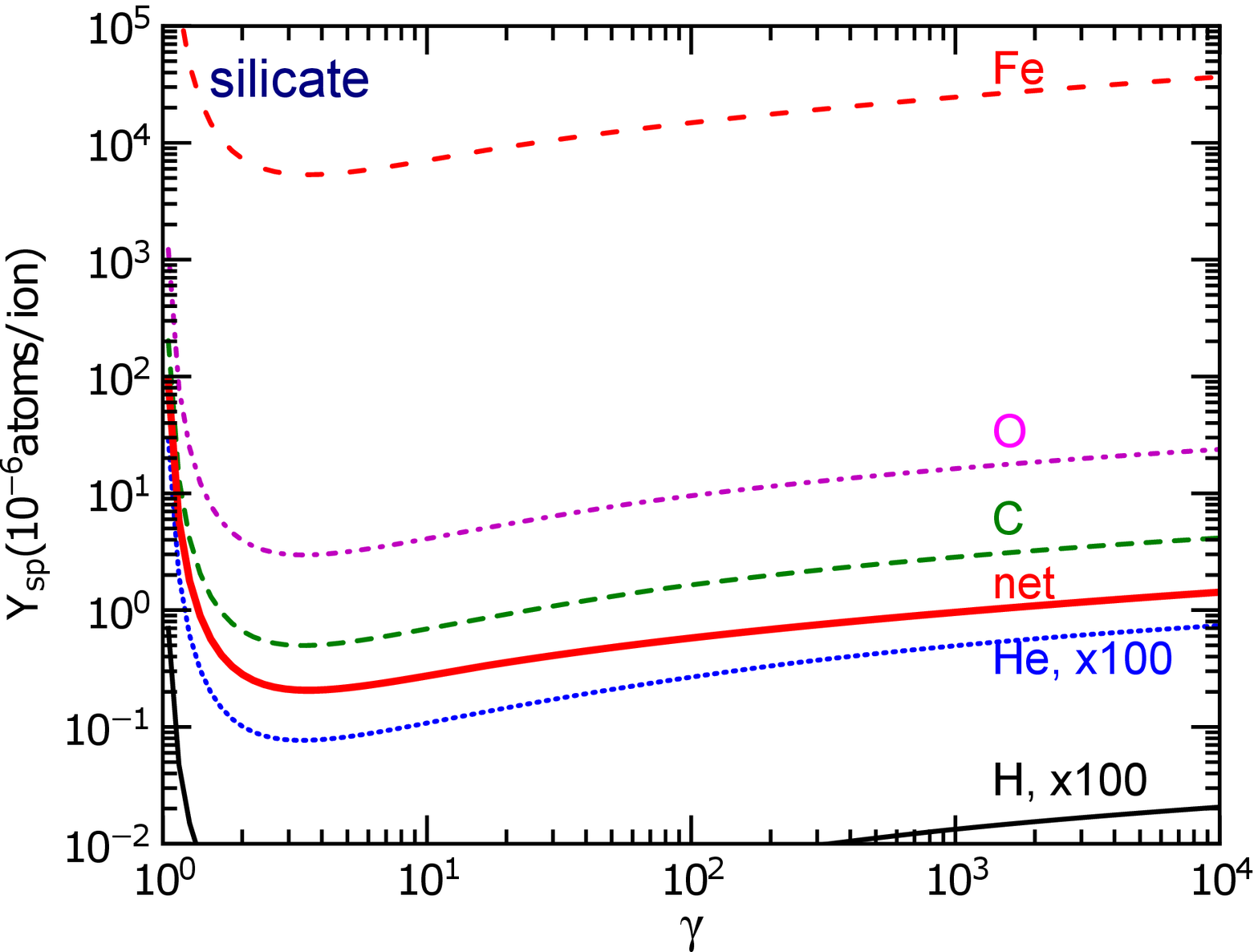}
\includegraphics[width=0.4\textwidth]{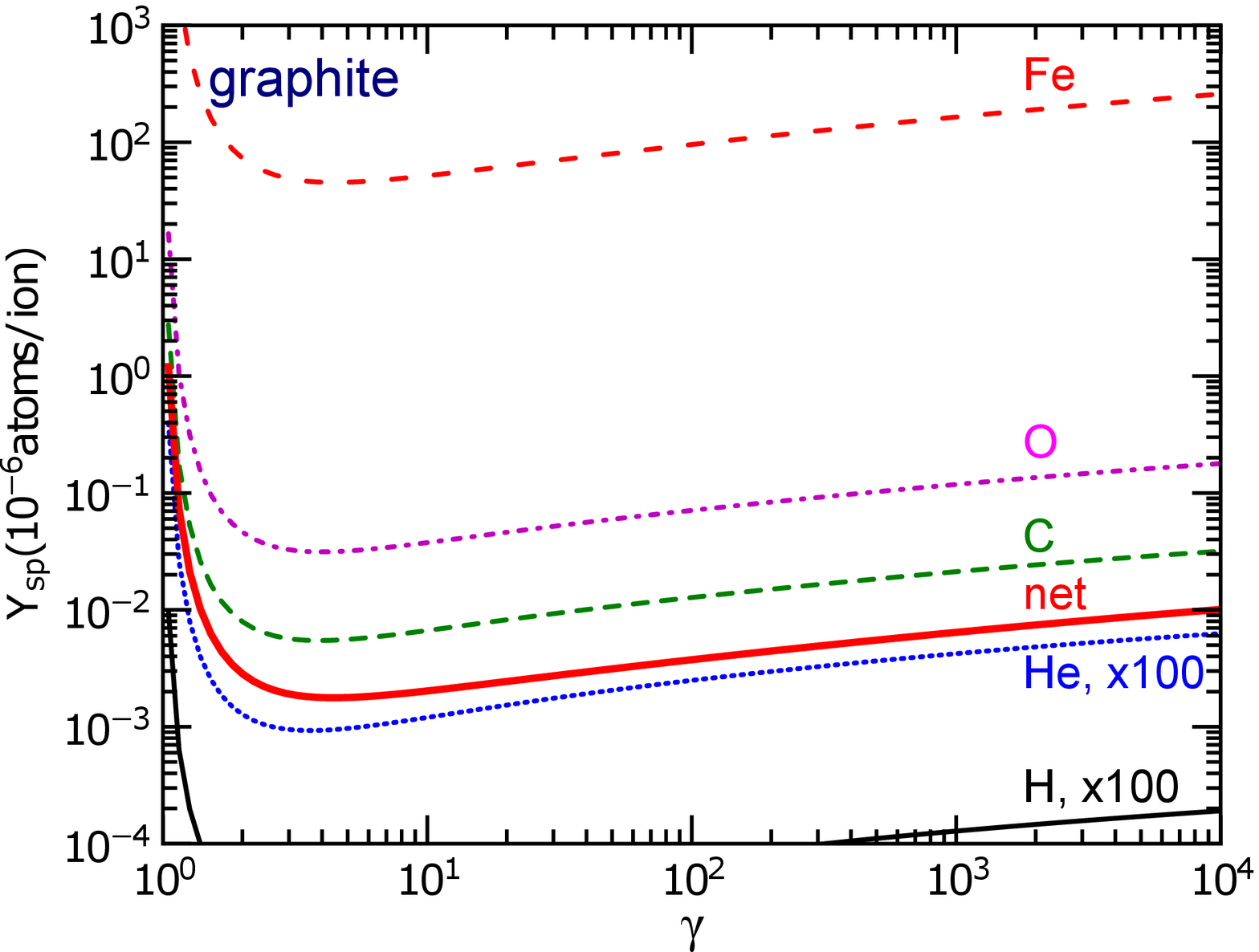}
\caption{Electronic sputtering yield $Y_{\sp}$ vs. $\gamma$ for silicate (left) and graphite grains (right) bombarded by H, He, C, O and Fe atoms. The net sputtering yield by all atoms accounting for their abundances is shown in the thick solid lines.}
\label{fig:Ysp}
\end{figure*}

\begin{figure*}
\centering
\includegraphics[width=0.4\textwidth]{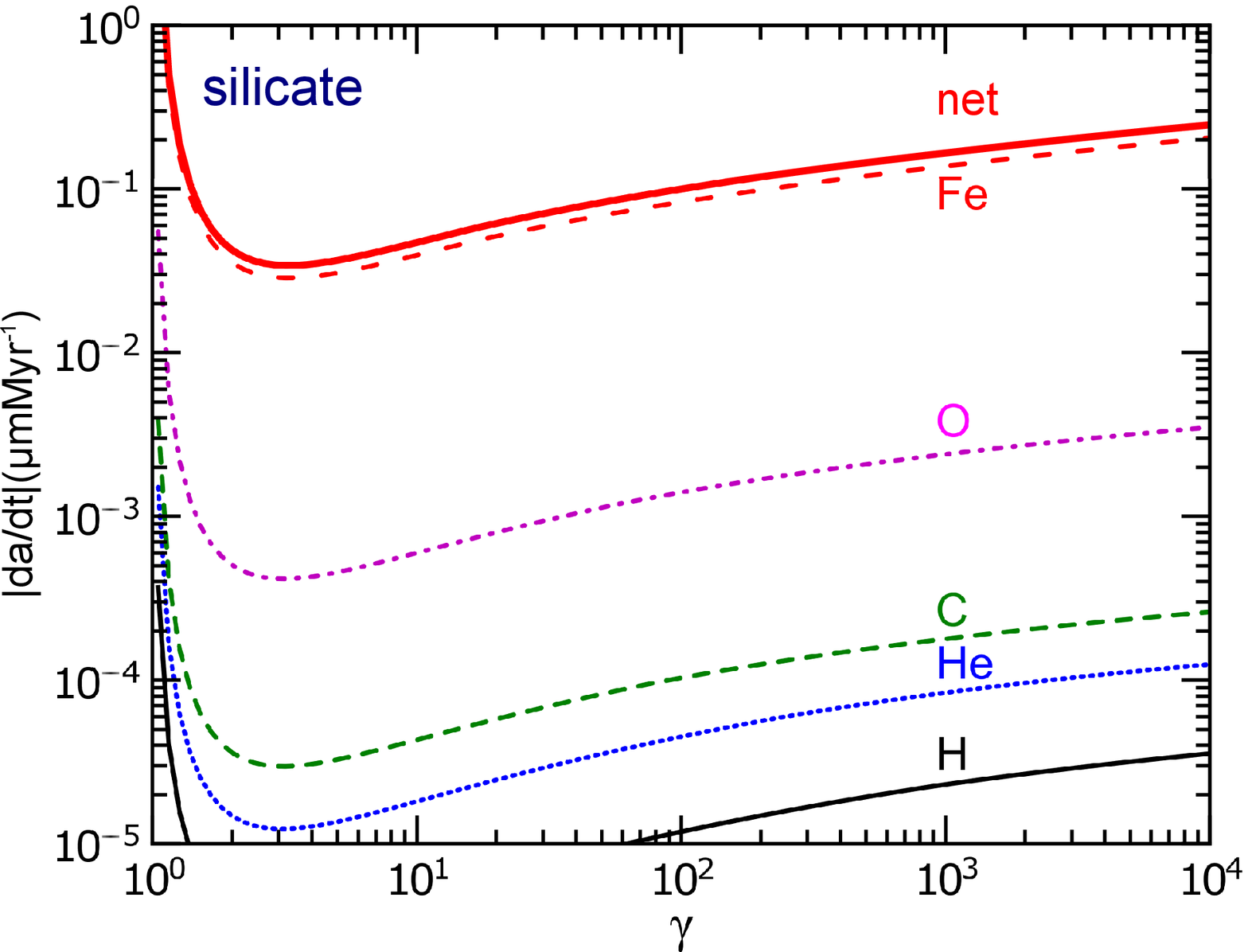}
\includegraphics[width=0.4\textwidth]{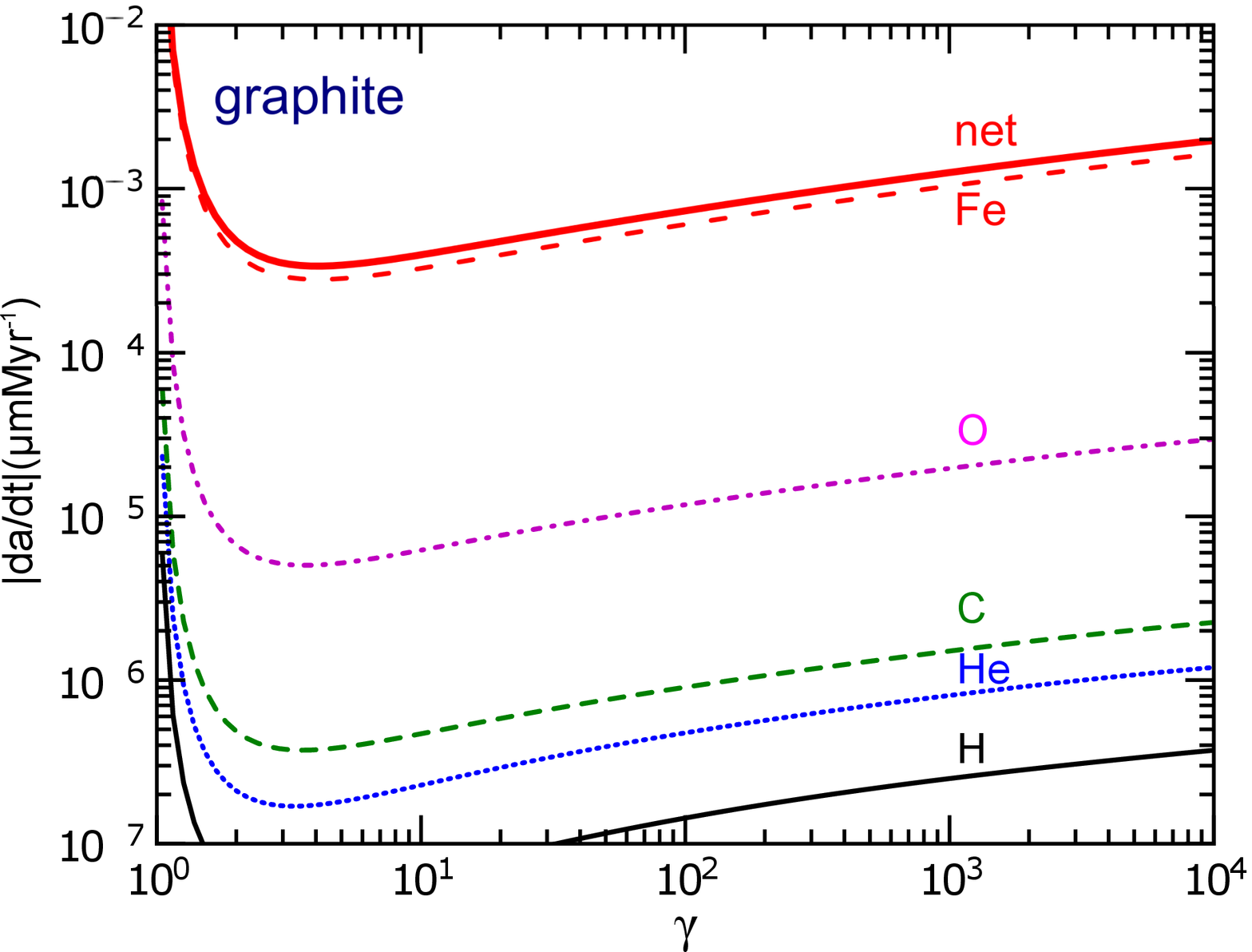}
\caption{Sputtering rate $|da/dt|$ vs. $\gamma$ for silicate (left) and graphite grains (right) bombarded by several atoms accounting for their abundances. The net sputtering rate is shown in thick solid lines. Sputtering by Fe is dominant and sputtering by light ions is negligible.}
\label{fig:dadtsp}
\end{figure*}

\subsection{Evaporation due to Grain-Grain Collisions}

{For relativistic speeds, grain-grain collision can be considered as the simultaneous bombardment of a target grain by a large number of neutral atoms from the projectile grain. Both target and projectile grains can rapidly be heated to high temperature by secondary electrons produced during the collision and evaporate. 

Let $a$ and $a_{T}$ be the radius of the projectile and target grain, respectively. Here the target grain is moving with $\gamma \gg 1$ through the ambient gas. The energy transferred from the projectile to the target grain is equal to 
\bea
\Delta E = N_{P}\times \frac{4a_{T}}{3}\frac{dE}{dx},\label{eq:N_dEdx}
\ena
where $dE/dx$ is the energy loss per projectile atom per pathlength in the target grain, and $N_{P}=n_{d}4\pi a^{3}/3$ is the total number of atoms in the projectile grain.

According to the Debye model at high temperature, the volume heat capacity is $C_{V}= 3 N_{T}k_{\B}$ where $N_{T}=n_{d}4\pi a_{T}^{3}/3$ is the total number of atoms in the target grain. Thus, the transient temperature of the target is
\bea
T_{d}&=&\frac{\Delta E}{C_{V}}=  \left(\frac{a}{a_{T}}\right)^{3}\frac{4a_{T}}{9}\frac{dE/dx}{k_{\B}},\nonumber\\
&\simeq & 2.1\times 10^{7}\left(\frac{a}{a_{T}}\right)^{3}a_{T,-5}\left(\frac{dE/dx}{4\times 10^{8}\eV\cm^{-1}}\right) \K,~~~\label{eq:Td_coll}
\ena
where $n_{d}\approx 10^{23}\cm^{-3}$ for both the target and projectile grains.

If $T_{d} \ge U_{0}/3k_{B}$ (i.e., the resulting thermal energy per atom is above the binding energy), then the solid grain suddenly transforms to vapor state, resulting in complete destruction of the grain (see e.g., \citealt{1994ApJ...431..321T}). The critical size $a_{c}$ of the projectile that heats the target grain to $T_{d}=U_{0}/3k_{\B}$ can be evaluated using Equation (\ref{eq:Td_coll}) as follows:
\bea
\frac{a_{c}}{a_{T}} &=&\left(\frac{3U_{0}}{4a_{T}dE/dx}\right)^{1/3},\nonumber\\
&\simeq& 0.1\left[a_{T,-5}^{-1}\frac{U_{0}}{6\eV}\left(\frac{4\times 10^{8}\eV}{dE/dx}\right)\right]^{1/3}.~~~\label{eq:ac_aT}
\ena

For $T_{\rm sub}< T_{d} < U_{0}/3k_{B}$, the sublimation occurs efficiently, resulting both in the loss of the grain mass and the evaporative cooling in addition to the radiative cooling. To study the dependence of mass loss of the grain versus the grain temperature, it is convenient to write
\bea
\frac{da}{dt} = \frac{da}{dT}\times \frac{dT}{dt}\label{eq:dadt_coll}
\ena
where $dT/dt$ is the decrease in grain temperature due to the cooling given by
\bea
\frac{dT}{dt} \equiv \frac{1}{3N_{T}k_{\B} }\left(\frac{dE_{\rm c, rad}}{dt} + \frac{dE_{\rm c, evap}}{dt}\right).\label{eq:dTdt}
\ena

Using $da/dt$ (absolute value) from Equation (\ref{eq:dasdt}) and $dT/dt$ from the above equation for (\ref{eq:dadt_coll}) one obtains 
\bea
\frac{da}{dT} &=&\frac{3N_{T}k_{\B}}{\left(dE_{\rm c, rad}/dt + dE_{\rm c, evap}/dt \right) }\left(n_{d}^{-1/3}\nu_{0}e^{-B/T_{d}}\right),\nonumber\\
&=&\left(\frac{a_{T}k_{\B}}{U_{0}}\right)\left(\frac{1}{\tau_{\rm c, rad}^{-1}/\tau_{\rm c, evap}^{-1} + 1}\right).\label{eq:dadT}
\ena 

Above, $\tau_{\rm c, rad}$ and $\tau_{\rm c, evap}$ are the radiative and evaporative cooling times given by
\bea
\tau_{\rm c, rad}^{-1} &=& \frac{dE_{\rm c,rad}/dt}{3N_{T}k_{\B}T_{d}}= \frac{\langle Q_{\abs}\rangle_{T_{d}}\sigma T_{d}^{4}}{n_{d} a_{T} k_{\B}T_{d}},\\
\tau_{\rm c, evap}^{-1} &=& \frac{dE_{\rm c,evap}/dt}{3N_{T}k_{\B}T_{d}}=\frac{n_{d}^{-1/3}\nu_{0}e^{-B/T_{d}}U_{0}}{a_{T}\kB T_{d}},
\ena
where $\langle Q_{\abs}\rangle_{T_{d}}$ is the Planck-averaged emission efficiency (see Eq. \ref{eq:Qabsavg}).

To calculate the fraction of the grain mass evaporated after a grain-grain collision, we first calculate the temperature ($T_{i}$) from Equation (\ref{eq:Td_coll}) with $dE/dx$ taken at $\gamma=2$ (i.e., the case of minimum energy loss, see Fig. \ref{fig:dEdx}). Then we integrate Equation (\ref{eq:dadT}) from $T_{i}$ to $T_{f}=1000\K$ to obtain the grain final size $a_{f}$. The value $T_{f}$ is chosen such that the sublimation is negligible (the grain mass gets stabilized). The fraction of grain mass loss is $f_{\rm sub}=\left(a_{i}^{3}-a_{f}^{3}\right)/a_{i}^{3}$. 

Figure \ref{fig:fm-evap} shows $f_{\rm sub}$ (solid lines) and $T_{d}$ (dashed lines) as functions of $a/a_{T}$ for both silicate (left) and graphite (right) grains with $a_{c}/a_{T}$ marked by filled circles. It can be seen that the mass loss is complete for $a/a_{T}\ge a_{c}/a_{T}$ and sharply declines below this limit as expected. In the following, we take $a_{c}/a_{T}$ from Equation (\ref{eq:ac_aT}) for the threshold of the complete evaporation by grain-grain collisions.

\begin{figure*}
\centering
\includegraphics[width=0.4\textwidth]{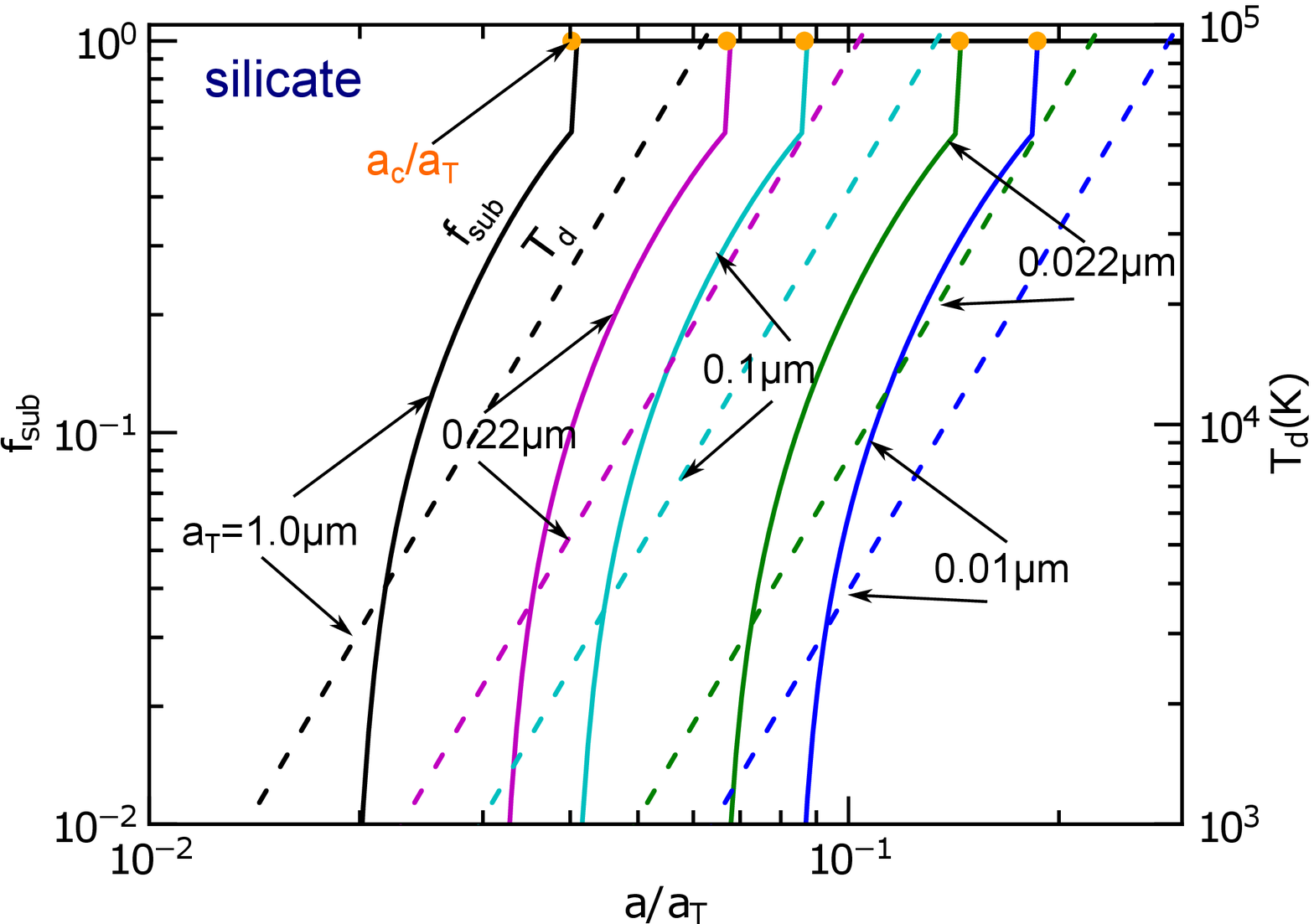}
\includegraphics[width=0.4\textwidth]{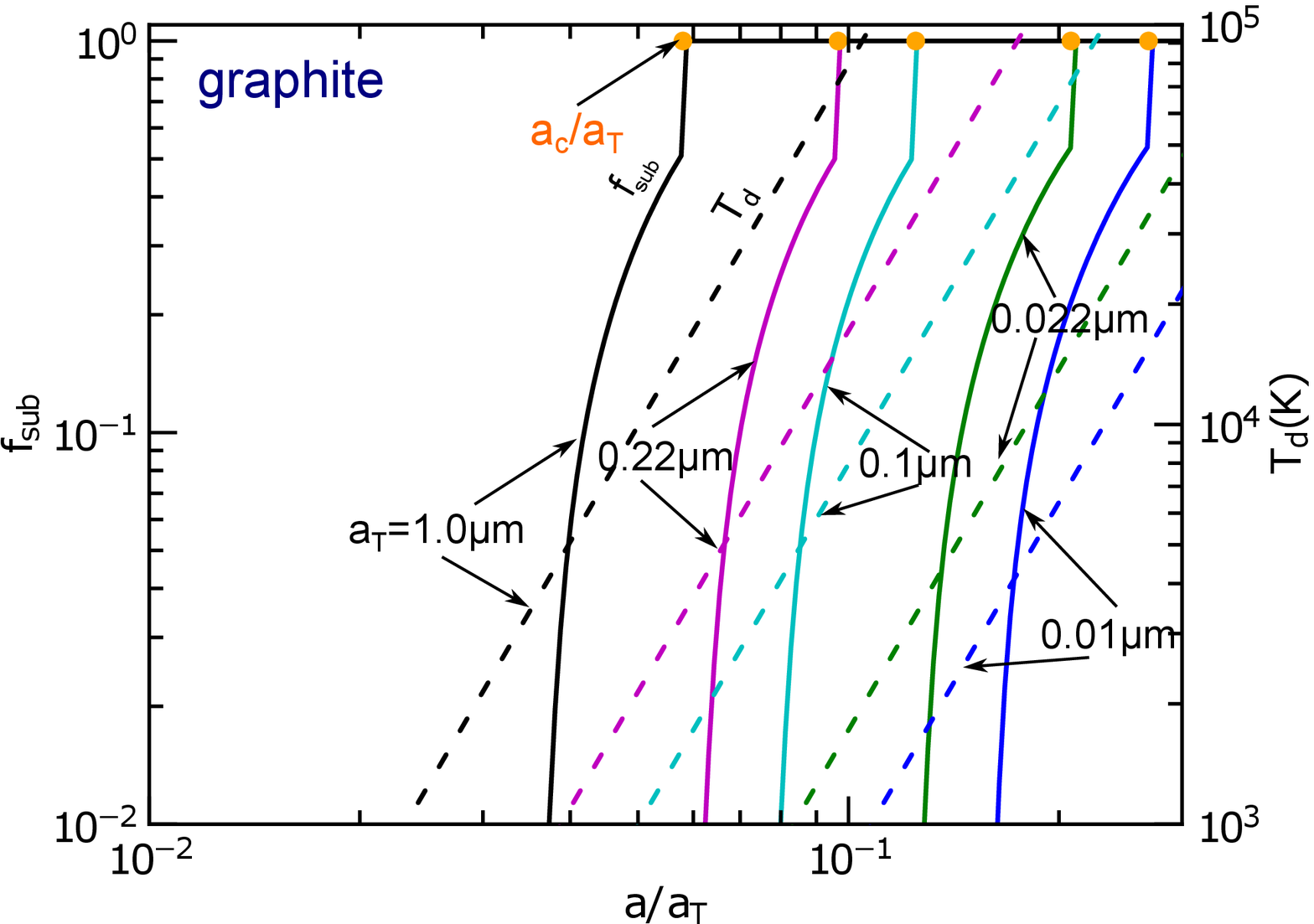}
\caption{Fraction of the grain mass evaporated after a single grain-grain collision versus the size ratio $a/a_{T}$ (solid lines) for quartz (left) and graphite (right). Dashed lines show $T_{d}$ obtained from Equation (\ref{eq:Td_coll}), and filled circles denote $a_{c}/a_{T}$ computed by Equation (\ref{eq:ac_aT}).}
\label{fig:fm-evap}
\end{figure*}

The rate of hard-sphere grain-grain collisions is given by
\bea
R_{\rm coll}(a,a_{T})=n(a)\beta c \pi (a^{2}+a_{T}^{2}),
\ena
where $n(a)=dn/da$ is the grain size distribution. The total collision rate that results in the complete evaporation of the target grain is obtained by integrating $R_{\rm coll}(a,a_{T})$ over the grain size distribution from $a_{c}$ to $a_{\max}$:
\bea
R_{\coll}(a_{T})= \int_{a_{c}}^{a_{\max}}\pi (a^{2}+a_{T}^{2})\beta c \frac{dn}{da} da,~~~~
\ena
where $a_{\max}$ is the upper cutoff of grain size distribution. 

For the size distribution $dn/da=n_{\gas} A_{\rm MRN}a^{-3.5}$ from \cite{Mathis:1977p3072} with $A_{\rm MRN}=10^{-25.16}\cm^{2.5}$, we obtain
\bea
R_{\coll}(a_{T})&=& n_{\gas}\beta c \pi A_{\rm MRN}\nonumber\\
&&\times\left[\frac{a_{T}^{2}}{2.5}\left(a_{c}^{-2.5}-a_{\max}^{-2.5}\right)-2\left(a_{c}^{-0.5}-a_{\max}^{-0.5} \right)\right].~~~~~\label{eq:Rcoll}
\ena

Assuming a constant gas-to-dust mass ratio, the column density of gas swept by the target grain before its complete destruction is
\bea
N_{\coll}&=&n_{\gas}\beta c R_{\coll}^{-1}\nonumber\\
&\simeq &\left(\pi A_{\rm MRN}\right)^{-1}\left[\frac{a_{T}^{2}}{2.5}\left(a_{c}^{-2.5}-a_{\max}^{-2.5}\right)-2\left(a_{c}^{-0.5}-a_{\max}^{-0.5} \right)\right]^{-1}.\label{eq:Ncollgr}
\ena

Using $a_{c}/a_{T}$ from Equation (\ref{eq:ac_aT}) for (\ref{eq:Ncollgr}) yields $N_{\coll}\sim 5\times 10^{19}- 5\times 10^{20}\cm^{-2}$ for $a_{T}\sim 0.01- 1\mum$.

\section{Discussion}\label{sec:discussion}

\subsection{Can dust grains be accelerated to Lorentz factor $\gamma \gg 1$?}
Let us begin our discussion with the original question whether dust grains can be accelerated to relativistic speeds with $\gamma \gg 1$.

First, radiation pressure by powerful radiation sources, such as quasars and Seyfert galaxies, is found to be able to accelerate grains to relativistic speeds as in earlier works. However, our improved calculations taking into account the redshift of the radiation spectrum in the grain's comoving frame and the explicit dependence of $\langle Q_{\rm pr}\rangle$ on the grain radius $a$ and $\gamma$ show that the terminal Lorentz factor cannot exceed $\gamma \sim 2$, even for grains initially located at distance $r_{\rm sub}$ from the central source. This result is much lower than earlier simple estimates which predict $\gamma$ up to $\sim 10$ with assumption of constant $Q_{\rm pr}=1$ (\citealt{1972Ap&SS..16..238H}; \citealt{1999APh....12...35B}). Meanwhile, radiation pressure from supernovae can only accelerate grains to $v\sim 0.1c$. 

Second, diffusive shocks from supernova remnants (SNRs), which are widely believed to be important for the acceleration of CRs \citep{1978ApJ...221L..29B}, are also found to be important for charged grains by theoretical study \citep{1997ApJ...487..197E} and numerical simulations \citep{2009ApJ...701.1865G}. \cite{1997ApJ...487..197E} estimated the maximum grain velocity achieved by the shock to be:
\bea
\beta_{\max}\simeq 0.024\eta^{-1/3}\left(\frac{V_{\rm sh}}{400 \km \s^{-1}}\right)^{2/3}a_{-5}^{-1/3}n_{\gas}^{-1/3}\phi_{10}^{1/3}B_{10}^{1/3},~~~\label{eq:vshock}
\ena
where the typical value $\eta=1$, $\phi_{10}=\phi/10$ V with $\phi$ the grain potential, $B_{10}=B/10\mu$G with $B$ the magnetic field strength, and $V_{\rm sh}$ is the shock velocity.

For a high shock speed $V_{\rm sh}=10^{4}\km \s^{-1}$, one obtains $\beta_{\max}\sim 0.2$ for $B\sim 10\mu$G. Since the magnetic field may be amplified in the shock by streaming instability (\citealt{2003A&A...412L..11B}; \citealt{2009ApJ...707.1541B}; \citealt{2010ApJ...717.1054R}), we expect the grain velocity acquired from shock acceleration would be enhanced, but it is unclear whether grains can achieve $\gamma \gg 1$.

Third, charged grains can be accelerated by resonance acceleration (\citealt{2002ApJ...566L.105L}; \citealt{Yan:2004ko}) and transit time damping (\citealt{Hoang:2012cx}) by fast modes of magnetohydrodynamic (MHD) turbulence. Very small dust grains (e.g., polycyclic aromatic hydrocarbons) can be accelerated to a few times thermal velocity by charge fluctuations (\citealt{Ivlev:2010hb}; \citealt{2012ApJ...761...96H}). Although the aforementioned acceleration mechanisms are only able to accelerate grains to subrelativistic speeds, they can provide sufficient injection energy for grains to enter diffusive shocks and to be accelerated to higher speeds. {It is worth noting that sputtering is expected to be important in shocks, which may destroy grains before being accelerated to relativistic speeds (see \citealt{2004ApJ...614..796S}).}

Fourth, the acceleration of charged grains may also occur due to magnetic reconnection that gets fast in turbulent environments (\citealt{1999ApJ...517..700L}; \citealt{2009ApJ...700...63K}; \citealt{2011ApJ...743...51E}; see \citealt{2014SSRv..181....1L} for a review). The corresponding first-order Fermi mechanism suggested for cosmic rays (\citealt{2005A&A...441..845D}; \citealt{2005AIPC..784...42L}) has been successfully tested numerically (\citealt{2012PhRvL.108x1102K}) and was invoked for explaining several space physics and astrophysical problems (see \citealt{2009ApJ...703....8L}; \citealt{2010ApJ...722..188L}). We are confident that a similar mechanism should be efficient for the acceleration of charged grains. 

Finally, relativistic shocks in extragalactic sources (e.g. AGN jets, gamma-ray bursts) may also be a potential source for relativistic grains. Detailed studies on this issue are beyond the scope of the present paper.

\subsection{Can relativistic dust survive in the ISM?}\label{sec:ISM}
Before entering the solar system, relativistic grains travel through the ISM, and their collisions with interstellar gas and dust are expected to be important for grain destruction. We have investigated in detail new destruction processes, including electronic sputtering by ions/neutrals and grain-grain collisions in Section \ref{sec:sputt}.

For electronic sputtering, we found that heavy Fe ions are the most important, which can produce a considerable yield ($Y_{\sp}\sim 10^{-2}$), whereas lighter ions (e.g., H, He, C) induce negligible sputtering yields. Using the solar abundances for gas elements and assuming that all Fe atoms are present in the gas phase, the net sputtering yield is estimated to be $Y_{\sp}\sim 10^{-6}$ and $10^{-8}$ for silicate and graphite grains, respectively. As a result, electronic sputtering only destroys the relativistic grain after it has swept a large column of gas, with $N_{\coll}\sim 10^{25}a_{-5}\cm^{-2}$ for silicate and $N_{\coll}\sim 10^{27}a_{-5}\cm^{-2}$ for graphite grains. 

For grain-grain collisions, we found it an efficient process for the destruction of relativistic grains. A grain-grain collision can impulsively heat the target grain to high temperature due to the energy transfer from secondary electrons to the entire grain. As a result, fast evaporation following a single collision can completely destroy the grains if the projectile grain is sufficiently large. For the MRN size distribution, we found that grain-grain collisions can destroy relativistic grains after sweeping a gas column density $5\times 10^{19}\cm^{-2}<N_{\coll} < 5\times 10^{20}\cm^{-2}$ for $0.01\mum < a< 1\mum$. The lifetime against grain-grain collisions is $\tau_{\coll, gg} = N_{\coll}/c\beta n_{\gas} < 55\beta^{-1}(10\cm^{-3}/n_{\gas})$ yr for $a< 1\mum$.

Let us now consider grain destruction related to grain heating by the interstellar gas and radiation field (ISRF). Assuming the isotropic ISRF with the energy spectrum $u(\nu)$ estimated in the solar neighborhood \citep{1983A&A...128..212M}, the density of radiation energy in the GF is related to $u(\nu)$ through Equation (\ref{eq:unu_iso}). Collisional heating by the ambient gas is  calculated by Equation (\ref{eq:collheat}). Thus, we can solve Equation (\ref{eq:Teq}) for grain equilibrium temperature $T_{d}$ for relativistic dust moving in the ISM.

Figure \ref{fig:TdISM} shows $T_{d}$ obtained for silicate and graphite grains traversing different regions of the ISM. For $\gamma=1$ (i.e., grains at rest in the gas), the grain temperature is dominated by radiative heating from interstellar photons due to inefficient collisional heating. For $\gamma> 1$, collisional heating becomes dominant, and $T_{d}$ tends to increase with increasing $\gamma$. The sublimation time $\tau_{\rm sub}$ for the $0.1\mum$ grains is presented. As shown, silicate grains with $\gamma<100$ can survive collisional heating even in dense regions with $n_{\gas}=10^{4}\cm^{-3}$ for $>100$ yr, but those with $\gamma>10^{3}$ will be evaporated in regions with $n_{\gas}>10^{3}\cm^{-3}$ {in less than $\sim 100$ yr. Graphite is scarcely affected by collisional heating due to its higher sublimation limit as expected.}

\begin{figure*}
\centering
\includegraphics[width=0.415\textwidth]{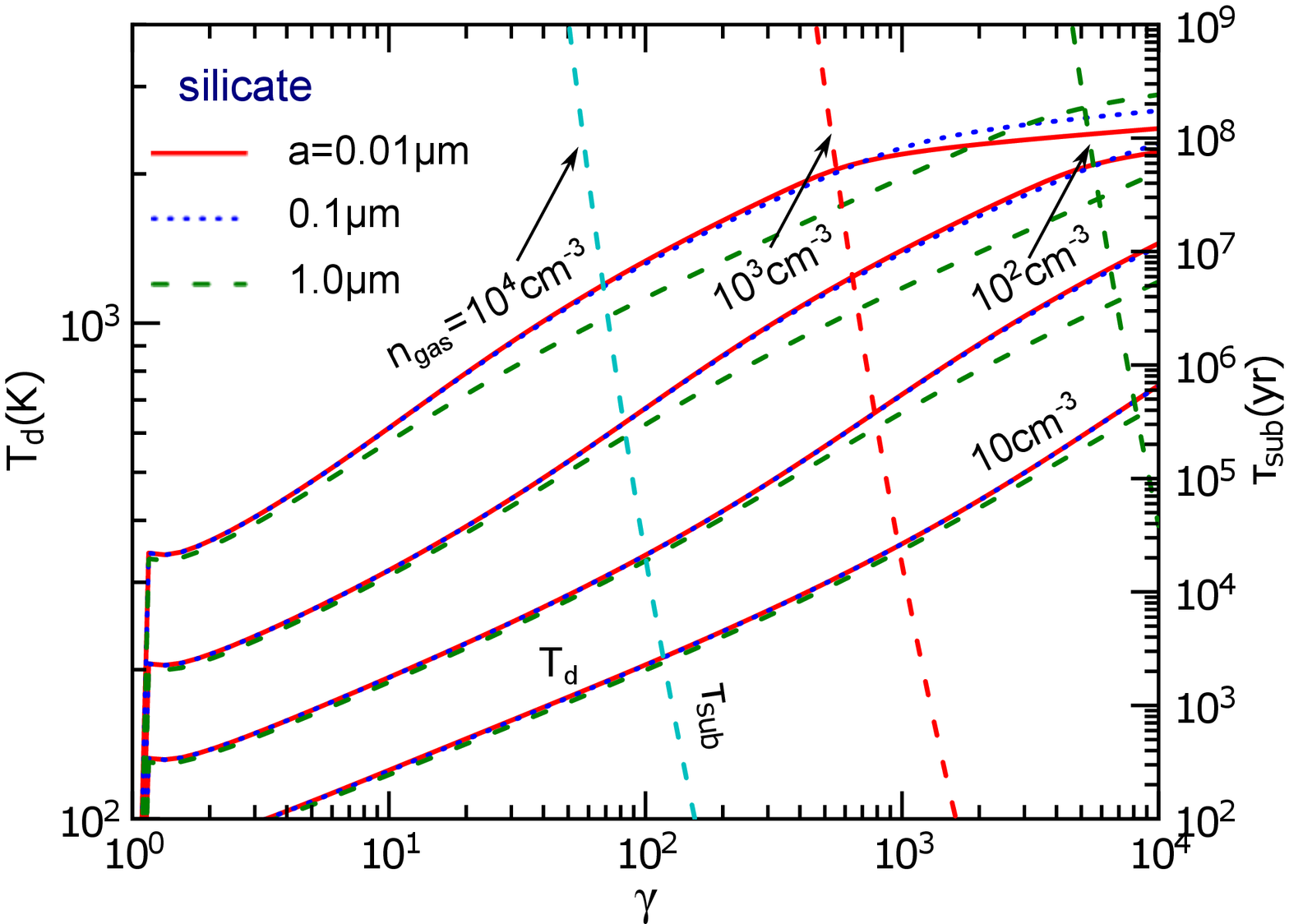}
\includegraphics[width=0.4\textwidth]{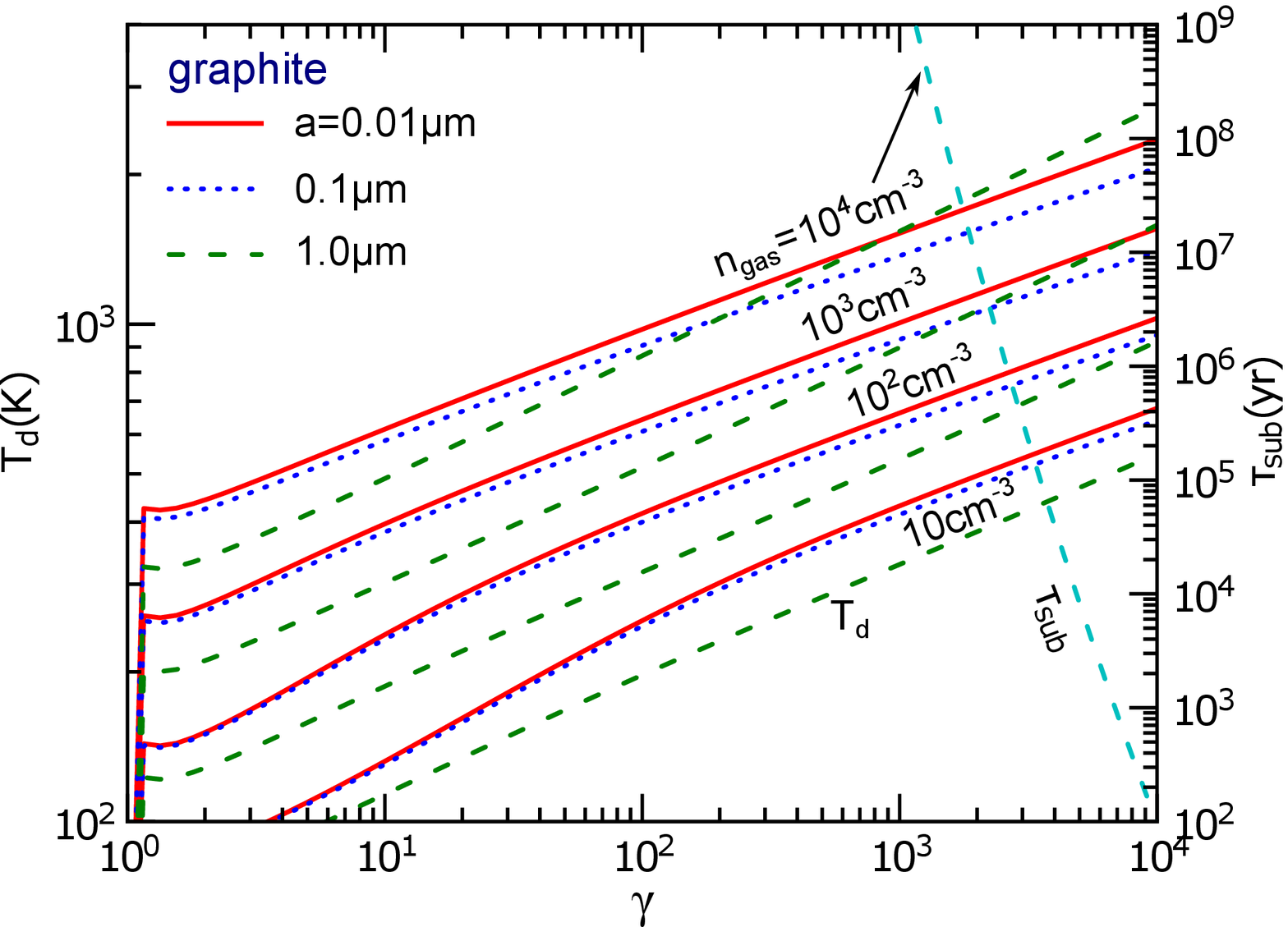}
\caption{Grain equilibrium temperature $T_{\d}$ and sublimation time $\tau_{\rm sub}$ vs. $\gamma$ for silicate (left) and graphite (right) grains heated by the ISRF and gas collisions in the ISM. Different values of $a$ and $n_{\gas}$ are considered.}
\label{fig:TdISM}
\end{figure*}

To study the effect of grain charging on Coulomb explosions in the ISM, we calculate the maximum distance that a relativistic grain has traveled before exploding by integrating Equation (\ref{eq:dZdt}) from $Z=0$ to $Z=Z_{\max}$:
\bea
L_{\max}=  \gamma \beta c\int_{0}^{Z_{\max}}\frac{dZ}{J_{\rm phe}(a, Z) + J_{\coll}(a, Z)}.\label{eq:Lmax}
\ena

Figure \ref{fig:Lmax} shows $L_{\max}$ and $N_{\rm Coul}=n_{\gas}L_{\max}$ as functions of $\gamma$ for different grain sizes. {It turns out that relativistic grains are electrically disrupted after traversing a short distance $L_{\max}\sim 10^{-3}-10^{-2}$pc or column gas density $N_{\rm Coul}\sim 10^{16}-10^{18}\cm^{-2}$, depending on $a$ and $\gamma$. Therefore, relativistic dust of typical strength adopted $S_{\max}=10^{10}\rm dyn cm^{-2}$ has no chance to survive its trip in the ISM, as pointed out by our anonymous referee.}

{If the grain material is very strong such that Coulomb explosions have not occurred when the grain potential reaching $\phi_{\max}=3\times 10^{8}{\rm Vcm^{-1}}$, then ion field emission will act to reduce the grain mass in which in each subsequent ionization will eject one atom from the grain. In this case, the gas column traversed until complete destruction by ion field emission is much larger, as given by (see Appendix \ref{apdx:Jcoll}) }
\bea
N_{\rm ife} \simeq 2\times 10^{20}\beta^{2}\left(\frac{10}{Z_{T}}\right)\left(\frac{\mathcal{S}_{\max}}{10^{10}{\rm dyn} \cm^{-2}}\right)^{1/2}a_{-5}\cm^{-2}.~~~~~
\ena

{In conclusion, relativistic dust will scarcely survive its trip in the ISM unless they are accelerated to $\gamma \gg 1$ within $0.01$pc from Earth. Ion field emission may help ideal material to survive the Coulomb explosions on a distance $\sim 6.5a_{-5}(10\cm^{-3}/n_{\gas})$ pc.}

\begin{figure*}
\centering
\includegraphics[width=0.4\textwidth]{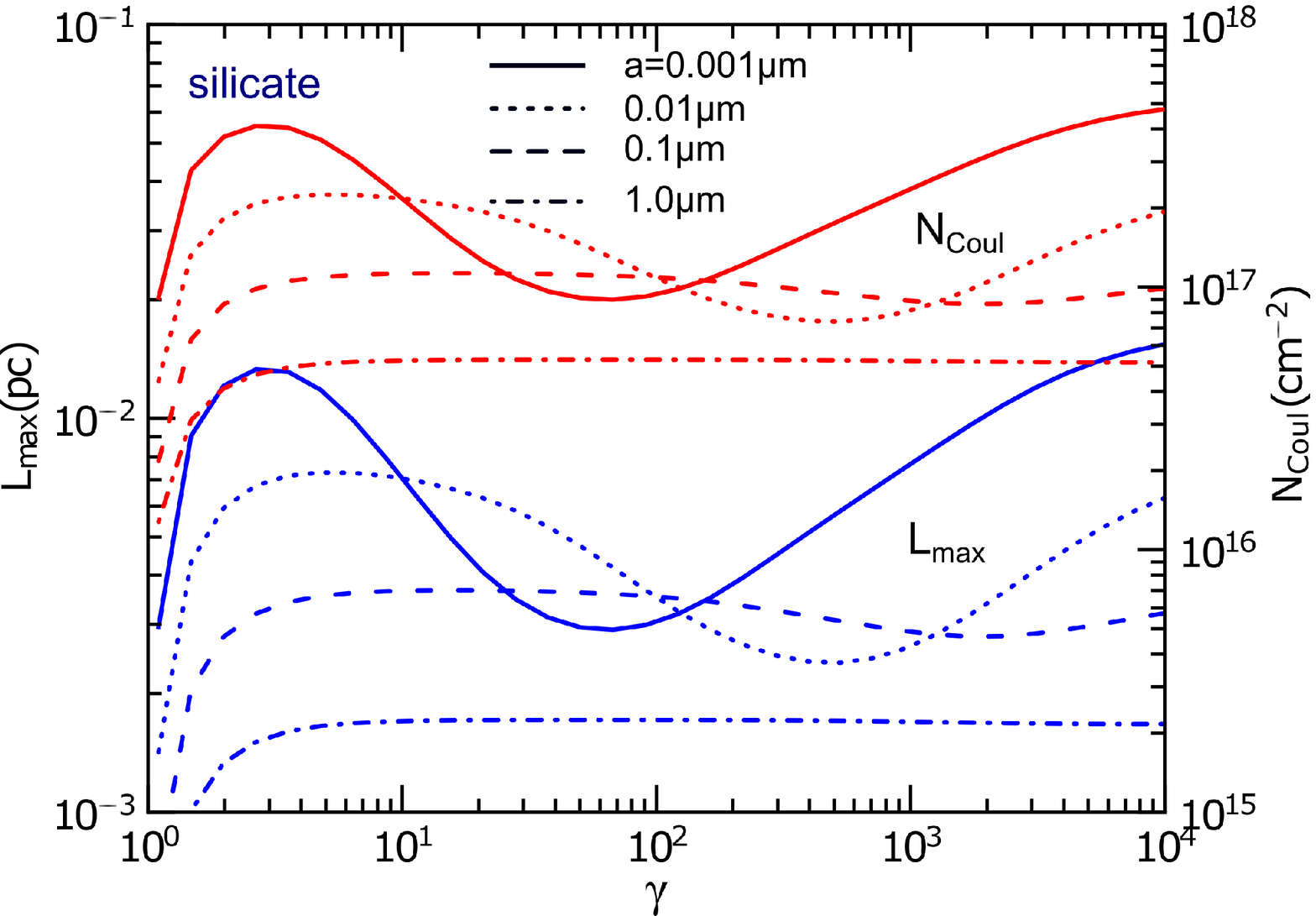}
\includegraphics[width=0.4\textwidth]{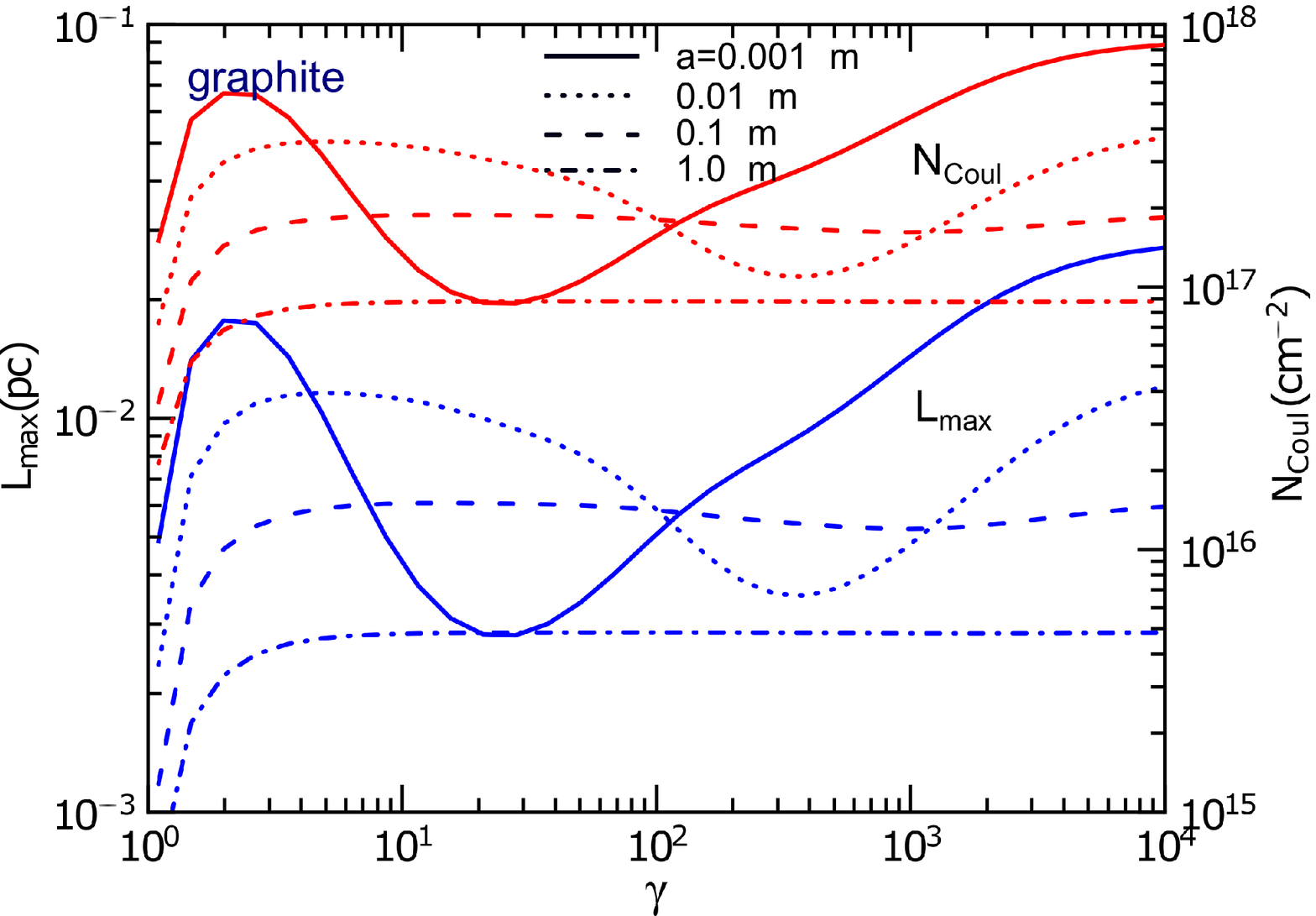}
\caption{Maximum distance $L_{\max}$ (blue lines) vs. $\gamma$ for silicate (left) and graphite (right) grains traveled before reaching $Z_{\max}$ in the ISM with $n_{\gas}=10\cm^{-3}$. Gas column swept up by dust to Coulomb explosions $N_{\rm Coul}$ is shown in red lines.}
\label{fig:Lmax}
\end{figure*}

\subsection{Can relativistic dust survive in the IGM?}

The IGM has typical number density of gas $n_{\IGM} \sim 10^{-3}-10^{-4}\cm^{-3}$. With the sputtering yield $Y_{\rm sp}\sim 10^{-6}$, Equation (\ref{eq:dadt}) shows that it would take $\tau_{\sp}=a/|da/dt|\approx 10a_{-5}-100a_{-5}$ Gyr to destroy the relativistic grains via sputtering. Thus, the sputtering is negligible for the destruction of relativistic grains in the IGM. The presence of intergalactic dust and circumgalactic dust certainly disfavors the survival of relativistic dust since a single grain-grain collision can destroy the grain completely.

{The same as in the ISM, Coulomb explosion by collisional charging is expected to be important for the destruction of relativistic dust in the IGM after sweeping up a gas column $N_{\rm Coul}\sim 10^{17}\cm^{-2}$. For $n_{\rm IGM}\sim 10^{-4}\cm^{-3}$, the maximum distance traversed by the grain is $L_{\max}\sim N_{\rm Coul}/n_{\IGM}\sim 300(n_{\rm IGM}/10^{-4}\cm^{-3})$ pc, which is small for intergalactic scales.}

Photoelectric emission by CMB photons is unlikely important due to low photon energy. For Coulomb explosions, the CMB photon energy must be boosted to $ h\nu' \sim \gamma h\nu \ge e\phi_{\max} \simeq 10^{3}a_{-5}$ eV in the GF. The average energy of CMB photons in the IGM at redshift $z$ defined by the ratio of the total radiation energy density to the total number of photons is $h\bar{\nu}_{\rm CMB}=2.7k_{\B}T_{\rm CMB}\approx 6.34(1+z)\times 10^{-4}\eV$ where $T_{\rm CMB}=2.726(1+z)$ K, which corresponds to $h\nu' \sim \gamma h\bar{\nu}_{\rm CMB} \sim 6.34(1+z)\times 10^{-4}\gamma\eV$. Thus, CMB photons are important for grain disruption if $\gamma > 1.6\times 10^{6}(1+z)^{-1}a_{-5}$. 

{The IGM is also filled with cosmic infrared radiation (CIB), optical background (OB), and ultraviolet background (UVB), which originate from galaxies, quasi stellar objects, and quasars (see \citealt{2000A&A...360....1G}). 

CIB radiation can be described by thermal dust emission with $\beta_{\rm CIB} \sim 1.4$ and $T_{\rm CIB} \sim 13\K$ \citep{2000A&A...360....1G}, such that $h\bar{\nu}_{\rm CIB}\sim 0.005$eV. Thus, these photons are relevant for Coulomb explosions for $\gamma > 10^{3}a_{-5}/0.005\sim 2\times 10^{5}a_{-5}$.

OB-UVB radiation should be important for photoelectric emission and Coulomb explosions. For quantitative estimates, we consider the IGM at $z\sim 3$ where the OB-UVB spectrum can roughly be described by a power law $J_{\nu}= J_{0}(\nu/\nu_{0})^{-1.9}$ with $h\nu_{0}=13.6$ eV and $J_{\nu_{0}}\sim 10^{-21} \erg\cm^{-2}\s^{-1}\Hz^{-1}\sr^{-1}$ (see \citealt{2005MNRAS.358..379B}). 

The number density of photons that can induce Coulomb explosions is given by
\bea
n_{\nu >\nu_{\rm Coul}}=\int_{\nu_{\rm Coul}}^{\nu_{\max}}\frac{4\pi J_{\nu}}{c h\nu}d\nu,\label{eq:nu_Coul}
\ena
where $h\nu_{\rm Coul}=e\phi_{\max}/\gamma \sim 10^{3}a_{-5}/\gamma{~\rm eV}$, and $h\nu_{\max} = 54.4$ eV is the sharp cutoff of UVB spectrum (see \citealt{2005MNRAS.358..379B}).

Plugging $J_{\nu}$ into Equation (\ref{eq:nu_Coul}}) one obtains}
\bea
n_{\nu > \nu_{\rm Coul}}
\simeq 4.7\times 10^{-3}\left[\left(10^{-3}\gamma/a_{-5}\right)^{1.9} - 5\times 10^{-4}\right]\cm^{-3}.~~~\label{eq:nu_IGM}
\ena

{To obtain $n_{\nu > \nu_{\rm Coul}} > 0$, Equation (\ref{eq:nu_IGM}) yields $\gamma > 19a_{-5}$. For photoelectric yield $Y\sim 0.1$ and $Q_{\abs}\sim 0.1$, it is straightforward to show that the $0.1\mum$ relativistic dust with $\gamma > 20$ would be electrically disrupted after moving a distance $L_{\max} \sim500$ kpc, less than the distance to the closest Andromeda galaxy.

If relativistic dust is made of ideal material, then ion field emission will prevent Coulomb explosions to occur and help the grains to survive over a gas column $N_{\rm gas, ife}\sim 10^{20}\cm^{-2}$ or distance $L_{\max}\sim 324(n_{\rm IGM}/10^{-4}\cm^{-3})$ kpc. Therefore, relativistic dust hardly survives in the IGM.
}

\subsection{Can relativistic dust survive in the solar radiation field?}
The solar radiation field is the most important, final barrier that relativistic grains must overcome to reach the Earth's atmosphere as primary particles of CRs. 

We have calculated the grain equilibrium temperature as a function of solar distance and evaluated the fraction of grain size lost by sublimation by the time the grain reaches $1\AU$. The destruction by Coulomb explosions is also revisited by using an improved treatment of photoelectric charging that takes into account the photoemission of Auger and secondary electrons. The same as the grain temperature, the grain equilibrium charge as a function of solar distance is also calculated. Thereby, the grain size $a$ and Lorentz factor $\gamma$ of relativistic grains that survive the solar radiation field can be determined.

In Figure \ref{fig:gamexp} we summarize the obtained parameter space $(a,\gamma)$ for the survival and destruction of silicate (left) and graphite (right) grains in solar radiation. The grains with $(a,\gamma)$ located in the gray areas in Figure \ref{fig:gamexp} would survive both Coulomb explosions and thermal sublimation to enter the Earth's atmosphere. The grains with $(a,\gamma)$ located above the solid lines (with circle symbols) would be destroyed efficiently by sublimation, gradually releasing heavy atoms/nuclei when approaching the Earth. 

\begin{figure*}
\centering
\includegraphics[width=0.4\textwidth]{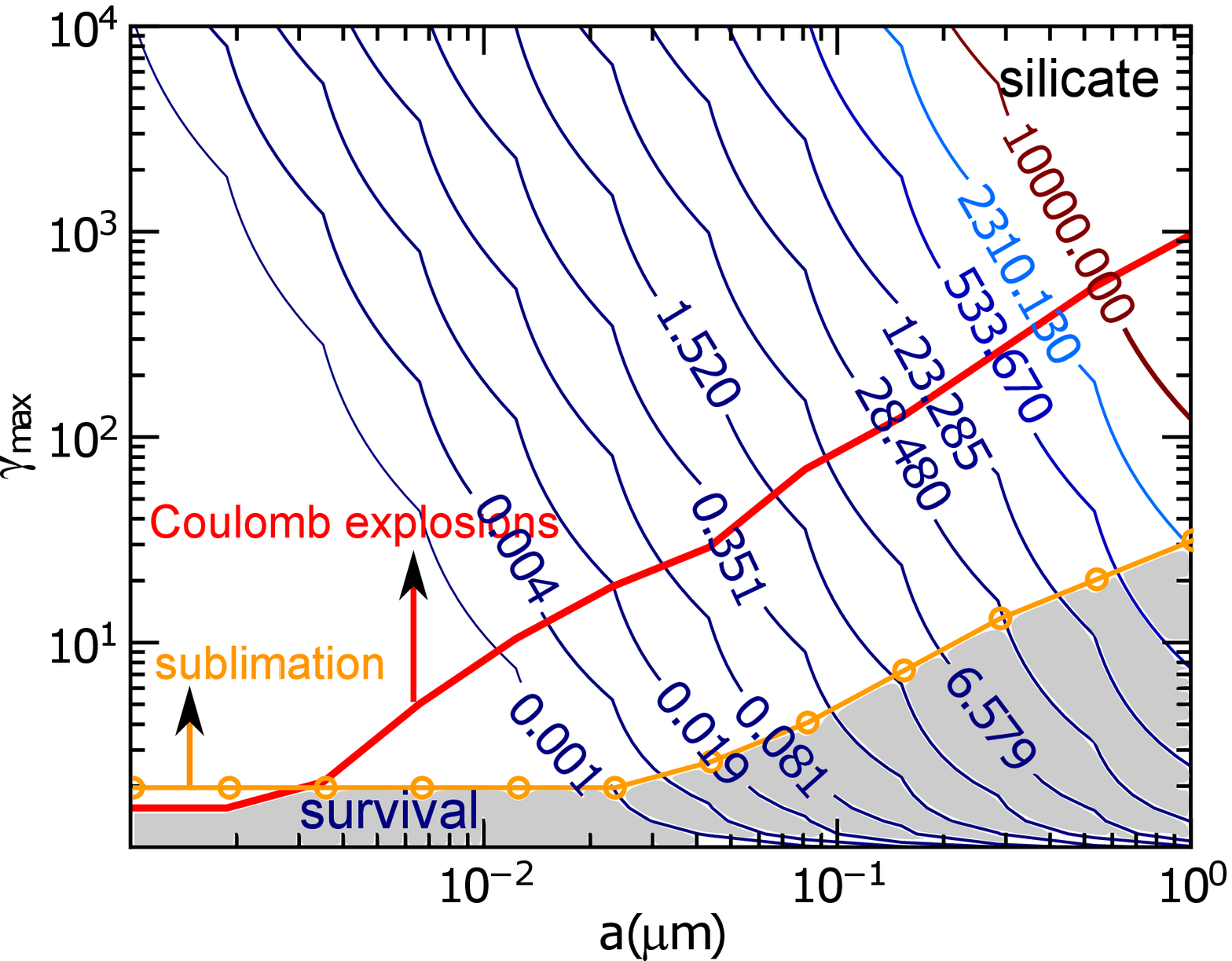}
\includegraphics[width=0.4\textwidth]{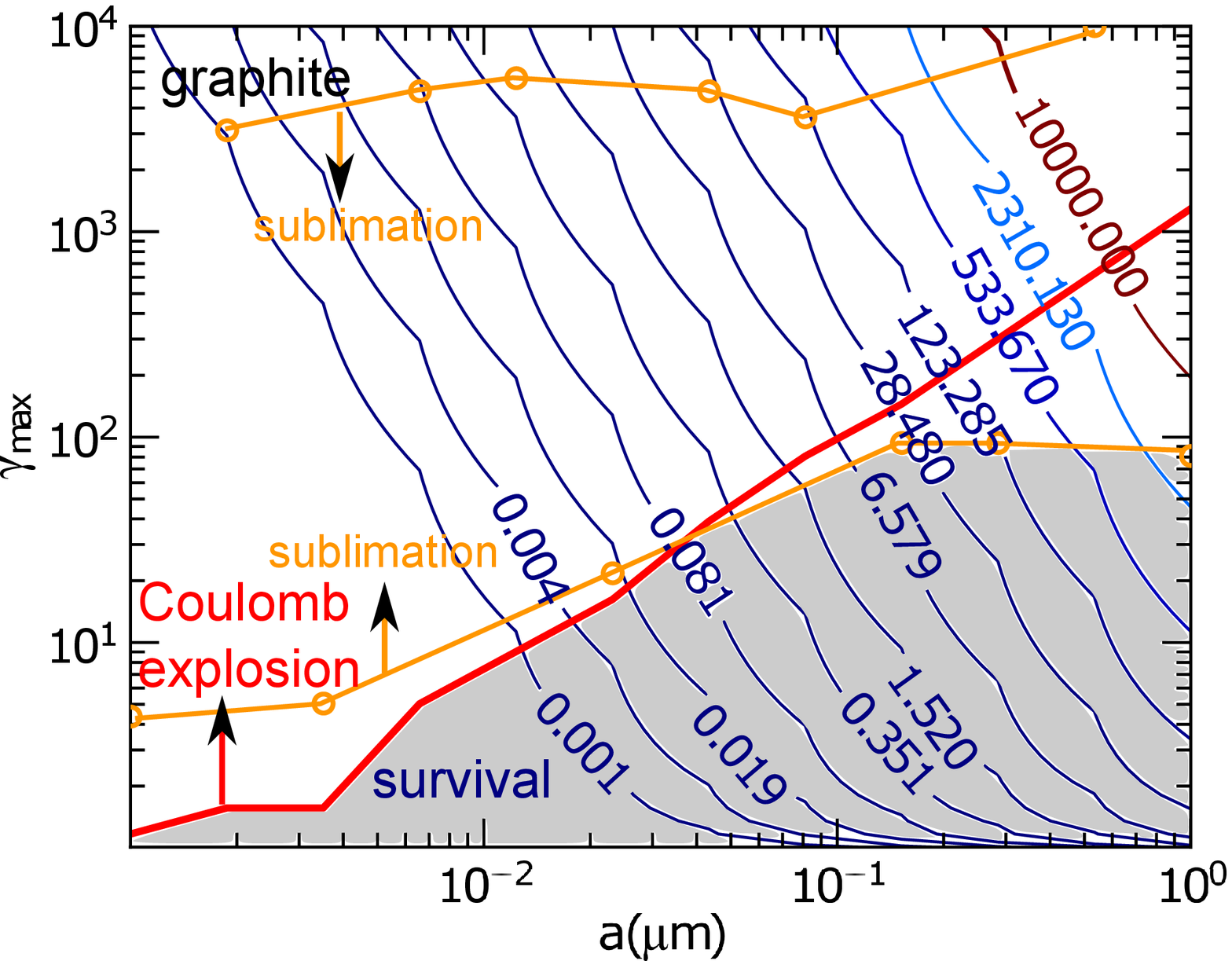}
\caption{Maximum Lorentz factor $\gamma_{\max}$ for grain survival in solar radiation against Coulomb explosions (thick solid line) and sublimation (solid line with circle symbols) vs. grain size $a$ for silicate (left) and graphite (right). Contours show $E_{\gr}(a,\gamma_{\max})$ levels in units of $10^{20}$ eV. Grains with $(a,\gamma)$ located above the red line (gold line) are destroyed by Coulomb explosions (sublimation) and those with $(a,\gamma)$ located below these lines will survive. Gray areas show the grains that survive both Coulomb explosions and sublimation.}
\label{fig:gamexp}
\end{figure*}

Thermal sublimation appears to be a powerful mechanism in destructing relativistic grains, which results in the complete destruction of grains with $\gamma > 10-100$, whereas Coulomb explosion is dominant for small graphite grains only (see Figure \ref{fig:gamexp}). It is noted that earlier studies mostly focused on melting as a principal destruction process of relativistic dust in solar radiation (see e.g., \citealt{1993Ap&SS.205..355M}), although it is unclear whether melting has any significant effect on the destruction of refractory dust. Moreover, our obtained threshold for grain survival $\gamma < 10$ (silicate) is lower than $\gamma \sim 360$ found in \cite{1977Ap.....13..432E} for irons and much lower than the result $\gamma \sim 10^{4}-10^{5}$ by \cite{1999APh....12...35B}. Table \ref{tab:sumwork} summarizes the Lorentz factor $\gamma$ achieved by radiation pressure, various destruction processes, and the survival of relativistic dust in the solar radiation field investigated previously and in this paper, where {\it NA} indicates that the process is not answered and {\it NQ} indicates that the process is mentioned but not studied in detail.

\subsection{Observational Constraints of UHECRs and Relativistic Dust Hypothesis}
Our detailed calculations reveal that, if there exists relativistic dust entering the solar system, then there will exist a window $(a, \gamma)$ that the grains could survive solar radiation and reach the Earth's atmosphere.

After entering the Earth's atmosphere, a relativistic grain is expected to produce a huge EAS corresponding to the superposition of $\sim 10^{10}$ shower events initiated by an energetic nucleon. The depth of the maximum shower $X_{\max}$ induced by relativistic dust is estimated to be rather low, which was shown unable to account for the observational data \citep{1980ApJ...235L.167L}. Latest measurements essentially provide the maximum depth of EAS $ X_{\max} \sim 700-800 \g\cm^{-2}$ for $E\sim 10^{18}-10^{20}$ eV, which can be reproduced by heavy nuclei lighter than irons (see \citealt{2012APh....39..129A}; \citealt{2014PhRvD..90l2005A}). Such heavy nuclei must have huge Lorentz factor ($\gamma \sim 10^{9}$) to account for UHECRs though.

{It is worth to mention that Monte Carlo simulations by \cite{2000PhRvD..61h7302A} show that the air shower event recorded at the Yakutsk array on May 7, 1989 might be explained by relativistic dust with $\log(\gamma)=3.8-4$. However, as shown by us, dust grains moving with such high $\gamma$ hardly survive traversing the ISM and solar system to reach the Earth's atmosphere (see Figure \ref{fig:gamexp}).}

Following \cite{1980ApJ...235L.167L} the maximum depth $X_{\max}$ is approximately given by
\bea
X_{\max}= A + (1-B)X_{0}\ln \gamma< A + X_{0}\ln \gamma,
\ena
where $X_{0}=37.7\g\cm^{-2}$, $ A=110\pm 20$ and $0<B<1$. For relativistic dust with $\gamma<100$, it yields $X_{\max}<300\g\cm^{-2}$, which obviously cannot reproduce the observational data aforementioned. 

\begin{table*}
\centering
\caption{Summary of previous works on relativistic dust and our study}\label{tab:sumwork}
\begin{tabular}{l l l l l l l l }\hline\hline\\
Authors & acceleration & melting & explosion & sublimation & sputtering & grain & survival\\
& & & & & & collision &  \cr
\hline\\
{\cite{Spitzer:1949bv}} & $v\sim 0.1c$ & NA &  NA & NA  & NQ & NQ & NQ \cr

{\cite{1972Ap&SS..16..238H}} & $\gamma\sim 10$ & NA &  NA & NA  & NA  & NA & NA \cr

{\cite{1973Ap&SS..21..475B}} & NA & NA &  $\gamma>30-50$ & NA  & NA  & NA & $\gamma<30-50$ \cr
{\cite{1974ApJ...187L..93G}} & NA & $\gamma\gtsim 10^{3}$ & $\gamma> 5a_{-5}(h\nu/1{\eV})$ & NA  & NA  & NA & NQ \cr
{\cite{1977Ap.....13..432E}} & NA & NA & $\gamma>360$  & NA  & NA  & NA & iron, $\gamma<360$\cr

{\cite{1993Ap&SS.205..355M}} & NA & iron, $\gamma>1$ &  NA & NA  & NQ & NA & tungsten \cr

{\cite{1999APh....12...35B}} & $\gamma\sim 10$ & NA & $\gamma>10^{4}$ & NA  & NA  & NA & $\gamma<10^{4}$\cr
{This work} & $\gamma< 2$ & see~text &  quantified & quantified & inefficient  & efficient & $\gamma<10-100$\cr
\cr
\hline
\cr
\end{tabular}
\end{table*}

\subsection{Effects of magnetic fields on grain motion}
{One remaining issue is the effect of ambient magnetic fields on the acceleration of dust by radiation pressure near the radiation sources.} The ratio of magnetic pressure $p_{\rm mag}=B^{2}/8\pi$ to radiation pressure is given by
\bea
\frac{p_{\rm mag}}{p_{\rm rad}}= \frac{B^{2}}{8\pi}\frac{4\pi r^{2}c}{\langle Q_{\rm pr}\rangle L_{\bol}}\simeq 3.7\times 10^{-10}B_{10}^{2} r_{\pc}^{2}\left(\frac{10^{13}L_{\odot}}{\langle Q_{\rm pr}\rangle L_{\bol}}\right),~~~\label{eq:pratio}
\ena
which indicates that magnetic pressure is negligible compared to radiation pressure during the acceleration stage. 

Grain motion in the magnetized ISM is inevitably constrained by the Lorentz force if it does not move along the mean magnetic field lines. The gyroradius of a charged grain moving across $\Bv$ with perpendicular velocity $v_{\perp}$ is
\bea
r_{g} = \frac{m_{\gr}cv_{\perp}}{\langle Z\rangle eB}\simeq 0.6\left(\frac{v_{\perp}}{c}\right)\langle Z\rangle^{-1}B_{10}^{-1}\hat{\rho}a_{-5}^{3} {~\rm kpc},\label{eq:rgLarmor}
\ena
where $\langle Z\rangle e$ is the equilibrium charge of the grain.

One should not assume that charged grains move along the field lines. However, due to the magnetic field {wandering} of Alfv$\acute{e}$nic component of MHD turbulence,\footnote{A discussion and practical procedures of decomposing magnetic fields into Alfv$\acute{e}$nic, slow and fast modes are given in \cite{{2002PhRvL..88x5001C},{2003MNRAS.345..325C}} and \cite{2010ApJ...720..742K}.} the grain will move across the mean magnetic field, while following the local magnetic field direction. The corresponding field wandering is described in \cite{1999ApJ...517..700L}, and it results in superdiffusive behavior at scales less than the turbulence injection scale (see \citealt{2014ApJ...784...38L}) and diffusive behavior at scales above the turbulence injection scale. The latter scale in our galaxy is associated with supernovae explosions and around 100 pc. According to Equation (\ref{eq:rgLarmor}) the superdiffusive behavior may be important for grains with $a< 0.2\mum$ and $\langle Z\rangle \sim 30$ {for nonrelativistic grains in the ISM. For relativistic grains with much higher mean charge, the superdiffusion will be achieved for larger grains.}

\subsection{Other related effects of relativistic dust}

In the ISM, relativistic grains heated to temperature $T_{d}$ (dominantly by gas collisions) will isotropically reemit thermal radiation with the average energy per photon $(3+\beta_{d})k_{\B}T_{d}$ in the GF {where $\beta_{d}\sim 1.5-2$ \citep{PlanckCollaboration:2014dz} is the spectral index of dust opacity}. Dust grains with $\gamma > 10^{3}$ can be heated to $T_{d}> 10^{3}\K$ in regions with $n_{\gas}\ge 10^{3}\cm^{-3}$ (see Fig. \ref{fig:TdISM}). In the observer's frame, reemitted photons will have energy $h\bar{\nu} \sim \gamma(3+\beta_{d})k_{\B}T_{d}) > 258(1+\beta_{d}/3)$ eV, which are X-rays. A consequence of collisional heating by the gas is grain's deceleration due to the conversion of its kinetic energy into thermal radiation, although this slowing down is likely inefficient for the diffuse media. In another context, \cite{1972Ap&SS..19..173H} suggested that the scattering of CMB by relativistic grains with high $\gamma$ may produce metagalactic X-rays.
 
Relativistic dust was also suggested to explain gamma-rays by \cite{1974ApJ...187L..93G}. According to this study, an iron grain moving with $\gamma \sim 10^{3}$ will explode suddenly within $10^{5}$ AU, releasing Fe ions/atoms. Blueshifted solar photons can then excite K-shell electrons of the Fe ion/atom to higher energy levels, and followed radiative transitions to the ground level will produce X-ray photons of energy $E_{\nu, {\rm GF}}$. Therefore, in the observer's frame, we will see photons of energy $E_{\nu, {\rm SF}} = E_{\nu, {\rm GF}}\gamma(1+\cos\phi) < \gamma E_{K\alpha}$ with $\phi$ the emission angle of photon relative to the grain motion direction. For instance, an Fe ion released from the relativistic grain moving with $\gamma\sim 10^{2}$ can produce gamma-rays with maximum energy $E_{\nu}\sim 700$ keV. Hence, the grains with $(a,\gamma)$ above the thick solid lines would explode within distance less than $10^{5}\AU$ from the Sun, simultaneously releasing a huge number of relativistic heavy ions/atoms. Such relativistic ions/atoms may create gamma-ray bursts. 

In addition to grain heating and charging, strong solar radiation may slow down relativistic grains when they enter the solar system. Using Equation (\ref{eq:tacc}) {and accounting for the Doppler blueshift} we can estimate the deceleration time at 1 AU to be $\tau_{\rm dec} \sim 2.6\times 10^{10}\gamma^{-1}(1+\beta)^{-2}a_5 \s$ for $v\sim c, <Q_{\rm pr}>=1$, and $L=L_{\odot}$. Thus, the deceleration by solar radiation pressure is negligible for relativistic grains. 

The effect of infrared (IR) emission from Earth and Moon appears to be unimportant for grain destruction. Indeed, the ratio of energy density at distance $d_{\rm E}$ from the Earth to the solar energy density at distance $\sim 1\AU$ is $u_{\rm rad,E}/u_{\rm rad,\odot} \sim (T_{\rm E}/T_{\odot})^4(1\AU/R_{\odot})^2(R_{\rm E}/d_{\rm E})^{2}$ where $R_{\odot}$ is the Sun radius, $T_{\rm E} \sim 300\K$ is the Earth temperature, and $R_{\rm E}$ is the Earth radius. Let assume that the Earth emission begins to be important when $u_{\rm rad,E}/u_{\rm rad,\odot}\sim 0.1$. Then, we get $d_{\rm E}\sim 2R_{\rm E}$ and the arrival time $2R_{\rm E}/c\sim 0.02\s$. The latter time is much shorter than the sublimation time given by Equation (\ref{eq:tausub_sil}). The effect of Moon radiation is apparently much weaker due to the much smaller ratio of its radius to the distance to Earth, i.e., $(R_{\rm M}/d_{\rm M})^{2}$. 

\section{Summary}\label{sec:sum}
We revisited the idea of relativistic grains as primary particles of UHECRs by studying in detail various destruction mechanisms, including thermal sublimation and Coulomb explosions, electronic sputtering, and grain-grain collisions. Our principal results are summarized as follows:
\begin{itemize}

\item[1.] Grain acceleration by radiation pressure force from AGNs is revisited taking into account the redshift of radiation spectrum in the grain comoving frame and explicit dependence of the pressure efficiency on grain size and Lorentz factor $\gamma$. Our results show that the maximum grain speeds achieved are $\gamma < 2$, i.e., several times lower than earlier simple estimates.

\item[2.]{In the solar radiation field, we found that thermal sublimation is a {powerful} mechanism for the destruction of relativistic grains, which can destroy silicate grains with $\gamma>10$ and graphite grains of $\gamma>100$. Coulomb explosions due to photoelectric emission are found to be efficient for relativistic dust with $\gamma>10^{2}-10^{3}$. }

\item[3.] A parameter space $(a,\gamma)$ that relativistic grains could survive thermal sublimation and Coulomb explosions in the solar radiation field is identified. Compared to previous studies, our results reduce the survival chance of relativistic dust and thus its possibility as primary particles of UHECRs.

\item[4.] {In the ISM, relativistic grains would be destroyed via Coulomb explosions after traversing a gas column density $N_{\rm Coul} \sim 10^{16}-10^{18}\cm^{-2}$ due to collisional charging. For near-ideal strength material ion field emission would help relativistic dust to traverse a larger gas column $N_{\rm ife}\sim 2\times 10^{20}\cm^{-2}$.}

\item[5.] We studied electronic sputtering by ions and its effect on the destruction of relativistic dust. Although heavy Fe ion is found to produce considerable sputtering yield ($Y_{\rm sp}\sim 10^{-2}$), its low abundance makes the net sputtering yield rather small. The destruction by electronic sputtering is inefficient for relativistic grains.

\item[6.] The evaporation induced by grain-grain collisions in the ISM is an efficient destruction mechanism, for which relativistic grains with $a\le 1\mum$ can be completely destroyed after sweeping a gas column density $N_{\rm coll} \sim 5\times 10^{20}\cm^{-2}$.

\item[7.] {Relativistic dust from extragalactic origins appears to have extremely narrow chance to survive moving in the IGM due to Coulomb explosions induced by OB-UVB radiation and gas collisions. Ion field emission would help relativistic dust made of very strong material to survive Coulomb explosion, but collisions with intergalactic dust should destroy them quickly.}
\\

\item[8.] Relativistic dust of ideal strength with $\gamma<10-100$ arriving from a distance with $N_{\rm gas} \sim 10^{20}\cm^{-2}$ in the Galaxy would survive both the ISM and solar radiation to reach the Earth's atmosphere. Given unknown acceleration mechanisms that can drive grains to $\gamma \gg 1$ and the low $X_{\max}$ predicted for these surviving grains, the chance for relativistic dust to account for primary particles of UHECRs is much lower than previously thought.

\end{itemize}

\acknowledgments
We thank the anonymous referee for an extremely thorough review and her/his insightful comments that significantly improved our paper. T.H. thanks Dr. Anthony Jones for useful discussion and comments. T.H. is supported by the Alexander von Humboldt Fellowship at the Ruhr-Universit$\ddot{\rm a}$t Bochum. A.L. acknowledges the financial support of NASA grant NNX11AD32G and the Center for Magnetic Self-Organization and thanks IIP/UFRN (Natal) for hospitality. R.S. acknowledges partial support by the Deutsche Forschungsgemeinschaft grant Schl 201/29-1 and the Mercator Research Center Ruhr (MERCUR) grant Pr-2012-0008.

\appendix
\section{A. Grain physics}
\subsection{A.1. Absorption and Radiation Pressure Cross-Section}\label{sec:Qabs_Qpr}

For a spherical grain of radius $a$, the absorption and scattering efficiency of radiation of wavelength $\lambda$ is given by $Q_{\abs}=C_{\abs}/\pi a^{2}$, $Q_{\sca}=C_{\sca}/\pi a^{2}$, where $C_{\abs}$ and $C_{\sca}$ are the absorption and scattering cross-sections, respectively. The radiation pressure efficiency is defined as $Q_{\rm pr}=Q_{\abs} + (1-\langle \cos\theta\rangle) Q_{\rm sca}$
where $\langle \cos\theta\rangle$ is the scattering asymmetrical factor with $\theta$ scattering angle.
 
Throughout this paper (unless specified otherwise), we consider a mixed-dust model that consists of silicate and graphite grains. Their complex refractive index $m$ and dielectric functions $\epsilon$ are taken from \cite{2003ApJ...598.1026D}. Using the publicly available Mie code \citep{1980ApOpt..19.1505W}, we compute $Q_{\abs}, Q_{\rm sca}$ and $\langle \cos\theta\rangle$ for $x=2\pi a/\lambda< 2\times 10^{4}$. 

For $x>2\times 10^{4}$, the anomalous diffraction theory \citep{1957lssp.book.....V} is employed, which provides
\bea
Q_{\abs}=1+\frac{\exp\left[-4x{\rm Im}(m)\right]}{2x{\rm Im}(m)}+\frac{\exp\left[-4x {\rm Im}(m)\right]-1}{8x^{2}{\rm Im}^{2}(m)},~~~~~
\ena
where ${\rm Im}(m)$ denotes the imaginary part of $m$. In this case $Q_{\rm pr} = Q_{\abs}$ due to negligible scattering effect.

To compute $Q_{\abs}$ and $Q_{\rm pr}$ for graphite grains, we consider two cases in which the electric field of radiation $\bE$ is parallel and perpendicular to the grain optical axis ($c$-axis) with the corresponding dielectric function $\epsilon_{\|}$ and $\epsilon_{\perp}$. Using the $(1/3)-(2/3)$ approximation (i.e., 1/3 of graphite grains have $\epsilon=\epsilon_{\|}$ and 2/3 of them have $\epsilon=\epsilon_{\perp}$), one can obtain the extinction cross-section $C_{\ext}=\left[C_{\ext}(\bE\| c)+2C_{\ext}(\bE\perp c)\right]/3$ for a randomly oriented grain. Figure \ref{fig:Qabs} shows $Q_{\abs}$ as a function of photon energy computed by Mie theory for silicate (left) and graphite (right) grains.

\begin{figure*}
\centering
\includegraphics[width=0.4\textwidth]{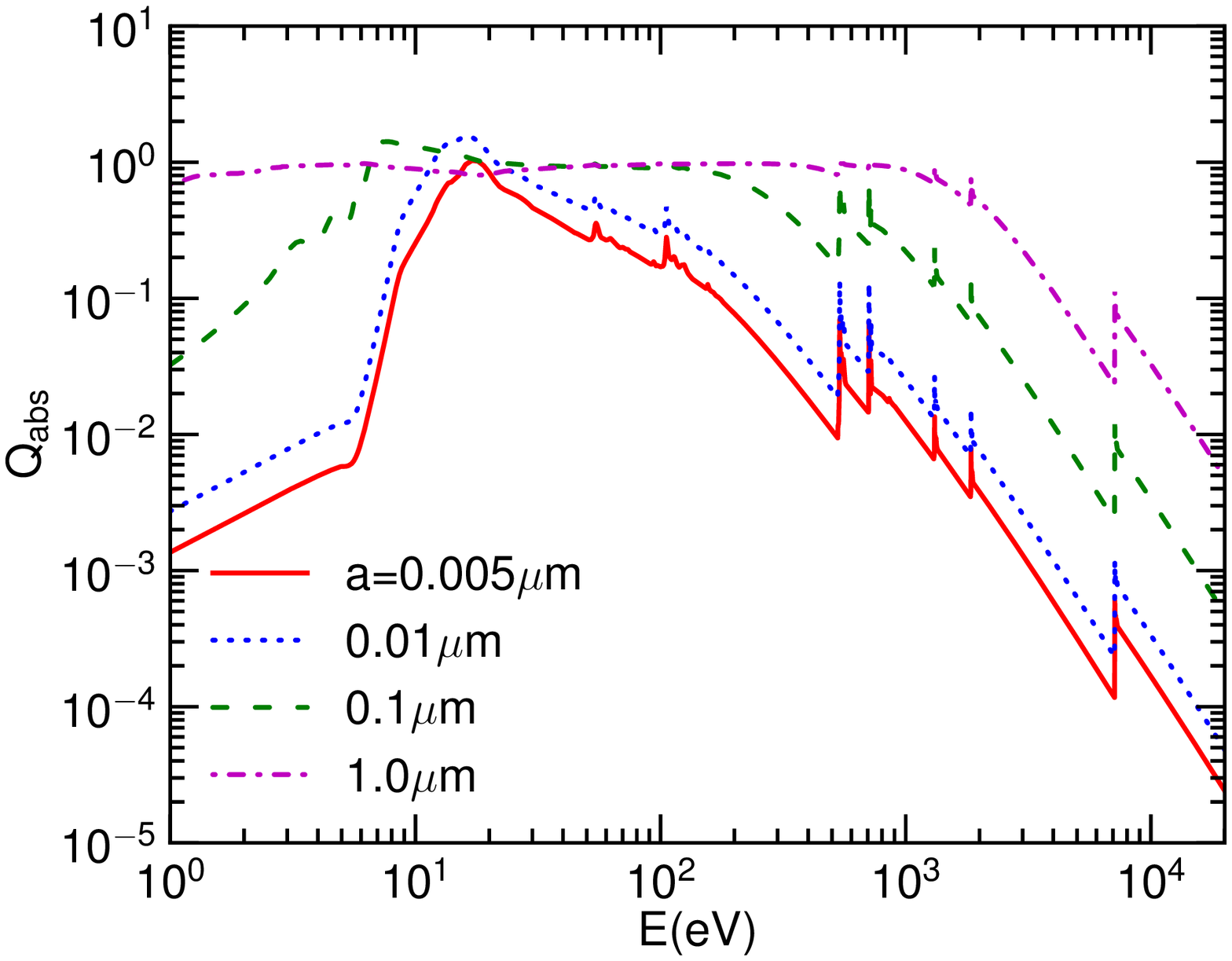}
\includegraphics[width=0.4\textwidth]{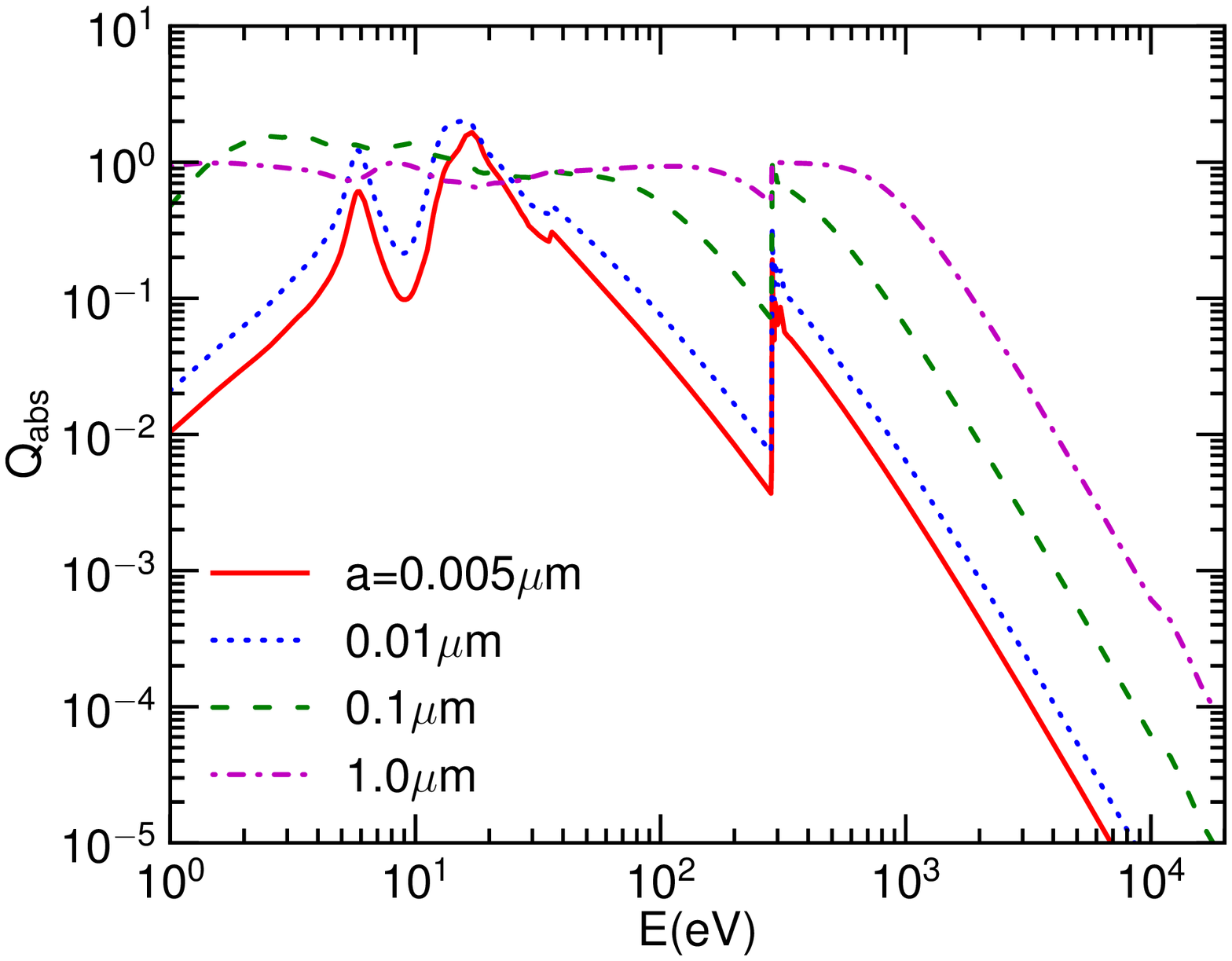}
\caption{Absorption efficiency $Q_{\abs}$ as a function of photon energy for different grain sizes for silicate (left) and graphite (right) grains. A sharp spectral feature at $E\approx 291 \eV$ in the right panel corresponds to the excitation of K-shell of C atom. Several spectral features in silicate grain (left panel) are also due to K-shell excitations of Si, Mg, O and Fe atoms.}
\label{fig:Qabs}
\end{figure*}

\subsection{A.2. Dust survival radius near a point source of radiation}\label{apdx:sub}
The grain equilibrium temperature at distance $r$ from the central radiation source can be approximated by
\bea
T_{d}(r)=1800\left[\left(\frac{L_{\rm UV}}{5\times 10^{12}L_{\odot}}\right)^{1/2}\left(\frac{1\pc}{r}\right)^{2}\right]^{1/5.6} \K,\label{eq:Tgr}
\ena
where $L_{\rm UV}$ is the luminosity in the optical and UV, which is about a half of the bolometric luminosity (see e.g., \citealt{1995ApJ...451..510S}). By setting $T_{d}$ equal to sublimation temperature $T_{\rm sub}$, the closest distance that dust can survive is equal to
\bea
r_{\rm sub}=\left(\frac{L_{\rm UV}}{5\times 10^{12}L_{\odot}}\right)^{1/2}\left(\frac{T_{\rm sub}}{1800\K}\right)^{-5.6/2} \pc.\label{eq:Rs}
\ena

For SN explosions with $L_{\rm UV}\sim 5\times 10^{8}L_{\odot}$, we get $r_{\rm sub}\sim 0.01\pc$. For AGN with $L_{\rm UV}\sim 10^{13}L_{\odot}$, we get $r_{\rm sub}\sim 1\pc$. 

Due to radiation pressure, grains are rapidly accelerated to high velocities from $r_{\rm sub}$. Comparing $\tau_{\rm acc}$ with $\tau_{\rm sub}$, we see that grains initially at $r_{i}=r_{\rm sub}$ are likely destroyed by sublimation before achieving their terminal velocities, while grains with $r_{i}\ge 1.5r_{\rm sub}$ will survive the sublimation and get their highest velocities.

The interstellar medium surrounding AGNs is fully ionized by its strong UV radiation. The grain equilibrium charge due to photoelectric emission and collisional charging is
\bea
\langle Z \rangle = 1.3\times 10^{3}a_{-5}\left(\frac{L_{\gamma 13}}{n_{e8}r_{pc}^{2}}\right)^{2/11}T_{e4}^{1/11},
\ena
where $L_{\gamma}$ is the total UV and X-ray luminosity and $L_{\gamma 13}=L_{\gamma}/10^{13}L_{\odot}$, $n_{e8}=n_{e}/10^{8}\cm^{-3}$ is the electron number density, and $T_{e4}=T_{e}/10^{4}\K$ is the gas temperature (see more details in \citealt{1995ApJ...451..510S}).

For the diffuse medium that is important for grain acceleration $n_{e}<10^{4}\cm^{-3}$, $\langle Z \rangle<10^{4}a_{-5}\left(\frac{L_{\gamma 13}}{n_{e4}r_{pc}^{2}}\right)^{2/11}T_{e4}^{1/11}$. Comparing to $Z_{\max}$, we can see that the $a>0.01\mum$ grains would be unlikely destroyed by Coulomb explosions. At distance $r\sim 100\pc$ from quasars with $L_{\gamma}\sim 10^{13}L_{\odot}$ where grains have achieved terminal velocity, we get $\langle Z\rangle = 4285a_{-5}$ for the ISM with $n_{e}=10\cm^{-3}$ and $T_{e}\sim 5000\K$.

\subsection{A.3. Effect of ion sputtering during the grain acceleration stage}\label{apdx:sputt}
To evaluate the impact of grain erosion by ion sputtering during the radiation acceleration stage from AGNs, we write ${da}/{dv} ={da}/{dt}\times {dt}/{dv}$. Using Equations (\ref{eq:dvdt_rad}) and (\ref{eq:dadt}) for $dv/dt$ and $da/dt$, one obtains 
\bea
\frac{da}{a} = \left(\frac{16\pi r^{2}c}{3L_{\bol}\langle Q_{\rm pr}\rangle}\right)\left[\frac{n_{\gas}M_{d}Y_{\sp}(v)v^{2}}{4\rho} \right]\frac{dv}{v} \simeq 2.09\times 10^{-4}n_{3}\frac{r_{\pc}^{2}}{L_{13}\langle Q_{\rm pr}\rangle}\left(\frac{M_{d}}{m_{C}}\right)\hat{\rho}^{-1}\frac{Y_{\sp}(v)}{0.1}\left(\frac{v}{10^{3}km}\right)^{2}\frac{dv}{v},
\ena
where $n_{3}=n_{\gas}/10^{3}\cm^{-3}, L_{13}=L_{\bol}/10^{13}L_{\odot}$. It can be seen that even a high sputtering yield $Y_{\sp}\sim 0.1$ (maximum yield by He, see \citealt{1994ApJ...431..321T}) is adopted, the destruction by sputtering is still negligible compared to the acceleration for $v \le 1000$ km. Above $\sim 1000$ km, $Y_{\sp}(v)$ rapidly declines with $v$ as $1/v^{2}$ due to the decrease of Coulomb collision cross-section, which substantially reduces the sputtering yield (see \citealt{1994ApJ...431..321T}).

\section{B. Lorentz Transformations of Radiative Quantities}\label{apdx:trans}
Some quantities are invariant under Lorentz transformations, so-called Lorentz invariants, which are the same in the different inertial frame of reference. These include the total number $N$ of photons or atoms, 4-volume $dVdt$, $d^{3}p/E$, and the phase volume $d\mathcal{V}=d^{3}xd^{3}p$.

Assuming that a dust grain is moving with $\gamma \gg 1$ in a radiation field. Let $\mu=\cos\theta$ be the cosine angle between the direction of grain motion and that of photon propagation, and let $u(\nu,\Omega)=dE/d\nu d\Omega dV$ be the specific spectral energy density of radiation in the SF. In the GF, the photon energy $h\nu'$ and specific spectral energy density $u'(\nu',\Omega')$ are governed by the following transformations:
\bea
\frac{u(\nu,\Omega)}{\nu^{3}}=\frac{u'(\nu',\Omega')}{\nu'^{3}},~~
\frac{n(\nu,\Omega)}{\nu^{2}}=\frac{n'(\nu',\Omega')}{\nu'^{2}},~~
\nu'=\gamma\nu(1-\beta \mu),~~
\mu'=\frac{\mu-\beta}{1-\beta\mu},~~
d\mu'=\frac{d\mu}{\gamma^{2}(1-\beta\mu)^{2}}=\frac{d\mu}{\mathcal{D}^{2}},
\ena
where $n=u/\nu$ and $n'=u'/\nu'$ are the photon density in the SF and GF, respectively (see \citealt{2002ApJ...575..667D}). 
 
The spectral energy density is given by
 \bea
u(\nu)=\int d\Omega u(\nu,\Omega).
 \ena
 
For an {\it isotropic} radiation field, $u(\nu,\Omega)={u(\nu)}/{4\pi}$, and we obtain
\bea
u'(\nu',\Omega')=\frac{u(\nu)}{4\pi}\mathcal{D}^{3}, {\rm ~ or~ } u'(\nu',\mu')=\frac{u(\nu)}{2}\mathcal{D}^{3},
\ena
where $\mathcal{D}=\gamma(1-\beta\mu)$ is the Doppler parameter, and the symmetry of emission over the azimuthal angle is assumed.

The  energy density of radiation from an isotropic source in the GF is then equal to
\bea
u'(\nu') d\nu'&=& \int d\mu' u(\nu',\mu') d\nu'=d\nu' \int_{-1}^{1} \frac{d\mu}{\mathcal{D}^{2}} \frac{u(\nu)}{2}\mathcal{D}^{3}=d\nu \frac{u(\nu)}{2}\int_{-1}^{1}{d\mu} \mathcal{D}^{2},\nonumber\\
&=&u(\nu) d\nu \times \frac{1}{2}\int_{-1}^{1}d\mu \gamma^{2}\left(1-\beta\mu\right)^{2}= u(\nu)d\nu \gamma^{2}\left(1+\frac{\beta^{2}}{3}\right),\label{eq:unu_iso}
\ena
where the integration over $\mu$ is taken from $-1$ to $1$, over the entire space of isotropic incident radiation. Cosmic microwave background (CMB) radiation and interstellar radiation field (ISRF) are examples for the isotropic radiation field. 

Let $u_{\rad}=\int d\nu u(\nu)$. The total energy density from the isotropic radiation field in the GF is then equal to
\bea
u'_ {\rad}&=& \int d\nu' u'(\nu')=\int d\nu u(\nu) \gamma^{2}\left(1+\frac{\beta^{2}}{3}\right)
= u_{\rad} \gamma^{2} \left(1+\frac{\beta^{2}}{3}\right).\label{eq:urad_isoa}
\ena

For a {\it unidirectional} radiation field from a point source (e.g., Sun) with luminosity $L=\int L_{\nu}d\nu$,\footnote{The spectral luminosity from a star of radius $R_{\star}$ and surface temperature $T_{\star}$ is $L_{\nu}=4\pi R_{\star}^{2}\pi B_{\nu}(T_{\star})$ where $B_{\nu}(T_{\star})$ is the Planck function, and $L=\int L_{\nu} d\nu = 4\pi R_{\star}^{2}\sigma T_{\star}^{4}$.} the specific spectral energy density $u(\nu,\mu)$ can be represented through a $\delta$ function as follows:
\bea
u(\nu,\mu)=\frac{L_{\nu}}{4\pi d^{2}c}\delta(\mu-\mu_{\rm gr}),
\ena
where $f={L_{\nu}}/{4\pi d^{2}}$ is the flux of radiation energy at the grain's position, and $\delta(\mu-\mu_{\rm gr})$ denotes the fact that the flux of radiation energy is defined as the energy propagating through an surface unit perpendicular to the the radiation direction (i.e., $\mu_{\rm gr}=\pm 1$). The spectral energy density from a point source  is given by the usual expression:
\bea
u(\nu)=\int d\mu u(\nu,\mu) =\frac{L_{\nu}}{4\pi d^{2}c}. \label{eq:unu_p}
\ena

In the GF, the incident radiation becomes focused, and we obtain $u(\nu')d\nu'= \int d \mu' u(\nu',\mu') d\nu'$ as the following:
\bea
u'(\nu')d\nu'&=& \int d\mu' u(\nu,\mu)\mathcal{D}^{3} (\mathcal{D}d\nu)= \int d\mu' \mathcal{D}^{4}\frac{L_{\nu}}{4\pi d^{2}c}\delta(\mu-\mu_{\rm gr})d\nu= \frac{L_{\nu}}{4\pi d^{2}c}d\nu\int_{-1}^{1} \frac{d \mu}{\mathcal{D}^{2}}\delta(\mu-\mu_{\rm gr}) \mathcal{D}^{4},\nonumber\\
&=& \frac{L_{\nu}}{4\pi d^{2}c} d\nu \int_{-1}^{1} d \mu \delta(\mu-\mu_{\rm gr}) \gamma^{2}(1-\beta\mu)^{2}=u(\nu)d\nu \int_{-1}^{1}d \mu\gamma^{2}(1-\beta\mu)^{2}\delta(\mu-\mu_{\rm gr})
=u(\nu)d\nu \gamma^{2}(1-\beta\mu_{\rm gr})^{2}.~~~\label{eq:unu_uni}
\ena
where $\mu_{\rm gr}=-1$ for the case the grain arrives from the direction opposite to the Sun. 

The total energy density from a point source in the GF is then equal to
\bea
u'_ {\rad}&=& \int d\nu' u'(\nu')=\int d\nu u(\nu) \gamma^{2}(1-\beta\mu_{\rm gr})^{2}
= u_{\rad} \gamma^{2} (1-\beta\mu_{\rm gr})^{2}.\label{eq:urad_unia}
\ena

In the assumption of monochromatic approximation, the spectral energy density of radiation can be represented through $u_{\rad}$ by a $\delta$ function $u(\nu)=u_{\rad}\delta(\nu-\bar{\nu})$,
where $h\bar{\nu}$ is the average energy of a photon, and $u_{\rad}(\erg \cm^{-3})$ is the total energy density. 

\section{C. Photoemission of electrons from dust by X-ray and collisional ionization}\label{apdx:pe}
\subsection{C1. Ionization Potential and Photoelectric Yield}\label{apx:A}
In the GF, optical or near-UV photons appear as X-ray. The interactions of photons with relativistic dust grains is similar to those of X-ray photons with stationary grains. 

The energy threshold for photoemission from the band structure in the solid is given by
\bea
h\nu_{\rm pet}= \IP(Z) ~for~ Z\ge -1,~~and~~
h\nu_{\rm pet}= \IP(Z) + E_{\min} ~for~ Z<-1.
\ena
where $\IP(Z)$ is the valence band ionization potential of a grain of charge $Z$, which is given by
\bea
{\IP}(Z)=W+\left(Z+\frac{1}{2}\right)\frac{e^{2}}{a}+(Z+2)\frac{e^{2}}{a}\frac{0.3\AA}{a} \label{eq:IPz}
\ena
with $W$ the work function (i.e., the energy needed to bring an electron from the highest occupied level in the neutral solid to infinity with zero kinetic energy),
and
\bea
E_{\min}=0 ~{\rm for}~ Z \ge -1,~{\rm and} ~
E_{\min}=\theta(\nu=|Z+1|)\left(1-0.3\left(\frac{a}{10\AA}\right)^{-0.45}|Z+1|^{-0.26}\right) ~{\rm for}~ Z<-1.
\ena
where $\theta$ is given in \cite{1987ApJ...320..803D}.

The detailed treatment of photoelectric emission by high-energy photons is presented in WDB06, where the photoelectric yield is represented through three functions $y_{0}, y_{1}$ and $y_{2}$:
\bea
Y(h\nu,Z,a)=y_{2}(h\nu,Z,a)\min[y0(\Theta)y_{1}(a,h\nu),1]. \label{eq:Y}
\ena

To compute $y_{0}, y_{1},$ and $y_{2}$, we need to known the absorption cross-section $\sigma_{i,s}$ for shell $s$ of element $i$ (DS96), which can be obtained using the FORTRAN code phfit2 via the link {\it http://www.pa.uky.edu/~verner/fortran.html}. Ionization potentials for inner electronic shells are listed in Table 1 of WDB06. 

\subsection{C2. Collisional ionization and Charging Rate}\label{apdx:Jcoll}
A relativistic incident particle can be assumed to move in a straight trajectory through the grain and impulsively transfers part of its momentum to atomic electrons that are assumed to be at rest during the interaction (Bohr approach). In such an impact approximation, the transverse momentum $\Delta p_{\perp}$ and kinetic energy transferred from the incident particle moving with velocity $v$ and impact factor $b$ to the atomic electron are given by
\bea
\Delta p_{\perp} = \frac{2Z_{P}e^{2}}{bv},~~T = \frac{(\Delta p_{\perp})^{2}}{2m_{e}}=\frac{2Z_{P}^{2}e^{4}}{m_{e}v^{2}b^{2}}.
\ena

The cross-section of the collision for kinetic energy in $T, T+dT$ becomes
\bea
d\sigma = \frac{2\pi Z_{P}^2 e^4}{m_{e}v^{2}}\left(\frac{1}{T^{2}}\right) dT.
\ena

For an isolated atom, the electron kinetic energy must be larger than the atom ionization potential $I$ to become a secondary electron. We integrate this equation from $I$ to $T_{\max}$ to obtain the total ionization cross-section:
\bea
\sigma_{I}\equiv \sigma_{T> I} = \frac{2\pi Z^2 e^4}{m_{e}v^{2}}\left(\frac{1}{I}-\frac{1}{T_{\max}}\right).\label{eq:signa_ion}
\ena


In order for a secondary electron to escape from the positively charged grain, its kinetic energy must be sufficiently large such that it will not be stopped within the grain before reaching the surface and to overcome the grain attractive surface potential. The first criterion is given by $T > {\rm IP}(Z)$, whereas the second criterion can be derived as follows.

Following \cite{1979ApJ...231...77D}, the energy loss and electron range can be represented by power laws:
\bea
\frac{dE}{dx} = -AE^{1-n},~~~R_{e}(E_{0})=(An)^{-1}E_{0}^{n},\label{eq:dEdx_e}
\ena
where $n=1.5$ and $(An)^{-1}\simeq 300{\rho}^{-0.85}(keV)^{-1.5}\AA$ for electron energy from $0.5-100$ keV. 

The stopping length must be larger than the grain radius:
\bea
l_{e}\sim \frac{E_{0}}{dE/dx}\ge a, {\rm so~} \frac{dE}{dx} \le \frac{E_{0}}{a}.
\ena

Using Equation (\ref{eq:dEdx_e}) for the above equation we get
\bea
E_{\rm loss}({\rm keV}) = (aA)^{1/n}=\left(a \rho^{0.85}/300n\AA\right)^{1/n}\simeq 3.17\hat{\rho}^{1.7/3}a_{-5}^{2/3}.
\ena

Therefore, the kinetic energy of secondary electrons required to escape from the grain is roughly given by $T_{\rm esc}={\rm max}(E_{\rm loss}, {\rm IP}(Z))$. For a projectile $Z_{P}$ bombarding a grain of size $a$ and charge $Z$, the ionization cross-section can be written as
\bea
\sigma_{I}(Z_{P}, a, Z) = \frac{2\pi Z_{P}^2 e^4}{m_{e}v^{2}}\left(\frac{1}{T_{\rm esc}}-\frac{1}{T_{\max}} \right)\approx \frac{2\pi Z_{P}^2 e^4}{m_{e}v^{2}}\left(\frac{1}{T_{\rm esc}}\right),\label{eq:sigmaion_Z}
\ena
where the dependence of $\sigma_{I}$ on $v$ is omitted and $T_{\max} \gg T_{\rm esc}$ for the relativistic case.

For a dust grain moving with velocity $v\sim c$ through the ambient gas, the total charging rate due to collisions with nuclei and electrons is respectively given by
\bea
J_{i}(v, a, Z) =\gamma  \sum_{P={\rm H, He}} n_{P}v\sigma_{I}(Z_{P}, a, Z)\times Z_{d}N_{d},~~
J_{e}(v, a, Z) = \gamma n_{e}v\sigma_{I}(1, a, Z)\times Z_{d}N_{d},\label{eq:dJe_dt}
\ena
where $Z_{d}$ is the mean atomic charge of the target atom and $N_{d}=n_{d}4\pi a^{3}/3$. It is noted that for the same relativistic velocity, the ionization cross-section for H proton and electron is the same (see Equation \ref{eq:sigmaion_Z}). Here, the charging by heavy nuclei is disregarded because their abundances are essentially more than one order of magnitude lower than $1/Z_{P}^{2}$. 


For the typical ISM comprising hydrogen ($90\%$) and He ($10\%$), the total number density of electrons in the gas and atoms is $n_{e} = n({\H}) + 2n({\rm He})=0.9n_{\gas} + 2\times 0.1\times n_{\gas} \simeq 1.21n_{\H}$ where the H proton density $n_{\H} = n(\H) =0.9n_{\gas}$ has been used. For this case, it is easily to show that $J_{i}\simeq 1.44\gamma  n_{\H}v\sigma_{I}(Z_{P}=1, a, Z)\times Z_{d}N_{d}$ and $J_{e}\simeq 1.22\gamma  n_{\H}v\sigma_{I}(Z_{P}=1, a, Z)\times Z_{d}N_{d}$. Thus, 
\bea
J_{\coll}(v, a, Z) = J_{i} + J_{e}= 2.66\gamma  n_{\H}v\sigma_{I}(Z_{P}=1, a, Z)\times Z_{d}N_{d}=2.39\gamma  n_{\gas}v\sigma_{I}(Z_{P}=1, a, Z)\times Z_{d}N_{d}
\ena

If collisional charging is dominant, then we can evaluate the column density of gas traversed by the grain upon Coulomb explosions ($Z=Z_{\max}$) as the following:
\bea
N_{\rm Coul}=n_{\gas}L_{\max}= n_{\gas}\gamma v \int_{0}^{Z_{\max}}\frac{dZ}{J_{\coll}(v, a, Z)}= \int_{0}^{Z_{\max}}\frac{dZ}{2.39\sigma_{I}(Z_{P}=1, a, Z)Z_{d}n_{d}4\pi a^{3}/3}
\ena

Assuming that grains are rapidly charged to $Z_{\max}\propto a^{2}$ and we take $T_{\rm esc}={\rm IP}(Z_{\max})=e\phi_{\max}$ then
\bea
J_{\coll}(v, a, Z_{\max})=2.39\gamma n_{\gas}v\left(\frac{2\pi e^{4}}{m_{e}v^{2}}\right)\frac{Z_{T}n_{d}4\pi a^{3}/3}{e\phi_{\max}}=2.39\gamma n_{\gas}v\left(\frac{8\pi^{2} e^{4}Z_{T}n_{d} }{3m_{e}v^{2}}\right)\frac{a^{4}}{e^{2}Z_{\max}}
\ena
where $\phi_{\max} =(4\pi \mathcal{S}_{\max})^{1/2}a \sim eZ_{\max}/a$.

Using $Z_{\max} = 7.4\times 10^{4}\left(\mathcal{S}_{\max}/10^{10}{\rm dyn~ cm}^{-2}\right)^{1/2}a_{-5}^{2}$ one obtains
\bea
N_{\rm Coul}= \left(\frac{3m_{e}v^{2}}{8\pi^{2}e^{4}Z_{T}n_{d} }\right)\frac{(7.4\times 10^{4}ea_{-5}^{2})^{2}}{2.39\times 10^{-20}a_{-5}^{4}}\simeq 3.1\times 10^{16}\beta^{2}
n_{23}^{-1}\left(\frac{10}{Z_{T}}\right)\left(\frac{\mathcal{S}_{\max}}{10^{10}{\rm dyn} \cm^{-2}}\right)\cm^{-2},
\ena
which reveals the independence of $N_{\rm Coul}$ on the grain size. 

If $T_{\rm esc}=E_{\rm loss}\propto a^{2/3}$, then we get $N_{\rm Coul}\propto Z_{\max}E_{\rm loss}/a^{3}\propto a^{-1/3}$, which indicates the increase of $N_{\rm Coul}$ with the decreasing grain size.

For ideal material, ion field emission becomes important for $Z>Z_{\max}$, and each subsequent ionization will result in the emission of one atom. The gas column density traversed by the grain up to complete destruction is
\bea
N_{\rm ife} &=& \gamma n_{\gas}v\frac{N_{d}}{J_{\coll}(Z_{\max})}=\frac{1}{2.39}\left(\frac{3m_{e}v^{2}}{8\pi^{2}e^{4}Z_{T}n_{d} }\right)\frac{4\pi a^{3}n_{d}}{3}\left(\frac{e^{2}Z_{\max}}{a^{4}} \right),\nonumber\\
& \simeq & 1.8\times 10^{20}\beta^{2}\left(\frac{10}{Z_{T}}\right)\left(\frac{\mathcal{S}_{\max}}{10^{10}{\rm dyn} \cm^{-2}}\right)^{1/2}a_{-5}\cm^{-2}.
\ena

\section{D. Velocity, momentum and energy of relativistic particle}

Let $E$ be the kinetic energy of the particle, $\gamma$ be the Lorentz factor and $\beta=v/c$. In this paper, the particle energy is implicitly referred to as its kinetic energy. The momentum of the particle is $p=\gamma\beta mc$. Let recall that the total energy of a particle is
\bea
E_{\rm tot}^{2}&=&E_{0}^{2}+ p^{2}c^{2}=E_{0}^{2}+\beta^{2}\gamma^{2}E_{0}^{2}=E_{0}^{2}(1+\beta^{2}\gamma^{2})=
E_{0}^{2}\gamma^{2},
\ena
where $E_{0}=mc^{2}$ is the particle energy at rest, and $\gamma^{2}=1/(1-\beta^{2})$ has been used.

The kinetic energy is defined as $E=E_{\rm tot}-E_{0}=E_{0}\left(\gamma-1\right)$. Thus, we have the followings:
\bea
\gamma = 1+ \frac{E}{E_{0}}, ~~\beta=\frac{\sqrt{E(E+2E_{0})}}{E+E_{0}}.
\ena

\bibliography{ms.bbl}
\end{document}